%% file: PRC.tex
\documentclass[onecolumn,aps,prc,floatfix,superscriptaddress]{revtex4-1}

\usepackage{hyperref}
\usepackage[utf8]{inputenc}

\usepackage{graphicx}
\usepackage{dcolumn}
\usepackage{amsmath}
\usepackage{supertabular}
\usepackage{multirow}

\usepackage{color}
\usepackage{cancel}
\usepackage{ulem}
\usepackage{array}
\usepackage{tabularx}

\usepackage{amsmath}
\usepackage{amssymb}
\usepackage{xcolor}
\definecolor{myOrange}{rgb}{1,0.5,0.}
\definecolor{myGreen}{rgb}{0.0,0.6,0.1}
\newcommand{\removetext}[1]{}

\newcommand{\aaa}{\ensuremath{\mathrm{A}+\mathrm{A}}}
\newcommand{\AuAu}{\ensuremath{\mathrm{Au}+\mathrm{Au}}}
\newcommand{\PbPb}{\ensuremath{\mathrm{Pb}+\mathrm{Pb}}}
\newcommand{\pPb}{\ensuremath{\mathrm{p}+\mathrm{Pb}}}
\newcommand{\pp}{\ensuremath{{p+p}}} 

\newcommand{\sqrtsNN}{\ensuremath{\sqrt{s_\mathrm{NN}}}}
\newcommand{\sqrts}{\ensuremath{\sqrt{s}}}
\newcommand{\invnb}{\ensuremath{\mathrm{nb}^{-1}}}
\newcommand{\invpb}{\ensuremath{\mathrm{pb}^{-1}}}

\newcommand{\pizero}{\ensuremath{\pi^0}}

\newcommand{\gammadir}{\ensuremath{\gamma_\mathrm{dir}}}
\newcommand{\gammarich}{\ensuremath{\gamma_\mathrm{rich}}}

\newcommand{\pT}{\ensuremath{p_\mathrm{T}}}

\newcommand{\pTtrack}{\ensuremath{p_\mathrm{T,track}}}

\newcommand{\pTjet}{\ensuremath{p_\mathrm{T,jet}}}

\newcommand{\pTraw}{\ensuremath{p_\mathrm{T,jet}^\mathrm{raw}}}
\newcommand{\pTrawi}{\ensuremath{p_\mathrm{T,jet}^{\mathrm{raw},i}}}

\newcommand{\pTreco}{\ensuremath{p_\mathrm{T,jet}^\mathrm{reco,ch}}}
\newcommand{\pTrecoi}{\ensuremath{p_\mathrm{T,jet}^{\mathrm{reco},i}}}

\newcommand{\dpT}{\ensuremath{\delta{p}_\mathrm{T}}}

\newcommand{\pTjetpart}{\ensuremath{p_\mathrm{T,jet}^\mathrm{part,ch}}}
\newcommand{\pTjetdet}{\ensuremath{p_\mathrm{T,jet}^\mathrm{det,ch}}}

\newcommand{\Ajet}{\ensuremath{A_\mathrm{jet}}}
\newcommand{\Ajeti}{\ensuremath{A_\mathrm{jet}^{i}}}

\newcommand{\rhoA}{\ensuremath{\rho\Ajet}}

\newcommand{\ET}{\ensuremath{E_\mathrm{T}}}
\newcommand{\ETtrig}{\ensuremath{E_\mathrm{T}^\mathrm{trig}}}

\newcommand{\ETmin}{\ensuremath{E_\mathrm{T}^\mathrm{min}}}
\newcommand{\ETmax}{\ensuremath{E_\mathrm{T}^\mathrm{max}}}

\newcommand{\pTtrig}{\ensuremath{p_\mathrm{T}^\mathrm{trig}}}
\newcommand{\pTjetch}{\ensuremath{p_\mathrm{ T,jet}^\mathrm{ch}}}

\newcommand{\gev}{\ensuremath{\mathrm{GeV/}c}}

\newcommand{\zvtx}{\ensuremath{z_\mathrm{vtx}}}

\newcommand{\kT}{\ensuremath{k_\mathrm{T}}}
\newcommand{\antikT}{\ensuremath{\mathrm{anti-}k_\mathrm{T}}}

\newcommand{\rr}{\ensuremath{R}}

\newcommand{\dphi}{\ensuremath{\Delta\varphi}}

\newcommand{\fME}{\ensuremath{f^\mathrm{ME}}}

\newcommand{\Ntrig}{\ensuremath{N_\mathrm{trig}}}
\newcommand{\NtrigAA}{\ensuremath{N_\mathrm{trig}^\mathrm{A+A}}}
\newcommand{\Njet}{\ensuremath{N_\mathrm{jet}}}

\newcommand{\AAtoTrig}{\ensuremath{\mathrm{A+A}\rightarrow\mathrm{trig}}}
\newcommand{\AAtoTrigjet}{\ensuremath{\mathrm{A+A}\rightarrow\mathrm{trig+jet}}}

\newcommand{\dNjetdpTdphiAA}{\ensuremath{\frac{{\rm d}N_\mathrm{jet}^\mathrm{A+A}}{\mathrm{d}\pTjetch\mathrm{d}\dphi}}}

\newcommand{\etajet}{\ensuremath{\eta_\mathrm{jet}}}
\newcommand{\etatrack}{\ensuremath{\eta_\mathrm{track}}}

\newcommand{\YpTjetchR}[1]{\ensuremath{Y^{#1}(\pTjetch,\rr)}}

\newcommand{\YpTjetchRsmall}{\ensuremath{Y^{\mathrm{A+A}}(\pTjetch,\mathrm{small\ }\rr)}}
\newcommand{\YpTjetchRlarge}{\ensuremath{Y^{\mathrm{A+A}}(\pTjetch,\mathrm{large\ }\rr)}}

\newcommand{\Rbroadening}{\ensuremath{\mathfrak{R}^{\frac{{\rm small}-R}{{\rm large}-R}}}}

\newcommand{\Rbrtwofive}{\ensuremath{\mathfrak{R}^{0.2/0.5}}}
\newcommand{\Rinstr}{\ensuremath{R_\mathrm{instr}}}
\newcommand{\Rbkgd}{\ensuremath{R_\mathrm{bkg}}}
\newcommand{\Rtot}{\ensuremath{R_\mathrm{total}}}
\newcommand{\Rinstrarg}{\ensuremath{\Rinstr\left(\pTjetdet,\pTjetpart\right)}}
\newcommand{\Rbkgdarg}{\ensuremath{\Rbkgd\left(\pTreco,\pTjetdet\right)}}
\newcommand{\Rtotarg}{\ensuremath{\Rtot\left(\pTreco,\pTjetpart\right)}}
\newcommand{\Rpurity}{\ensuremath{R_\mathrm{impurity}}}
\newcommand{\zT}{\ensuremath{z_\mathrm{T}}}

\newcommand{\qT}{\ensuremath{q_\mathrm{T}}}
\newcommand{\qTjet}{\ensuremath{q_\mathrm{T,jet}}}

\newcommand{\vtwo}{\ensuremath{v_{2}}}
\newcommand{\ICP}{\ensuremath{I_\mathrm{CP}}}

\newcommand{\IAA}{\ensuremath{{I}_\mathrm{AA}}}

\newcommand{\Qsq}{\ensuremath{{Q}^{2}}}

\newcommand{\qhat}{\ensuremath{\hat{q}}}

\newcommand{\LtwoG}{\ensuremath{\mathrm{L2Gamma}}}

\newcommand{\Dgammadir}{\ensuremath{\mathcal{D}_{\gammadir}}}
\newcommand{\Dgammarich}{\ensuremath{\mathcal{D}_{\gammarich}}}
\newcommand{\Dpizero}{\ensuremath{\mathcal{D}_{\pizero}}}

\newcommand{\ztilde}{\ensuremath{\tilde{z}}}

\newcommand{\eTgenPar}{\ensuremath{E_{\mathrm{T}}^{\mathrm{part}}}}
\newcommand{\eTgenParPi}{\ensuremath{E_{\mathrm{T}}^{\pi^{0},\mathrm{part}}}}
\newcommand{\eTgenParGa}{\ensuremath{E_{\mathrm{T}}^{\gamma,\mathrm{part}}}}
\newcommand{\eTgenDet}{\ensuremath{E_{\mathrm{T}}^{\mathrm{det-clust}}}}
\newcommand{\eTgenDetPi}{\ensuremath{E_{\mathrm{T}}^{\pi^{0},\mathrm{det-clust}}}}
\newcommand{\eTgenDetGa}{\ensuremath{E_{\mathrm{T}}^{\gamma,\mathrm{det-clust}}}}
\newcommand{\eTgenMat}{\ensuremath{E_{\mathrm{T}}^{\mathrm{part-match}}}}
\newcommand{\eTgenMatPi}{\ensuremath{E_{\mathrm{T}}^{\pi^{0},\mathrm{part-match}}}}
\newcommand{\eTgenMatGa}{\ensuremath{E_{\mathrm{T}}^{\gamma,\mathrm{part-match}}}}

\newcommand{\Wargs}{\ensuremath{W\left(\eTgenPar;\eTgenDet\right)}}

\begin{document}

%\linenumbers

\title{Semi-inclusive direct photon+jet and \pizero+jet correlations measured in \pp\ and central \AuAu\ collisions at $\sqrtsNN=200$ GeV}
\bigskip

%%%% _______Author list______and Affiliation________

\input{./authorlist_short}

\collaboration{STAR Collaboration}\noaffiliation

%%%_______________

\date{ \today}

%\author{The STAR Collaboration}

%\affiliation{}

\begin{abstract}

The STAR experiment at RHIC reports new measurements of jet quenching based on the semi-inclusive distribution of charged-particle jets recoiling from direct photon ($\gamma_{\rm dir}$) and neutral pion ($\pi^{0}$) triggers in pp and central  Au+Au collisions at $\sqrt{s_{\rm NN}}=200$ GeV, for triggers in the range $9<E_{\rm T}^{\rm trig}<20$ GeV. The datasets have integrated luminosities of 3.9$ {\rm nb}^{-1}$ for \AuAu\ and 23$ {\rm pb}^{-1}$ for pp collisions.  Jets are reconstructed using the anti-$k_{\rm T}$ algorithm with resolution parameters $R$=0.2 and 0.5. The large uncorrelated jet background in central Au+Au collisions is corrected using a mixed-event approach, which enables precise charged-particle jet measurements at low transverse momentum $p_{\rm  T,jet}^{\rm ch}$ and large $R$. Recoil-jet distributions are reported in the range $p_{\rm  T,jet}^{\rm ch}<25$ \gev. Comparison of the distributions measured in pp and Au+Au collisions reveals strong medium-induced jet yield suppression for $R=0.2$, with markedly less suppression for $R=0.5$. Comparison is also made to theoretical models incorporating jet quenching. These data provide new insight into the mechanisms underlying jet quenching and the angular dependence of medium-induced jet-energy transport, and provide new constraints on modelling such effects. % for arxiv abstract 
%The STAR experiment at RHIC reports new measurements of jet quenching based on the semi-inclusive distribution of charged-particle jets recoiling from direct photon (\gammadir) and neutral pion (\pizero) triggers in \pp\ and central \AuAu\ collisions at $\sqrtsNN=200$ GeV, for triggers in the range $9<\ETtrig<20$ GeV. The datasets have integrated luminosities of 3.9 \invnb\ for \AuAu\ and 23 \invpb\ for \pp\ collisions.  Jets are reconstructed using the \antikT\ algorithm with resolution parameters $\rr=0.2$ and 0.5. The large uncorrelated jet background in central \AuAu\ collisions is corrected using a mixed-event approach, which enables precise charged-particle jet measurements at low transverse momentum \pTjetch\ and large \rr. Recoil-jet distributions are reported in the range \pTjetch$<25$ \gev. Comparison of the distributions measured in \pp\ and \AuAu\ collisions reveals strong medium-induced jet yield suppression for $\rr=0.2$, with markedly less suppression for $\rr=0.5$. Comparison is also made to theoretical models incorporating jet quenching. These data provide new insight into the mechanisms underlying jet quenching and the angular dependence of medium-induced jet-energy transport, and provide new constraints on modelling such effects. 
\end{abstract}

%\tableofcontents

\maketitle

%%%%%%%%%%%%%%%%%%%%%%%%%%     Introduction --- Prarambha 
\section{Introduction}
\label{sect:Intro}

Matter under conditions of extreme temperature and density forms a Quark-Gluon Plasma (QGP), a state of matter in which the predominant degrees of freedom are sub-hadronic~\cite{Collins:1974ky,Shuryak:1977ut,Busza:2018rrf,Harris:2023tti}. QGP filled the universe a few microseconds after the Big Bang, and is generated and studied today in collisions of heavy atomic nuclei at the Relativistic Heavy Ion Collider (RHIC) and the Large Hadron Collider (LHC). Measurements at these facilities, and their comparison to theoretical calculations, show that the QGP is a fluid with very low specific viscosity~\cite{Heinz:2013th,JETSCAPE:2020mzn,JETSCAPE:2020shq,Nijs:2020roc} that is opaque to the passage of energetic color charges~\cite{Cunqueiro:2021wls}.

Jets in hadronic collisions are generated by hard (high momentum transfer \Qsq) interactions of incoming quarks and gluons (partons). The scattered partons are initially highly virtual, and evolve by radiating gluons to produce a collimated parton shower. Theoretical calculations based on perturbative Quantum Chromodynamics (pQCD) are in excellent agreement with jet measurements in \pp\ collisions~\cite{Abelev:2006uq,Adamczyk:2016okk,Abelev:2013fn,Acharya:2019jyg,Aad:2014vwa,Khachatryan:2016mlc}.

Jets are generated in nuclear collisions and interact with the QGP, with components of the jet shower scattering elastically and radiatively~\cite{Majumder:2010qh,Cunqueiro:2021wls,Apolinario:2022vzg}. Such interactions modify observed jet production rates and properties relative to those of jets generated in vacuum (``jet quenching'')~\cite{Bjorken:1982tu,Wang:1994fx,Baier:1998yf,Gyulassy:2000fs}: jet energy loss due to transport of energy out of the jet cone, corresponding to yield suppression at fixed transverse momentum (\pT); modification of intra-jet structure; and deflection of the jet centroid (acoplanarity). 

Measurements of reconstructed jets in heavy-ion collisions are challenging, due to the large and complex background from uncorrelated processes~\cite{Cunqueiro:2021wls}. Jet quenching was initially observed in measurements of inclusive production and correlations of high-\pT\ hadrons, the leading fragments of jets, which are more readily measurable over the large background in heavy-ion collisions ~\cite{Adler:2002xw,Adler:2002tq,Adams:2003kv,Adams:2006yt,Adamczyk:2013jei,Adcox:2001jp,Adare:2012wg,Adare:2010ry,Aamodt:2011vg,Abelev:2012hxa,ALICE:2016gso,ATLAS:2015qmb,CMS:2012aa,Chatrchyan:2012wg}. The comparison of high-\pT\ hadron yield suppression measurements with theoretical calculations constrains the QGP transport parameter \qhat, which characterizes the momentum exchange  of jets with the QGP medium~\cite{Armesto:2009zi,Burke:2013yra,JETSCAPE:2021ehl,Apolinario:2022vzg}. High-\pT\ hadron yields are sensitive primarily to the magnitude of in-medium jet energy loss. More detailed understanding of the mechanisms underlying jet quenching requires measurements based on reconstructed jets, e.g.
~\cite{Abelev:2013kqa,Adam:2015ewa,Adam:2015doa,Acharya:2017goa,Acharya:2019jyg,Acharya:2019djg,ALICE:2023jye,ALICE:2023qve,Aad:2010bu,ATLAS:2012tjt,Aad:2014bxa,ATLAS:2014dtd,Aaboud:2018anc,ATLAS:2023iad,CMS:2011iwn,CMS:2012ulu,Chatrchyan:2012nia,Chatrchyan:2012gt,Khachatryan:2016jfl,Sirunyan:2017bsd,Sirunyan:2017qhf,CMS:2018jco,Sirunyan:2018qec,STAR:2016dfv,Adamczyk:2017yhe,Adam:2020wen}.

An important channel for such measurements is the coincidence of a neutral vector boson (direct photon \gammadir, $Z$) and recoiling jet~\cite{David:2019wpt}. Direct photons are defined as photons produced directly in hard interactions, rather than produced through hadronic decays or partonic fragmentation. Vector bosons are colorless and do not interact significantly with the QGP~\cite{PHENIX:2012jbv,ALICE:2015xmh,ATLAS:2015rlt,CMS:2020oen,
CMS:2011zfr,ATLAS:2012qdj}; their transverse energy (\ET) thereby provides a reference scale for precise measurements of jet quenching~\cite{Wang:1996yh}. For theoretical calculation of \gammadir+jet production at RHIC energies, at leading perturbative order (LO) the predominant mechanism is QCD Compton scattering ($qg\rightarrow\gamma{q}$), in which the direct photon \ET\ and recoiling quark jet \pT\ are balanced~\cite{Wang:1996yh}. 
However, Next-to-Leading Order (NLO) contributions, which generate photon-jet \pT-imbalance, are sizable even in vacuum~\cite{Dai:2012am}. Measurement of  the \gammadir+jet channel is nevertheless a key element of the program for precise jet quenching measurements at RHIC and the LHC. 

Coincidence  \gammadir+jet measurements have been carried out for \pp\ and \PbPb\ collisions at the LHC~\cite{Chatrchyan:2012gt,Sirunyan:2017qhf,Aaboud:2018anc,ATLAS:2023iad}. The \pp\ measurements exhibit significant \pT\ imbalance, consistent with NLO calculations~\cite{Dai:2012am}. Jet quenching in \aaa\ collisions is found to generate additional \pT\ imbalance, which however is smaller than in-vacuum NLO effects~\cite{Sirunyan:2017qhf,Aaboud:2018anc}. 
$Z$-boson+hadron correlations have also been measured in \pp\ and \PbPb\ collisions at the LHC~\cite{ATLAS:2020wmg,CMS:2021otx}. While \gammadir+hadron measurements have been reported for \pp\ and \AuAu\ collisions at $\sqrtsNN=200$ GeV~\cite{STAR:2009ojv,STAR:2016jdz,PHENIX:2009cvn,PHENIX:2010vgy,PHENIX:2012aba,PHENIX:2020alr}, there are currently no \gammadir+jet measurements reported at RHIC.

Measurements of the semi-inclusive distribution of charged-particle jets recoiling from a high-\pT\ hadron trigger have been carried out for \pp, \PbPb, and \pPb\ collisions at the LHC~\cite{Adam:2015doa, ALICE:2017svf}, and for \AuAu\ collisions at RHIC~\cite{Adamczyk:2017yhe}. In the semi-inclusive approach, selection bias is induced solely by the choice of trigger. Model studies suggest that, due to the effects of jet quenching, observed high-\pT\ hadrons in nuclear collisions are generated predominantly at the surface of the QGP, headed outwards~\cite{Baier:2002tc,Zhang:2007ja,Renk:2012ve,Bass:2008rv}.
For semi-inclusive observables with a hadron trigger, the recoiling jet population may therefore have on average larger in-medium path length than an unbiased population~\cite{Adam:2015doa,Adamczyk:2017yhe}. Semi-inclusive measurements incorporating both \gammadir\ and \pizero\ triggers can provide direct comparison of recoil-jet populations with different quark/gluon relative populations and different in-medium path-length distributions~\cite{STAR:2016jdz}.

Calorimetric jet measurements have limited spatial resolution due to instrumental effects, and jet reconstruction based on charged particles (also called ``track jets'') provides greater measurement precision (e.g.~\cite{Lee:2023xzv}). Charged-particle jets are not Infrared and Collinear (IRC) safe, however, and analytic pQCD calculations of charged-particle jet distributions require non-perturbative Track Functions determined from data~\cite{Lee:2023xzv}. In \pp\ collisions, charged-particle jets have been used to measure inclusive and coincidence observables down to very low \pTjetch ($\approx$ 10 \gev), with calculations from Monte Carlo generators at NLO accuracy found to be in good agreement~\cite{ALICE:2023waz,ALICE:2023jye}. In \aaa\ collisions, charged-particle jet analyses incorporating a statistical approach to the mitigation of jet background ~\cite{Adam:2015doa,Adamczyk:2017yhe,ALICE:2023jye, ALICE:2023qve} have measured jet quenching at both RHIC and the LHC, over much broader phase space at low \pTjet\ and large \rr\ than achievable by current calorimetric jet measurements~\cite{Chatrchyan:2012gt,Sirunyan:2017qhf,Aaboud:2018anc,ATLAS:2023iad}. 

This manuscript and the companion Letter~\cite{STAR:2023pal} report the first \gammadir+jet and \pizero+jet coincidence measurements in \pp\ and central \AuAu\ collisions at $\sqrtsNN=200$ GeV. Charged-particle jets are reconstructed  by the \antikT\ algorithm ~\cite{Cacciari:2008gp} with resolution parameter $\rr=0.2$ and 0.5. Semi-inclusive distributions of charged-particle jets~\cite{Adam:2015doa,Adamczyk:2017yhe} recoiling from identified \gammadir\ and \pizero\ trigger particles are reported, for triggers in the range $9<\ETtrig<20$ GeV. The analysis is based on previous developments to discriminate \gammadir\ and \pizero\ at high \pT\ using calorimetric shower shape~\cite{STAR:2009ojv,STAR:2016jdz}, and to mitigate the complex jet background in central \AuAu\ collisions by means of mixed events (ME)~\cite{Adamczyk:2017yhe}. While calorimetric shower shape can identify \pizero\ triggers efficiently on an event--wise basis, direct-photon trigger yields are determined by a statistical subtraction of the decay and fragmentation photons. This subtraction utilizes a measurement of the direct-photon purity made with the assumption that direct photons have no near-side associated hadrons above a specified \pT\ threshold. 

This manuscript presents details of the analysis, experimental results, and their comparison to theoretical model calculations. The companion Letter highlights selected results and model comparisons, and discusses the constraints that these measurements impose on the physical mechanisms underlying jet quenching. 

The manuscript is organized as follows: 
Sect.~\ref{Sec:DetectorDataset} presents the detector, dataset, and offline reconstruction;
Sect.~\ref{Sec:Theory} presents the theoretical model calculations that are compared to data;
Sect.~\ref{sect:AnalysisOverview} presents an overview of the analysis;
Sect.~\ref{Sect:TrigMeasurement} presents the photon and \pizero\ measurements;
Sect.~\ref{sect:JetReco} presents the jet reconstruction;
Sect.~\ref{sect:RawDistr} presents the raw coincidence distributions;
Sect.~\ref{sect:GammajetConversion} presents the extraction of \gammadir+jet distributions;
Sect.~\ref{sect:TrigRes} presents trigger-resolution effects;
Sect.~\ref{sect:Corrections} presents corrections to the recoil \pTjet\ distributions;
Sect.~\ref{sect:SysUncert} presents systematic uncertainties;
Sect.~\ref{sect:Closure} presents closure tests;
%Sect.~\ref{sect:CorrDist} presents the corrected recoil \pTjet\ distributions;
Sect.~\ref{sect:Results} presents the physics results;
and Sect.~\ref{sect:Summary} presents a summary and outlook.

%%%%%%%%%%%%Dataset, offline analysis, and simulations}
\section{Dataset, offline analysis, and simulations}
\label{Sec:DetectorDataset}

The Solenoidal Tracker at RHIC (STAR) detector is described in Ref.~\cite{Ackermann:2002ad}. STAR has a large solenoidal magnet with uniform magnetic field of strength of 0.5 T aligned with the beam direction, and detectors for triggering, tracking, electromagnetic calorimetry, and particle identification. This analysis utilizes the Barrel Electromagnetic Calorimeter (BEMC)~\cite{Beddo:2002zx} for triggering and EM shower measurements; the Barrel Shower Maximum Detector (BSMD)~\cite{STAR:2016jdz} for $\gamma$/\pizero\ discrimination; the Time Projection Chamber (TPC)~\cite{Anderson:2003ur} for charged-particle tracking; and the forward Vertex Position Detector (VPD)~\cite{Llope:2014nva} and Zero-Degree Calorimeter (ZDC)~\cite{Adler:2000bd} for triggering. The BEMC, BSMD, and TPC have acceptance in pseudo-rapidity $|\eta|<$1.0, with full azimuthal coverage.  

The data for \pp\ collisions at $\sqrts=200$ GeV were recorded during the high luminosity phase of the 2009 RHIC run. The data were selected online using the \LtwoG\ trigger.
%, which enhances the recorded rate of energetic photons and \pizero. %This trigger is initiated at Level-0 timing based on the presence of at least one BEMC tower with transverse energy \ET\ greater than 4.2 GeV. 
The first layer of event selection, after receiving data from the detectors, passes the event to the \LtwoG\ trigger based on the presence of at least one BEMC ``high tower'' with transverse energy \ET\ greater than 4.2 GeV.
The \LtwoG\ selection 
%is applied as a filter to the triggered event, 
requires the presence of a contiguous cluster of $3\times3$ BEMC towers, whose two most energetic towers have sum $\ET>7.44$ GeV. In total, 11.3 million \pp\ collision events were recorded which satisfy the \LtwoG\ trigger requirements. After offline cuts for event quality, the \pp\ integrated luminosity accepted for analysis is 23 \invpb. 

The data for \AuAu\ collisions at $\sqrtsNN=200$ GeV were recorded during the 2014 RHIC run, likewise utilizing the \LtwoG\ selection.  In this case, the %level-0%
first layer of triggering is a high-tower \ET\ threshold of 5.9 GeV. No centrality-specific selection is applied at the online trigger level. Pileup is suppressed offline by comparing two different measurements of the event-vertex position along the beam direction: reconstructed using tracks (see below), and measured by VPD timing. Events are rejected if the difference between the vertex-position measurements is greater than 6 cm along the beam direction, or if the track-based vertex position has distance larger than 70 cm from the nominal center of STAR in the beam direction. Since the Time-of-Flight (TOF) detector has faster readout than the TPC, the fraction of TPC tracks matched to TOF hits is also used to reject uncorrelated events. Additionally, events with noisy trigger towers, defined as those having a hit frequency greater than 5$\sigma$ from the average of all towers, are rejected. After offline cuts for event quality, the integrated luminosity for \AuAu\ collisions accepted for the analysis is 3.9 \invnb. 

Charged-particle tracks are reconstructed offline using hits in the TPC. No rejection of charged leptons is implemented; the predominant source of leptons in the charged-track population is conversion in detector material, whose effects on the analysis are found to be negligible in detailed simulations. Global tracks are defined as having at least 10 hits in the TPC and a 3-dimensional distance of closest approach to the primary vertex (DCA) less than 3 cm. The primary vertex position along the beam direction, denoted \zvtx, is determined from the extrapolation of a selection of global tracks to the beamline. The jet analysis uses primary tracks, which include the primary vertex in the momentum fit, with the additional requirements of at least 15 hits in the TPC and DCA$<1$ cm.

For \AuAu\ collisions, centrality is determined using the multiplicity of global tracks within $|\etatrack| < 0.5$~\cite{STAR:2018zdy}. The luminosity dependence of the centrality determination is accounted for, using the average instantaneous coincidence rate in the ZDCs. %Variation in acceptance is corrected by determining the average of the uncorrected charged track multiplicity as a function of \zvtx\ and applying a re-weighting to the charged track multiplicity distribution, which renders the average charged track multiplicity uniform as a function of \zvtx.
The variation in acceptance with vertex position is corrected by 
%scaling the highest bin of multiplicity distributions as a function of \zvtx\ and 
weighting to the charged-track multiplicity distribution, to render the average charged-track multiplicity uniform as a function of \zvtx~\cite{STAR:2018zdy}. In this analysis, ``central \AuAu\ collisions'' refers to the event population with multiplicity in the highest 15\% interval of the multiplicity distribution for all \AuAu\ hadronic interactions. 

The presence of large jet activity in an event could bias event ordering using this centrality metric relative to unbiased event ordering based on the (unobservable) impact parameter of the collision. However, the excellent agreement of a Glauber model calculation with the global track multiplicity distribution shown in Ref.~\cite{STAR:2018zdy} indicates that such bias is negligible for this analysis.

Jet reconstruction uses primary tracks. The primary track acceptance is $|\etatrack|<1.0$ and  $0.2<\pTtrack<30$ \gev, over the full azimuth. Additional primary-track selection criteria require the track to have more than 15 TPC hits, and the ratio of the number of its measured hits over the maximum possible (depending upon track geometry) to be greater than 52$\%$. The primary-track yield with $\pTtrack>30$ \gev\ is negligible for both collision systems. The primary-track efficiency is 55\% for $\pTtrack=0.2$ \gev, 60\%  for $\pTtrack=0.5$ \gev, and 72\% for $\pTtrack>2$ \gev\ in central \AuAu\ collisions; and 70\% for $\pTtrack=0.25$ \gev\ and 82\% for $\pTtrack>1$ \gev\ in \pp\ collisions (see also ~\cite{STAR:2021kjt}). The primary-track momentum resolution is approximately $\sigma_{\pT}/\pT=0.01\cdot\pT$ [\gev], for both central \AuAu\ and \pp\ collisions.

Detector simulations are carried out using GEANT3~\cite{Brun:1119728} with a detailed model of the STAR detector that is adjusted for each running period to incorporate the detector configuration of the corresponding dataset. Simulations based on the PYTHIA Monte Carlo event generator are used both to calculate the instrumental response to correct for detector effects and for physics studies. Events produced by Monte Carlo generation without detector simulations are denoted ``particle-level,'' whereas such events including detector simulations are denoted ``detector-level.'' Jet reconstruction is performed on all final-state charged particles at the particle-level.

Several PYTHIA tunes are employed: PYTHIA-6~\cite{Sjostrand:2006za} Perugia 0~\cite{Skands:2010ak}; PYTHIA-6 Perugia 2012 STAR tune~\cite{STAR:2019yqm}; and PYTHIA-8 Monash~\cite{Sjostrand:2007gs}. The reason for using multiple PYTHIA tunes is historical, since different elements of the analysis, such as detector-level simulations, were carried out at different times by different sub-groups. The tune used for each analysis element is specified. Physics studies focus on comparison to  PYTHIA-6 Perugia 2012 STAR tune (denoted ``PYTHIA-6 STAR tune'').

The distribution of hadrons in reconstructed jets is compared between data and a calculation based on PYTHIA-6 STAR tune~\cite{STAR:2022hqg}, with good agreement found for both longitudinal and transverse hadron distributions. This agreement indicates that PYTHIA-6 STAR tune provides an accurate description of jet fragmentation into hadrons at RHIC energies.

%===================================================

\section{Theoretical calculations}
\label{Sec:Theory}

The following theoretical calculations which incorporate jet interactions with the QGP are compared with the data:

\begin{itemize}

\item Jet-fluid model~\cite{Chang:2016gjp}: Initial conditions and the jet shower evolution are generated by PYTHIA. The QGP is modeled using 2+1 dimensional VISHNU hydrodynamics~\cite{Song:2010aq}. Jet propagation takes into account both elastic and inelastic energy loss, with latter calculated using the Higher Twist (HT) formalism, as well as \pT-broadening. Only the jet shower is modified; jet-induced medium excitation (``back-reaction'') is not considered.

\item Linear Boltzmann Transport (LBT) model~\cite{Luo:2018pto}:   Jet generation and shower evolution are calculated using PYTHIA. The QGP is modeled using CLVisc 3+1 dimensional viscous hydrodynamics~\cite{Pang:2014ipa} with AMPT initial conditions~\cite{Lin:2004en}. Jet propagation in the QGP includes both elastic and inelastic processes, with the latter calculated using the HT formalism. Back-reaction to the QGP medium arises from rescattering of thermal recoil partons with jet shower partons. The linear approximation corresponds to such rescatterings only occurring between jet and thermal medium partons. Energy and momentum are explicitly conserved.

\item Coupled Linear Boltzmann Transport and hydro (CoLBT-hydro) model \cite{Zhao:2021vmu,Chen:2017zte}: modifies the LBT model by including jet-induced medium excitation in the hydrodynamic evolution, thereby relaxing the linear approximation. This extends the applicability of the model to the case in which the medium excitation is similar in magnitude to the local thermal parton density.

\item Soft Collinear Effective Theory (SCET) model~\cite{Dai:2012am,Kang:2017xnc}: Photon-jet distributions  are calculated using PYTHIA. The QGP has initial energy density  proportional on average to participant number density, with event-by-event fluctuations, and evolves according to 2+1 dimensional viscous hydrodynamics. Jet-medium inelastic interactions are calculated using the GLV approach at first order in opacity, with a soft gluon emission approximation. Elastic interactions are also included. Energy loss is controlled by a variable coupling strength, with relative energy loss of quarks and gluons proportional to their color charges. Excitation of the QGP medium is equated with the jet energy transported beyond the jet cone radius \rr. 
%\comment{PMJ June 26}{Nihar: what coupling strength $g$ is used in Fig 2? Maybe add that number to the legend?} 

\item Hybrid model~\cite{Casalderrey-Solana:2014bpa,Casalderrey-Solana:2018wrw}. Hard processes are generated by PYTHIA, with the jet shower evolved by PYTHIA down to a cutoff virtuality $Q_0=1$ GeV. The jet showers are embedded in a hydrodynamically expanding medium and interact with it non-pertubatively, using  a holographic expression for energy loss derived from gauge/gravity duality. Energy and momentum lost by the interacting jet are absorbed by the QGP, manifesting as a wake of soft particles surrounding the jet. Generation of the wake can optionally be switched off.

\end{itemize}

All theoretical calculations that are compared to data take into account the experimental acceptance, including the \rr-dependent recoil jet acceptance (Sect.~\ref{Sect:RecoilJet}).

\section{Analysis overview}
\label{sect:AnalysisOverview}

STAR has previously reported measurements of \gammadir+hadron and \pizero+hadron coincidence distributions in \pp\ and \AuAu\  collisions~\cite{STAR:2016jdz,STAR:2009ojv}, and semi-inclusive distributions of charged-particle jets recoiling from high-\pT\ charged-hadron triggers in \AuAu\ collisions~\cite{Adamczyk:2017yhe}, both at $\sqrtsNN=200$ GeV. These analyses required the development of new  procedures to enable systematically well-controlled measurements in the large-background environment of central \AuAu\ collisions. 

The analysis reported here combines the two approaches to measure semi-inclusive distributions of charged-particle jets recoiling from high-\pT\ \gammadir\  and \pizero\ triggers in \pp\ and \AuAu\ collisions at $\sqrtsNN=200$ GeV. While we follow closely the procedures 
in Refs.~\cite{STAR:2016jdz,Adamczyk:2017yhe}, this analysis utilizes different datasets, and its performance and systematic uncertainties must therefore be evaluated in full. This section presents an overview of the analysis strategy and approach, with details provided in subsequent sections.

%-------------------------------------------
\subsection{Semi-inclusive distributions}
\label{Sect:SemiInclDistr}

Jet measurements in the high-multiplicity environment of a heavy-ion collision are complex, due to the large uncorrelated background. Reconstructed jets in this environment therefore can have contributions from physical jets generated by the hard process of interest (inclusive or triggered-coincidence production); physical jets generated by other hard processes (multiple high-\Qsq\ partonic interactions, or MPIs, which are uncorrelated due to QCD factorization); and a combinatorial component comprising  hadrons from soft (low-\Qsq) interactions. The analysis must remove the jet yield component that is not correlated with the process of interest (which may not include MPIs, depending upon the application), and then correct the \pTjet\ smearing due to the residual uncorrelated contribution~\cite{Adam:2015doa,Adamczyk:2017yhe,STAR:2020xiv}.

Semi-inclusive measurements enable data-driven procedures for such corrections over broad phase space~\cite{Adam:2015doa,Adamczyk:2017yhe}. In this approach, the event selection and jet measurement are carried out in separate steps. Events are first selected using a simple trigger condition, with the trigger particle distribution sampled inclusively. Jet reconstruction is then carried out on the selected events and the number of jets in a defined recoil acceptance is counted, without any requirements on the jet population beyond acceptance. 

The measured semi-inclusive distribution, corresponding to the distribution of recoil jets normalized to the number of trigger particles, is equivalent in the absence of background to the ratio of two production cross sections~\cite{Adam:2015doa,Adamczyk:2017yhe}:

%---------
\begin{equation}
\left(\frac{1}{\NtrigAA}\cdot\dNjetdpTdphiAA\right)\Bigg\vert_{\pTtrig}
= \left(
\frac{1}{\sigma^{\AAtoTrig}} \cdot
\frac{{\rm d}\sigma^{\AAtoTrigjet}}{\mathrm{d}\pTjetch\mathrm{d}\dphi}\right)
\Bigg\vert_{\pTtrig},
\label{eq:hJetDefinition}
\end{equation}
%-------

\noindent
where \pTjetch\ is the recoil-jet transverse momentum; \dphi\ is the azimuthal angular separation between trigger and recoil jet; $\sigma^{\AAtoTrig}$ is the inclusive production cross section for the trigger particle; and $\sigma^{\AAtoTrigjet}$ is the production cross section for both trigger and recoil jet in the acceptance. While the $\eta$-acceptance of the recoil is $|\etajet|<1.0-\rr$ (Sect.~\ref{Sect:RecoilJet}), the distribution in Eq.~\ref{eq:hJetDefinition} is normalized to unit $\eta$ (notation not shown). 

Since event selection is based on the presence of a trigger particle, preferred trigger particles satisfy two criteria: they are precisely measurable in central \aaa\ collisions without complex event-wise background corrections; and the cross sections in Eq.~\ref{eq:hJetDefinition} can be calculated perturbatively. Preferred trigger particles are therefore high-\pT\ charged hadrons, \pizero, direct photons, and (at the LHC) $Z$-bosons. 

%----
\begin{figure}[htb!]
\centering
\includegraphics[width=0.75\textwidth]{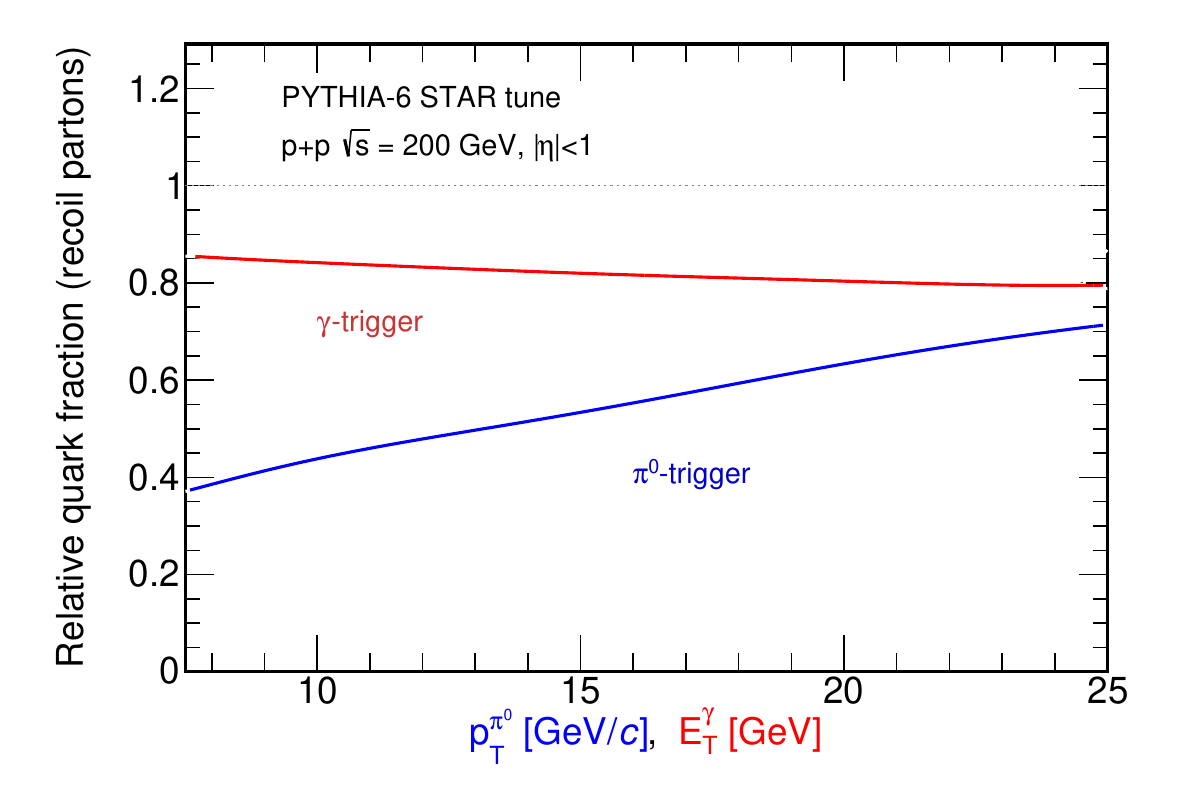}
\caption{Relative fraction of quarks ($q/(q+g)$) recoiling from  a \pizero\ or \gammadir\ trigger as a function of \pTtrig\ or \ETtrig, calculated using
 PYTHIA-6 STAR tune for \pp\ collisions at $\sqrts=200$ GeV. Acceptance is $|\eta|<1.0$ for both trigger particle and recoil partons. 
}
\label{Fig:qgFrac}
\end{figure}
%---

Since jets do not play a role in event selection in the semi-inclusive approach, selection bias is induced solely by the choice of trigger~\cite{Adam:2015doa,Adamczyk:2017yhe}. This analysis employs both \gammadir\ and \pizero\ triggers, whose event-selection biases are expected to differ. These differences include:

\begin{itemize}
    
\item {\bf Relative $q/g$ contribution to recoil-jet population:} High-\ET\ \gammadir\ and high-\pT\ \pizero\ are generated by different distributions of partonic scattering processes~\cite{PHENIX:2010vgy}. Figure~\ref{Fig:qgFrac} shows a calculation illustrating this,  using events generated by PYTHIA for \pp\ collisions at $\sqrts=200$ GeV. Events containing a high-\ET\ \gammadir\ or high-\pT\ \pizero\  are selected and the relative fraction of quarks in the population of recoil partons is determined.\footnote{This calculation assumes an LO ($2\rightarrow2$) production process and identifies the ``recoil parton'' in each PYTHIA event as the highest-energy parton within the same recoil acceptance as the data analysis. Accounting for additional QCD effects requires theoretical considerations that are beyond the scope of this study.} The quark fraction of recoil partons is seen to be significantly larger for \gammadir\ than for \pizero\ triggers at the same value of \ETtrig\ or \pTtrig, for the range considered in this analysis ($9<\ETtrig<20$ GeV, see Sect.~\ref{Sect:TrigMeasurement}).

\item {\bf In-medium path-length distribution of recoil jets:} Due to the interplay of the inclusive \pTjet-distribution, \pizero\ fragmentation function, and energy loss effects, observed high-\pT\ \pizero\ are expected to arise predominantly from the fragmentation of jets that have experienced little energy loss due to quenching, corresponding predominantly to jets generated at the periphery of the QGP fireball and headed outward~\cite{Baier:2002tc,Zhang:2007ja,Renk:2012ve,Bass:2008rv}. In that case, the in-medium path-length distribution of jets recoiling from \pizero\ triggers is biased towards larger values than that of the measured inclusive jet population~\cite{Adam:2015doa,Adamczyk:2017yhe,He:2024rcv}. Since \gammadir\ do not lose energy due to jet quenching, jets recoiling from \gammadir\ triggers do not have such a path-length bias; their in-medium path-length distribution therefore is the same as that of the measured inclusive jet population.

\item {\bf Event-Plane (EP) orientation:} A correlation between high-\pT\ hadron production and EP orientation (observed second-order plane of symmetry constructed from the soft hadron distribution) has been observed in \AuAu\ collisions~\cite{PHENIX:2014yml}. In this analysis, for \pizero\ triggers such a correlation could generate a dependence of background yield on the relative orientation of the trigger and EP, and possibly an in-medium path-length bias for the recoiling jet population. The EP-orientation dependence of the recoil spectrum was measured in the previous $h$+jet analyses, with negligible dependence found for central \aaa\ collisions at both RHIC and the LHC~\cite{Adam:2015doa,Adamczyk:2017yhe}. No such dependence is expected for \gammadir\ triggers.

\end{itemize}

\noindent
We note that the selection-bias effects we discuss here are expectations based on model calculations. We return to these expectations in Sect.~\ref{sect:Results}, when comparing the measured recoil-jet distributions for \gammadir\ and \pizero\ triggers.

%-------------------------------------------
\subsection{Recoil-jet measurement}
\label{Sect:RecoilJet}

For events accepted by the trigger selection, jet reconstruction is carried out with charged particles using the \antikT\ algorithm~\cite{Cacciari:2008gp} with resolution parameters $\rr=0.2$ and 0.5. The \antikT\ algorithm is utilized because of the insensitivity of its reconstructed jet shapes to underlying event fluctuations~\cite{Cacciari:2008gp,Adam:2015doa,Adamczyk:2017yhe}.

Detailed discussion of jet reconstruction is presented in Sect.~\ref{sect:JetReco}. The recoil jet acceptance is $\dphi \in [3\pi/4,5\pi/4]$ and $|\etajet|<1.0-\rr$, where \etajet\ is the pseudo-rapidity of the jet centroid. Corrections to the two-dimensional raw distribution (analogous to Eq.~\ref{eq:hJetDefinition}) are applied to account for uncorrelated jet yield and \pTjet-smearing due to background. These corrections are carried out separately (Sect.~\ref{sect:Corrections}): first a ``vertical'' correction which subtracts uncorrelated-jet yield, followed by a ``horizontal'' correction to account for \pTjet-smearing of the distribution of correlated yield.

The goal of the analysis is to measure the recoil-jet yield over broad phase space, including low \pTjet\ and large \rr\ where the correlated signal yield is small relative to background. The correction for uncorrelated yield must be fully data-driven, in order to achieve high systematic precision for such small signal/background. Since individual hadrons cannot be attributed uniquely to a trigger-correlated recoil jet or to uncorrelated background, such corrections must be carried out statistically, i.e. at the level of ensemble-averaged distributions~\cite{Adam:2015doa,Adamczyk:2017yhe}. 

Two related but distinct statistical correction approaches have been developed for this purpose. The ALICE collaboration has utilized two different ranges of \pTtrig\ (``Signal'' and ``Reference'')~\cite{Adam:2015doa} , while the STAR collaboration has utilized a distribution constructed from mixed events in place of the Reference distribution~\cite{Adamczyk:2017yhe}. In both approaches, the raw measured value of \pTjet\ is first shifted by the median level of uncorrelated \pT-density $\rho$ in each event, scaled by the jet area~\cite{Cacciari:2007fd}. The resulting distribution, $\pTreco$, is peaked near $\pTreco=0$, with approximately half of all jet candidates having $\pTreco<0$~\cite{Adam:2015doa,Adamczyk:2017yhe} (see Sect.~\ref{sect:JetReco}).

The Signal and Reference (ALICE) or ME (STAR) distributions in the region of large negative \pTreco\ are found to have identical shape with high precision, indicating that this region is strongly dominated by uncorrelated yield~\cite{Adam:2015doa,Adamczyk:2017yhe} and is therefore the optimal region for precise normalization of the Reference or ME distribution. After normalization, the correlated recoil-jet yield distribution is determined by subtracting the Reference or ME distribution from the Signal distribution. The difference distribution is subsequently corrected for smearing in \pTjet\ and \dphi\ by unfolding. This statistical approach has enabled systematically well-controlled measurements of recoil-jet yield over the entire physically-allowed phase space, without the need to impose a fragmentation bias on the recoil-jet population to suppress backgrounds~\cite{Adamczyk:2017yhe}.

The ALICE \pTtrig-difference and STAR ME approaches to uncorrelated background correction are compared in Ref.~\cite{Adamczyk:2017yhe}. The key distinction between them is that the trigger-difference approach subtracts all uncorrelated recoil yield, including physical jets generated by MPIs, whereas the ME approach does not subtract jets arising from MPIs. This difference was found to have negligible effect for $h$+jet correlations in central \AuAu\ collisions at $\sqrtsNN=200$ GeV~\cite{Adamczyk:2017yhe}.

%-------------------------------------------
\subsection{Observables}
\label{Sect:Observables}

The trigger-normalized jet yield in the recoil acceptance is defined as

\begin{equation}
\YpTjetchR{} = \frac{1}{\Ntrig}
\int_{3\pi/4}^{5\pi/4} \mathrm{d}\dphi \left[  \frac{\mathrm{d}^{2}\Njet(\rr)}{\mathrm{d}\pTjetch\mathrm{d}\dphi}
 \right]_{\ETtrig\in [\ETmin,\ETmax]},
\label{Eq:YieldExpression}
\end{equation}

\noindent
where \rr\ is the jet reconstruction resolution parameter. The yield is also differential in jet pseudo-rapidity $\eta_\mathrm{jet}$, which is not specified for clarity. This definition applies to both \pp\ and \AuAu\ collisions, and for both raw and fully corrected distributions. The collision system and level of correction are indicated by superscripts, where needed.

Jet energy loss due to quenching is measured by the ratio of corrected trigger-normalized jet yields in central \AuAu\ and \pp\ collisions,

%----
\begin{equation}
\IAA = \frac{\YpTjetchR{\AuAu}}{\YpTjetchR{\pp}}.
\label{Eq:IAA}
\end{equation}
%----

\noindent
Measurements of \IAA\ using $h$+jet correlations have been reported at RHIC and at the LHC ~\cite{Adam:2015doa,Adamczyk:2017yhe}. 

The modification of transverse jet structure due to quenching is measured by the ratio of trigger-normalized recoil jet yields at small and large \rr\ for the same collision system,

%----
\begin{equation}
\Rbroadening = \frac{\YpTjetchRsmall}{\YpTjetchRlarge},
\label{Eq:Intrajet}
\end{equation}
%----

\noindent
where $\mathrm{A+A}$ here refers to either \pp\ or \AuAu. Recoil jet distributions are observed to be uniform in $\eta$ within the \rr-dependent acceptance, $|\etajet|<1.0-\rr$, and the reported distributions are normalized per unit $\eta$. The different acceptance for different \rr\ consequently has negligible effect on the measurement of Eq.~\ref{Eq:Intrajet}. 

The observable \Rbroadening\ is of interest theoretically for probing jet shapes and jet transverse structure because common factors cancel in the ratio, enabling high-order pQCD calculations~\cite{Soyez:2011np,Dasgupta:2016bnd}. Experimentally, some systematic uncertainties likewise cancel in the ratio~\cite{ALICE:2013yva}. The value of \Rbroadening\ for inclusive jet production in \pp\ collisions is observed to be less than unity (defined with the smaller-\rr\ yield in the numerator)~\cite{ALICE:2013yva,ALICE:2019qyj,ALICE:2019wqv,ALICE:2023ama,CMS:2014nvq,CMS:2020caw}, reflecting jet shapes in vacuum and providing precise tests of pQCD calculations~\cite{Dasgupta:2016bnd}. For semi-inclusive $h$+jet correlations in \pp\ collisions, the value of \Rbroadening\ is also observed to be below unity, likewise reflecting jet shapes in vacuum~\cite{Adam:2015doa,ALICE:2023jye}. Calculations based on PYTHIA are found to be in good agreement with these measurements~\cite{Adam:2015doa,ALICE:2023jye}, though a pQCD calculation at NLO disagrees by a factor $\approx2$~\cite{Adam:2015doa}. Comparison of \Rbroadening\ for inclusive and semi-inclusive recoil jet populations in \pp\ collisions at the LHC reveals strikingly different behavior at low \pTjetch\ that is well--reproduced by PYTHIA, indicating a difference in jet production mechanisms below the \pT-scale of \pTtrig~\cite{ALICE:2023jye} (see also Sect.~\ref{subsect:Jetshape}).

The measurement of \Rbroadening\ has also been used to search for medium-induced changes in jet shape in heavy ion collisions, for an inclusive jet population at RHIC~\cite{STAR:2020xiv}, and for semi-inclusive recoil jet populations from $h$+jet correlations at both RHIC and the LHC~\cite{Adam:2015doa,Adamczyk:2017yhe,ALICE:2023jye}. While significant medium-induced modification was not observed initially~\cite{STAR:2020xiv,Adam:2015doa,Adamczyk:2017yhe}, the recent high-statistics ALICE measurement reports medium-induced modification of this ratio
~\cite{ALICE:2023jye}, which is discussed together with results from this analysis in Sect.~\ref{subsect:Jetshape}.
\section{Measurement of {$\gamma$} and {\pizero} triggers}
\label{Sect:TrigMeasurement}

The measurement of \gammadir\ and \pizero\ triggers in this analysis uses the procedures described in Ref.~\cite{STAR:2016jdz}. Events are selected online using the \LtwoG\ trigger. Offline analysis further selects events containing electromagnetic showers with $9<\ETtrig<20$~GeV measured in one or two adjacent BEMC towers. 

Offline discrimination of $\gamma$ and \pizero\ showers utilizes the transverse-shower shape measured by the BSMD. In the \ETtrig\ range, the \pizero-decay opening angle is sufficiently small that the two photons are likely to hit the same tower, but broad enough that the transverse shower profile measured in the BSMD is distinguishable from that of a single photon. The threshold value of $\ETtrig=9$ GeV is optimal for identifying \pizero\ triggers, while the \pizero\ identification efficiency for $\ETtrig>20$ GeV is significantly lower. The threshold value of 9 GeV is large enough that there is significant recoil-jet yield, while providing sufficient dynamic range for the measurement of three bins in \ETtrig.  

Each BEMC tower subtends $\Delta\eta \times \Delta\phi = 0.05 \times 0.05$ and overlaps with 7.5 BSMD strips in $\eta$ and 7.5 BSMD strips in $\phi$. For the shower-shape analysis, BSMD $\eta$ and $\phi$ strips are clustered~\cite{STAR:2016jdz}, and clusters whose central strips overlap geometrically with the BEMC trigger tower are identified. The intersection of the central $\eta$ and $\phi$ BSMD strips in a cluster, determined by the strip with the largest energy relative to $\pm 7$ adjacent strips in both $\eta$ and $\phi$, localizes the centroid of the shower with spatial resolution of 0.007 radians. The BEMC tower overlapping the centroid is identified as the central tower of the cluster if it also has an energy of at least 6 GeV. If the shower centroid falls within 0.018 radians of the edge of the triggering tower, then the nearest tower is also included in the energy measurement. Charged-particle tracks are projected to the face of the BEMC, and trigger towers that contain the projection of a track with $\pT>2$ \gev\ are rejected.

The shower shape is quantified by the Transverse Shower Profile (TSP),  defined as $ \mathrm{TSP} = E_{\rm tower}/\sum_{i} e_{i} r_{i}^{1.5}$, where $E_{\rm tower}$ is the BEMC trigger-tower energy; $e_{i}$ is the energy of the $i$-th BSMD strip; $r_{i}$ is the distance between the $i$-th strip and the center of the cluster, and the sum runs over the 15 strips in $\eta$ and $\phi$ which define the BSMD cluster. The value of $r_i$ for the central strip is taken to be half the distance between strips. The value of the exponent, 1.5, was determined by optimizing $\gamma$/\pizero\ discrimination using a GEANT simulation~\cite{STAR:2016jdz}. 

%---
\begin{figure}[htb!]
\centering
\includegraphics[width=0.85\textwidth]{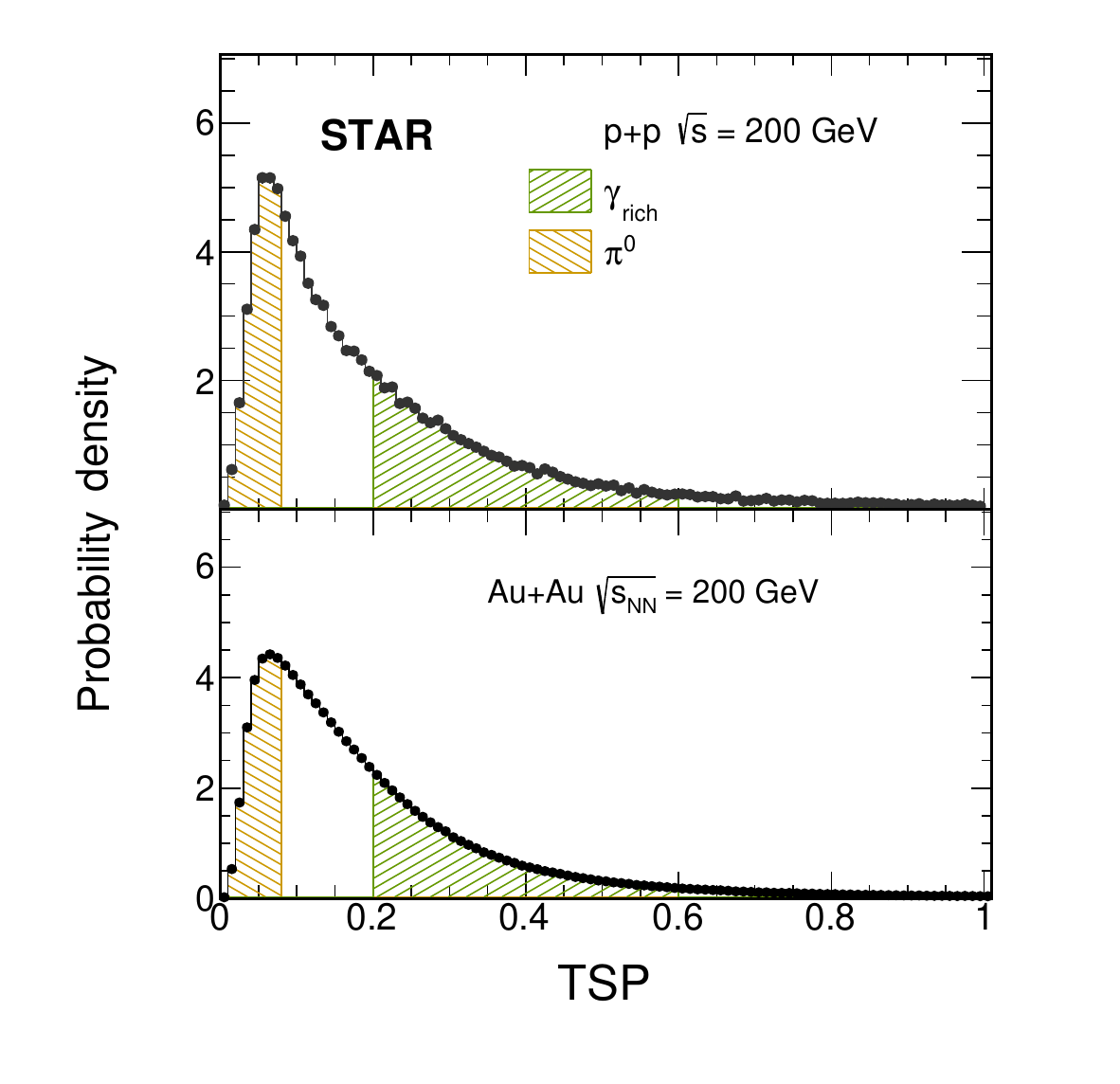}
\caption{TSP distributions for \pp\ (upper) and central \AuAu\ (lower) collisions. The orange and green regions show the selection for the \pizero\ and \gammarich\ triggers, respectively.}
\label{Fig:TSP}
\end{figure}
%---

Figure~\ref{Fig:TSP} shows the TSP distribution measured in central \AuAu\ and \pp\ collisions. The TSP cut labeled \pizero, corresponding to $\mathrm{TSP}<0.08$, generates a shower population with \pizero\ purity greater than 95\%. However, such high purity cannot be achieved for \gammadir\ based on the TSP observable. Rather, a \gammarich\ population is selected by $0.2 < \mathrm{TSP} < 0.6$, which contains direct photons together with an admixture of \pizero\ triggers and single photons from \pizero\ and $\eta$ decays. The fraction of non-direct photons in the \gammarich\ population is measured using the relative rate of near-side hadrons  as a function of \ETtrig, where proximity is determined from the relative azimuthal angle $\Delta\phi$ (Sect.~\ref{sect:GammajetConversion}). 

The central \AuAu\ dataset contains about 56000 \pizero\  and 134000 \gammarich\ triggers, while the \pp\ dataset contains about 18000 \pizero\  and 24000 \gammarich\ triggers. These trigger populations are then divided into \ETtrig\ bins with ranges [9,11], [11,15] and [15,20] GeV. Table~\ref{Tab:TrigStats} shows the number of triggers in each \ETtrig\ bin.

%---
\begin{table}[htb!]
\centering
\caption{Number of events containing a trigger passing all event- and trigger-selection criteria for \pizero\ and \gammarich\ selections, for the \ETtrig\ bins used in the analysis.}
\begin{tabular}{| c | c | c | c | c |}
\hline
\multirow{2}{*}{\ETtrig [GeV]} & \multicolumn{2}{c|}{\pp}  & \multicolumn{2}{c|}{central \AuAu} \\
\cline{2-5}
 &\pizero & \gammarich & \pizero & \gammarich \\
\hline
[9,11] & 12869 & 15232  & 40437 & 83804\\
{[}11,15{]} & 4918 & 7328 & 14262 & 42279\\
{[}15,20{]} & 699 & 1522 & 1553 & 8353\\
\hline 
%Total & 18426 & 24082 \\
%\hline
\end{tabular}
\label{Tab:TrigStats}
\end{table}
%---

The semi-inclusive recoil-jet distribution corresponding to pure \gammadir\ triggers is then determined statistically from the \pizero- and \gammarich-triggered recoil-jet distributions (Sect.~\ref{sect:GammajetConversion}).
\section{Jet reconstruction}
\label{sect:JetReco}

Jet reconstruction utilizes the approach described in Ref.~\cite{Adamczyk:2017yhe}, for both \pp\ and central \AuAu\ collisions. As discussed in Sect.~\ref{Sect:RecoilJet}, in central \AuAu\ collisions the correction for uncorrelated background yield is carried out statistically, i.e. at the level of ensemble-averaged distributions, without discrimination of individual jets as signal or background. The output of the jet reconstruction algorithm is therefore referred to as a population of ``jet candidates.''

The analysis is carried out in multiple steps, each with a different specification of \pTjet. To distinguish the different types of jet candidate we utilize the notation defined in Ref.~\cite{Adamczyk:2017yhe}: \pTraw\ refers to \pTjet\ generated by the jet reconstruction algorithm; \pTreco\ is \pTraw\ adjusted by \rhoA\ (Eq.~\ref{Eq:pTreco}), the estimated uncorrelated background contribution ; and \pTjetch\ is \pT\ of jet candidates after full correction for the effects of instrumental response and background fluctuations. For the simulation of \pp\ collisions, \pTjetpart\ is the reconstructed jet energy at the particle-level  and \pTjetdet\ is at the detector level.

Jets are reconstructed from accepted charged tracks using the \kT~\cite{Cacciari:2011ma} and \antikT~\cite{Cacciari:2008gp} algorithms with the $E$-recombination scheme~\cite{Cacciari:2011ma}. Jet area is measured using the FastJet active-area approach with ghost-particle area of 0.01~\cite{Cacciari:2008gn}.

In both \pp\ and central \AuAu\ collisions, jet reconstruction is performed twice. The first reconstruction pass uses the \kT\ algorithm with \rr\ = 0.2 and 0.5 to estimate the background transverse-momentum density in the event~\cite{Cacciari:2007fd},

%---
\begin{equation}
\rho=\mathrm{median}\left\{ \frac{\pTrawi}{\Ajeti} \right\},
\label{eq:rho}
\end{equation}
%---

\noindent
where $i$ labels the jets in the event with centroid $|\eta|<1-R$, and \pTrawi\ and \Ajeti\ are the transverse momentum and area of jet $i$. The median is calculated by excluding the hardest jet in the event for \pp\ collisions, and the two hardest jets for central \AuAu\ collisions~\cite{Adamczyk:2017yhe}.

The second jet reconstruction pass is then carried out using the \antikT\ algorithm with $\rr=0.2$ and 0.5. Jet candidates from the second pass are accepted for further analysis if their centroid satisfies $|\eta|<1-R$. The value of \pTraw\ from this pass is adjusted for the estimated background transverse-momentum density according to

%---
\begin{equation}
\pTrecoi = \pTrawi - \rho \times \Ajeti.
\label{Eq:pTreco}
\end{equation}
%---

\noindent
In \pp\ collisions, for $\rr=0.2$ the most probable value of $\rho$ is zero, while for $\rr=0.5$ the term $\rho \times \Ajeti$ is rarely greater than 1 \gev. For central \AuAu\ collisions the value of $\rho$ varies between 15  and 40 \gev\ (Fig.~\ref{Fig:RhoSEME}), largely independent of the choice of \rr\ for calculating $\rho$.

%------------------------------------------------
\subsection{Event mixing}
\label{sect:EvtMixing}

Correction for the uncorrelated jet yield in central \AuAu\ collisions is carried out by subtracting the normalized ME \pTreco\ distribution from that of the same-event (SE) population, following the procedure described in~\cite{Adamczyk:2017yhe}. The event mixing uses centrality-selected but otherwise unbiased events from the same dataset.

In brief, event mixing creates synthetic events made up of tracks from real events, but with each track in every mixed event originating from a different real event. The ME population thereby reproduces the detailed features of the real event population at the ensemble--averaged level but does not contain any multi--hadron correlations, including jets. The analysis is then carried out on the ME population, with the resulting distributions of ``jets'' providing the distribution of purely combinatorial jet candidates contributing to the jet candidate population in the analysis of real data. 

In this analysis, event mixing utilizes 5 bins in multiplicity, 18 bins in \zvtx, 2 bins in event-plane orientation, and 3 bins in run-averaged luminosity, for a total of 540 distinct mixing classes. The large number of bins in \zvtx\ is required to accommodate the effective variation of the STAR acceptance within the dataset, due to the broad RHIC interaction diamond. No jet candidates are excluded from the $\rho$ calculation for ME events~\cite{Adamczyk:2017yhe}.

%------------------------------------------------
\subsection{{$\rho$} alignment}
\label{sect:rhoalignment}

At low \pTreco\ the uncorrelated background is a large fraction of the total yield, and the ME subtraction procedure therefore corresponds to taking a small difference of two large numbers. Since the recoil-jet distribution varies rapidly as a function of \pTreco, for accurate subtraction it is therefore crucial that the \pTreco\ scales of the SE and ME distribution are well-aligned~\cite{Adamczyk:2017yhe}. 

%----
  \begin{figure}[htb!]
\centering
\includegraphics[width=0.7\textwidth]{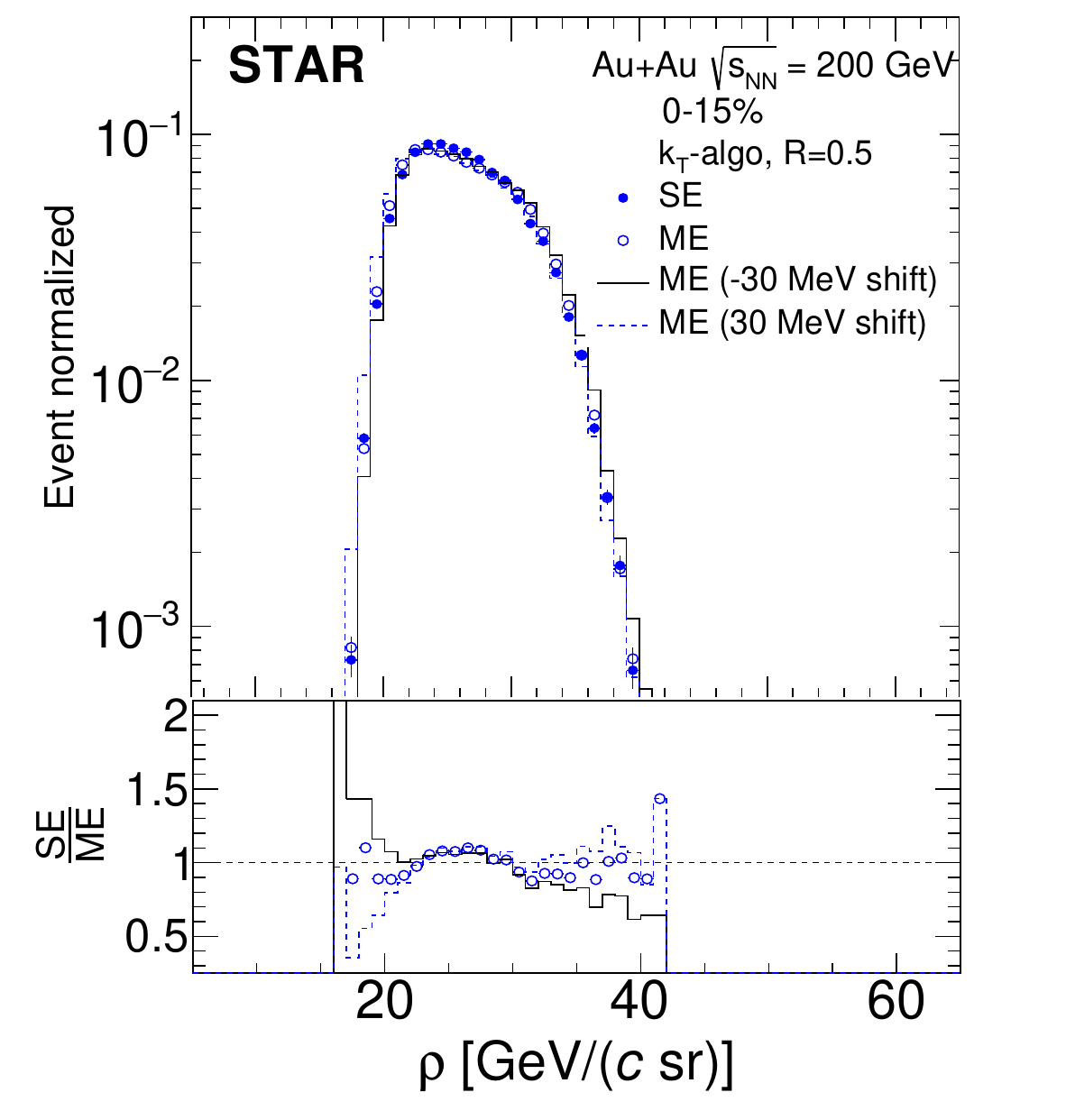}
\caption{Upper panel: distribution of $\rho$ ($\rr=0.5$) in \gammarich-triggered central \AuAu\ collisions for the SE and ME populations, for $\rho$-shift values 0 and $\pm30$ MeV (Sect.~\ref{sect:RawDistr}). Lower panel: ratio of the SE and ME distributions.}
\label{Fig:RhoSEME}
\end{figure}
%----

The definition of $\rho$ (Eq.~\ref{eq:rho}) contains arbitrary choices, and $\rho$ is not an absolutely defined physical quantity. In addition, the $\rho$ distributions for the SE and ME populations may differ, since the hard jet distribution of the SE and ME event populations is different. As shown in Ref.~\cite{Adamczyk:2017yhe}, precise alignment of the \pTreco\ scale for the SE and ME populations can be achieved in a data-driven way by shifting the ME $\rho$ distribution horizontally by a small amount. Figure~\ref{Fig:RhoSEME} shows the distribution of $\rho$ for the SE and ME populations in \gammarich-triggered central \AuAu\ collisions, with shifts of the ME $\rho$ distribution of zero and $\pm30$ MeV/$c$. The ratio of $\rho$ distributions for the SE and ME populations are seen to be tilted for non-zero shift, indicating a misalignment of the distributions. This study shows that a $\rho$-shift value of zero is close to optimal for this event selection, with a precision better than $\pm30$ MeV/$c$. 

%----
\begin{figure}[htb!]
\centering
\includegraphics[width=0.72\textwidth]{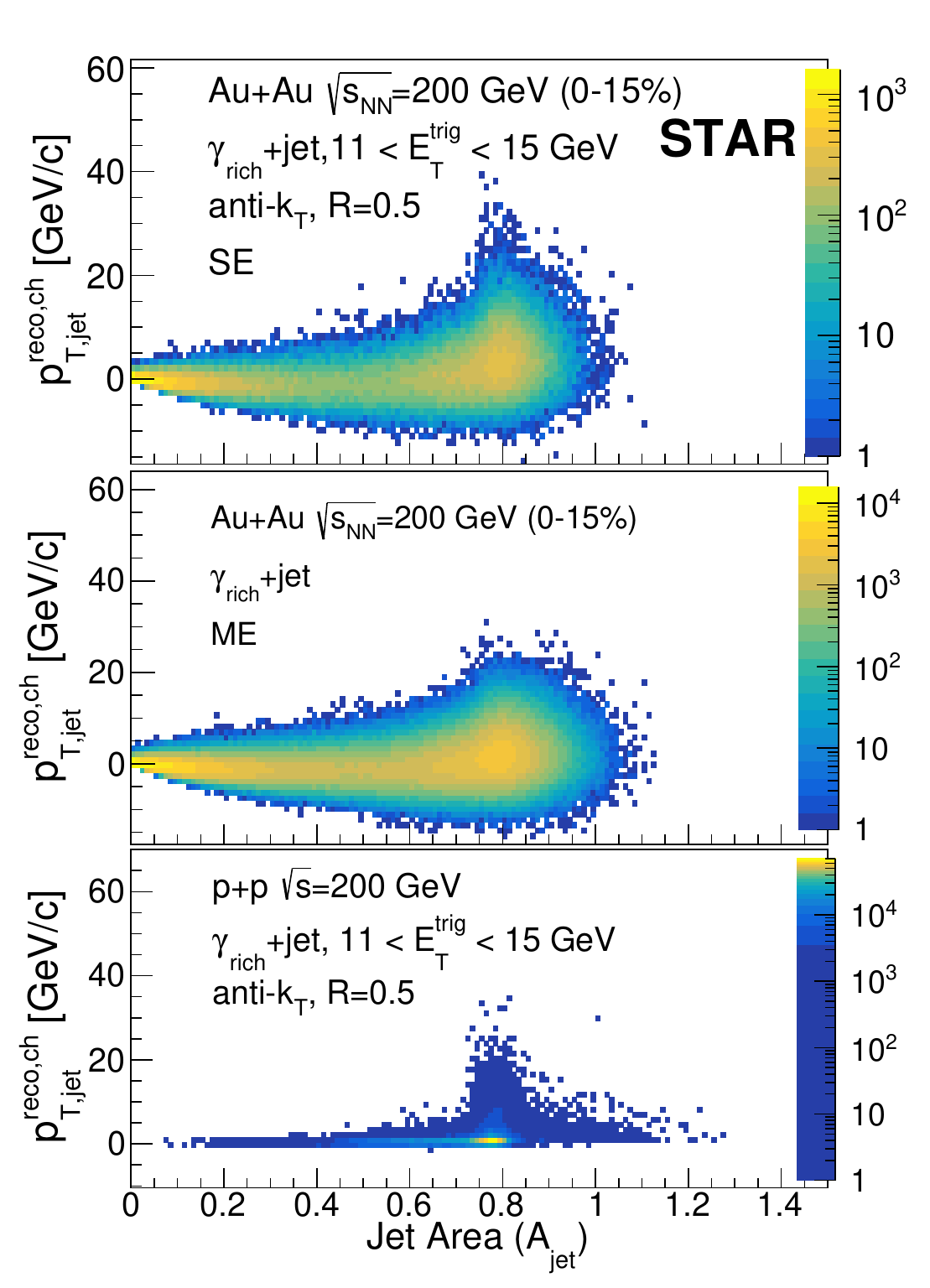}
\caption{Distribution of \pTreco\ vs. jet area for \gammarich-triggered recoil jets, $\rr=0.5$. Top: SE for central \AuAu\ collisions; middle: ME for central \AuAu\ collisions; bottom: \pp\ collisions.}
\label{Fig:pTrecoVsArea}
\end{figure}
%----

%------------------------------------------------
\subsection{Jet area}
\label{sect:jetarea}

%---
\begin{figure}[htb!]
\centering
\includegraphics[width=0.95\textwidth]{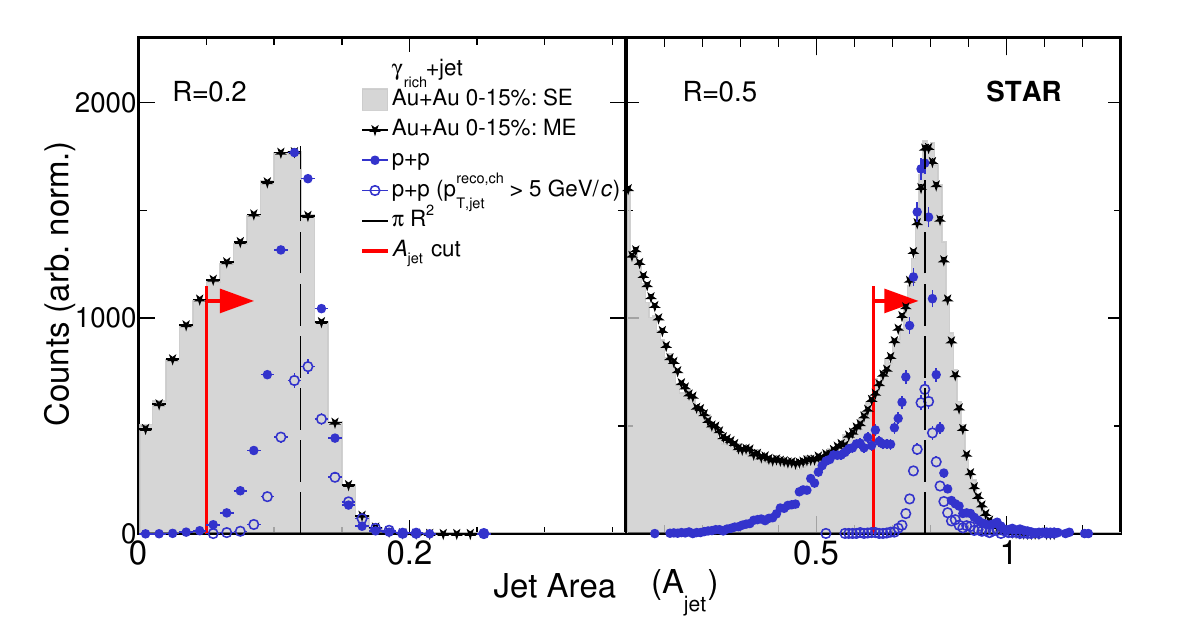}
\caption{Projection of distributions in Fig.~\ref{Fig:pTrecoVsArea} onto the \Ajet\ axis, for \gammarich-triggered recoil jets for SE and ME in \AuAu\ collisions and for \pp\ collisions, for $\rr=0.2$ (left) and $\rr=0.5$ (right). Dashed black lines indicate $\pi \rr^{2}$. Vertical red lines show the \Ajet\ acceptance cut, corresponding to $\Ajet>0.05$ for $\rr=0.2$ and 0.65 for $\rr=0.5$. } 
\label{Fig:ppAuAuJetAComp}
\end{figure}
\noindent
%----

As noted in Sect.~\ref{Sect:RecoilJet}, the \antikT\ algorithm is utilized in this analysis because of its insensitivity to underlying event fluctuations~\cite{Cacciari:2008gp}. Figure~\ref{Fig:pTrecoVsArea} shows the distribution of \pTreco\ vs. jet area for SE and ME events in central \AuAu\ collisions, and for \pp\ collisions. Figure~\ref{Fig:ppAuAuJetAComp} shows the projection of the distributions in Fig.~\ref{Fig:pTrecoVsArea} onto the \Ajet\ axis for jets with $\rr=0.5$ in \gammarich-triggered events. The SE and ME distributions for central \AuAu\ collisions agree in detail, consistent with~\cite{Adamczyk:2017yhe}. Both distributions, as well as those for \pp\ collisions, are peaked close to the value $\pi\rr^{2}=0.785$. Small-area clusters predominantly have $\pTreco\approx0$, as shown in Fig.~\ref{Fig:pTrecoVsArea}, and largely do not correspond to physical jets~\cite{Adamczyk:2017yhe}. This background is suppressed by imposing a cut $\Ajet>0.05$ for $\rr=0.2$ and $\Ajet>0.65$ for $\rr=0.5$, as shown in Fig.~\ref{Fig:ppAuAuJetAComp}.

The jet area cut may also remove signal from physical jets, especially at low \pTjetch, which would induce a jet--finding inefficiency. The effect of this cut can be assessed by comparing their values to the area distribution of jets with $\pTreco>5$ \gev\ in \pp\ collisions, because of the insensitivity of the area distribution of \antikT-reconstructed jets to underlying event fluctuations. The area cut is seen from this point of view to reject only a small fraction of physical jet yield, for both $\rr=0.2$ and 0.5. The residual inefficiency is accounted for by the procedure to determine the jet-matching efficiency used in correcting the spectra (Fig.~\ref{Fig:JetMatchEff}).\section{Raw distributions and Mixed Events}
\label{sect:RawDistr}

%---
\begin{figure}[htb!]
\centering
\includegraphics[width=0.85\textwidth]{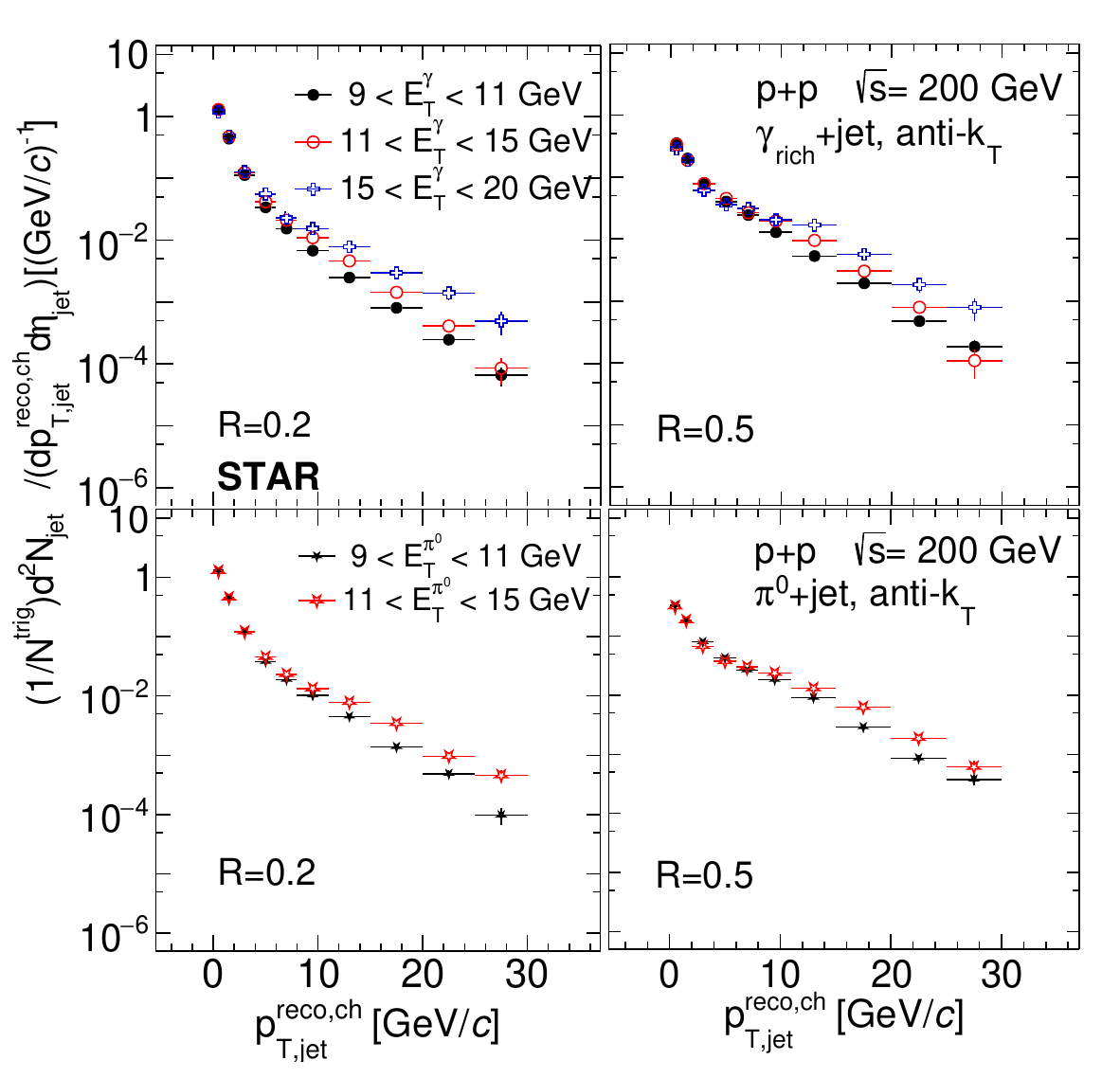}
\caption{(Color online) Uncorrected semi-inclusive recoil-jet distributions for selected \ETtrig\ intervals and recoil jet $\rr=0.2$ (left) and 0.5 (right), for \gammarich-triggered (top) and \pizero-triggered (bottom) \pp\ collisions at $\sqrts=200$ GeV.}
\label{Fig:ppRawGammaJetR2R5}
\end{figure}
%---

Figure~\ref{Fig:ppRawGammaJetR2R5} shows uncorrected semi-inclusive recoil-jet distributions for $\rr=0.2$ and 0.5 as a function of \pTreco, for \gammarich-triggered and \pizero-triggered \pp\ collisions at $\sqrts=200$ GeV in the selected \ETtrig\ bins. Larger \ETtrig\ corresponds to a harder recoil-jet spectrum, as expected. Recoil jet spectra are not provided  for \pizero\ triggers in \ETtrig=15-20~GeV because of limited trigger statistics (Tab.~\ref{Tab:TrigStats}), which causes the
unfolding procedure not to converge.

%---
\begin{figure}[htb!]
\centering
\includegraphics[width=0.90\textwidth]{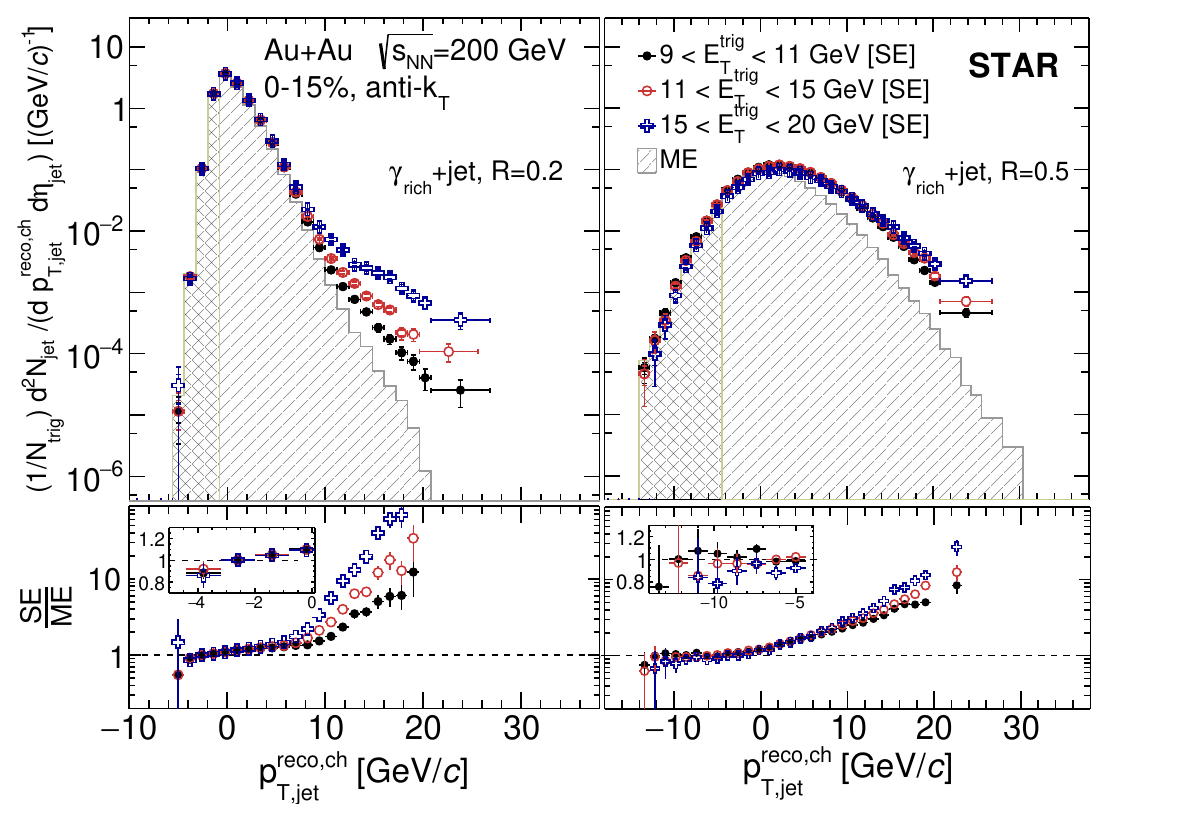}
\caption{(Color online) Upper panels: uncorrected semi-inclusive recoil-jet distributions for \gammarich-triggered central \AuAu\ collisions at $\sqrtsNN=200$ GeV in the \ETtrig\ bins of the analysis, for $\rr=0.2$ (left) and 0.5 (right). Colored markers show data (SE); histograms show normalized Mixed Event (ME) distributions. The hatched distribution is the ME normalization region. Lower panels: ratio of SE/ME. Insets show the ratio of SE/ME in the ME normalization region with linear vertical scale.} 
\label{Fig:SEMEGjet}
\end{figure}
%---

\begin{figure}[htb!]
\centering
\includegraphics[width=0.90\textwidth]{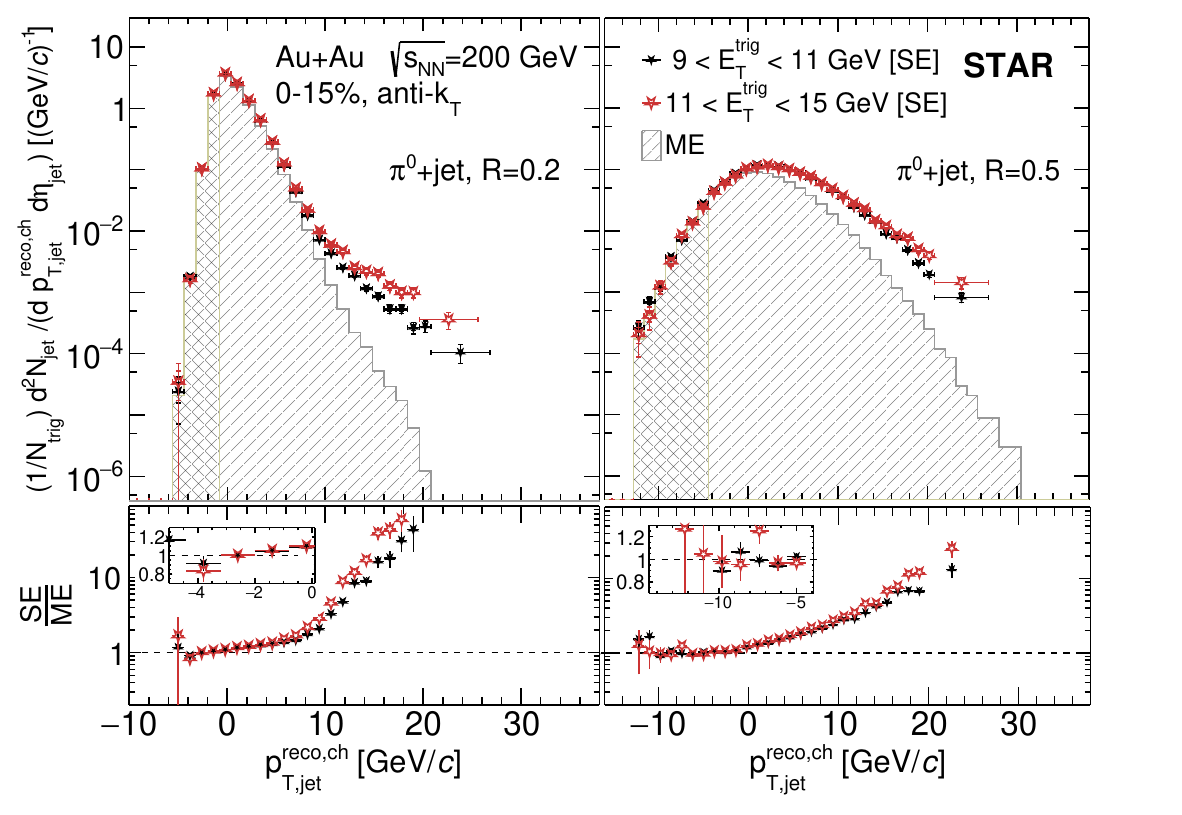}
\caption{(Color online) The same as Fig.~\ref{Fig:SEMEGjet}, for \pizero-triggered central \AuAu\ collisions.}
\label{Fig:SEMEPijet}
\end{figure}

Uncorrected semi-inclusive recoil-jet distributions for central \AuAu\ collisions are shown for \gammarich\ triggers in Fig.~\ref{Fig:SEMEGjet}, and for \pizero\ triggers in Fig.~\ref{Fig:SEMEPijet}. In all cases the recoil--jet yield at large positive \pTreco\ is larger for higher \ETtrig\ values, as expected for correlated jet yields, while at lower \pTreco\ the distributions are very similar for different \ETtrig\ intervals. These features were also observed in Ref.~\cite{Adam:2015doa,Adamczyk:2017yhe}.

%---
%---

Figures~\ref{Fig:SEMEGjet} and \ref{Fig:SEMEPijet} also show the \pTreco\ distribution for the ME populations. Figures~\ref{Fig:SEMEGjet} and \ref{Fig:SEMEPijet}, bottom panels, show that the yield ratio SE/ME in the far negative \pTreco\ region has weak (for $\rr=0.2$) or negligible (for $\rr=0.5$) dependence on \pTreco, as expected if this region is dominated by uncorrelated background yield~\cite{Adam:2015doa,Adamczyk:2017yhe}. This region (indicated as hashed in Figs.~\ref{Fig:SEMEGjet} and \ref{Fig:SEMEPijet}) is therefore used to normalize the ME distribution. The ensemble-averaged distribution of trigger-correlated recoil-jet yield is given by the difference of the SE and normalized ME distributions~\cite{Adam:2015doa,Adamczyk:2017yhe}.

In the semi-inclusive approach, a bias from EP orientation can only be induced by non-zero \vtwo\ and higher anisotropic flow harmonics of the trigger particle, though in practice such a bias is found to be negligible for this and related analyses (Sect.~\ref{Sect:SemiInclDistr}). However, event-by-event variation in EP orientation will also generate fluctuations in the SE distributions, which must be accounted for by the event mixing procedure. As noted in Sect.~\ref{sect:EvtMixing}, the ME population is constructed using two bins in EP orientation, and it is necessary to check whether this granularity is sufficient to account for such fluctuations.

The insets in the lower panels of Figs.~\ref{Fig:SEMEGjet} and \ref{Fig:SEMEPijet} compare the SE and ME distributions in detail in the far negative \pTreco\ region, showing that their shapes are within 10-20\% on a linear scale, over a range in \pTreco\ in which the distributions themselves vary by several orders or magnitude. This comparison demonstrates that the ME procedure with two bins in EP orientation is sufficiently granular to capture EP orientation fluctuations.

%------------
%\begin{center}
\begin{table*}
\caption{Integral of SE and ME distributions, ME normalization region, and ME normalization factor \fME\ for all \ETtrig\ bins for the \gammarich\ and \pizero\ triggers, and for recoil jet $\rr=0.2$ and 0.5.
\label{Tab:SEMENorm}}
\begin{tabular}{ |c|c|c||c|c||c|c| }
\hline
Trigger & \rr & \ETtrig [GeV]  & \multicolumn{2}{|c|}{Integral} &  ME norm [\gev] & \fME \\ \hline 
\multicolumn{3}{|c|}{} & SE & ME & &  \\ \hline \hline
\multirow{6}{*}{\gammarich} & \multirow{3}{*}{0.2} & [9,11] & 12.72 & 12.60 & [-10,2] & $0.80\pm0.03$ \\ 
&  & [11,15] & 12.73 & 12.60 & [-10,2] & $0.83\pm0.04$ \\ 
&  & [15,20] & 12.73 & 12.60 & [-10,2] & $0.79\pm0.03$ \\ \cline{2-7}
& \multirow{3}{*}{0.5} & [9,11] & 1.22 & 1.18 & [-20,-5] & $0.72\pm0.01$ \\ 
&  & [11,15] & 1.23 & 1.19 & [-20,-5] & $0.74\pm0.02$ \\ 
&  & [15,20] & 1.23 & 1.18 & [-20,-5] & $0.70\pm0.02$ \\ 
 \hline \hline
\multirow{4}{*}{\pizero} & \multirow{2}{*}{0.2} & [9,11] & 12.73 & 12.64 & [-10,2] & $0.83\pm0.01$ \\ 
&  & [11,15] & 12.72 & 12.64 & [-10,2] & $0.84\pm0.02$ \\ \cline{2-7}
& \multirow{2}{*}{0.5} & [9,11] & 1.23 & 1.17 & [-20,-5] & $0.73\pm0.02$ \\ 
&  & [11,15] & 1.23 & 1.17 & [-20,-5] & $0.70\pm0.03$ \\ 
\hline 
\end{tabular}
\end{table*}
%\end{center}
%----------

Table~\ref{Tab:SEMENorm} gives the recoil-jet yield integrals of the SE and ME distributions prior to normalization, which agree within $\approx1\%$. Such invariance of the recoil-jet yield integral (or equivalently, jet density) has been observed in other high-multiplicity analyses~\cite{Adam:2015doa,Adamczyk:2017yhe}. This invariance is consistent with the resilience of \antikT\ jet reconstruction to distortion by large backgrounds~\cite{Cacciari:2008gp}, and it plays an important role in this analysis approach. Table~\ref{Tab:SEMENorm} also gives the values of the ME normalization factor \fME, which are similar to those in Ref.~\cite{Adamczyk:2017yhe}.

%---
\begin{figure}[htbp!]
\centering
\includegraphics[width=0.85\textwidth]{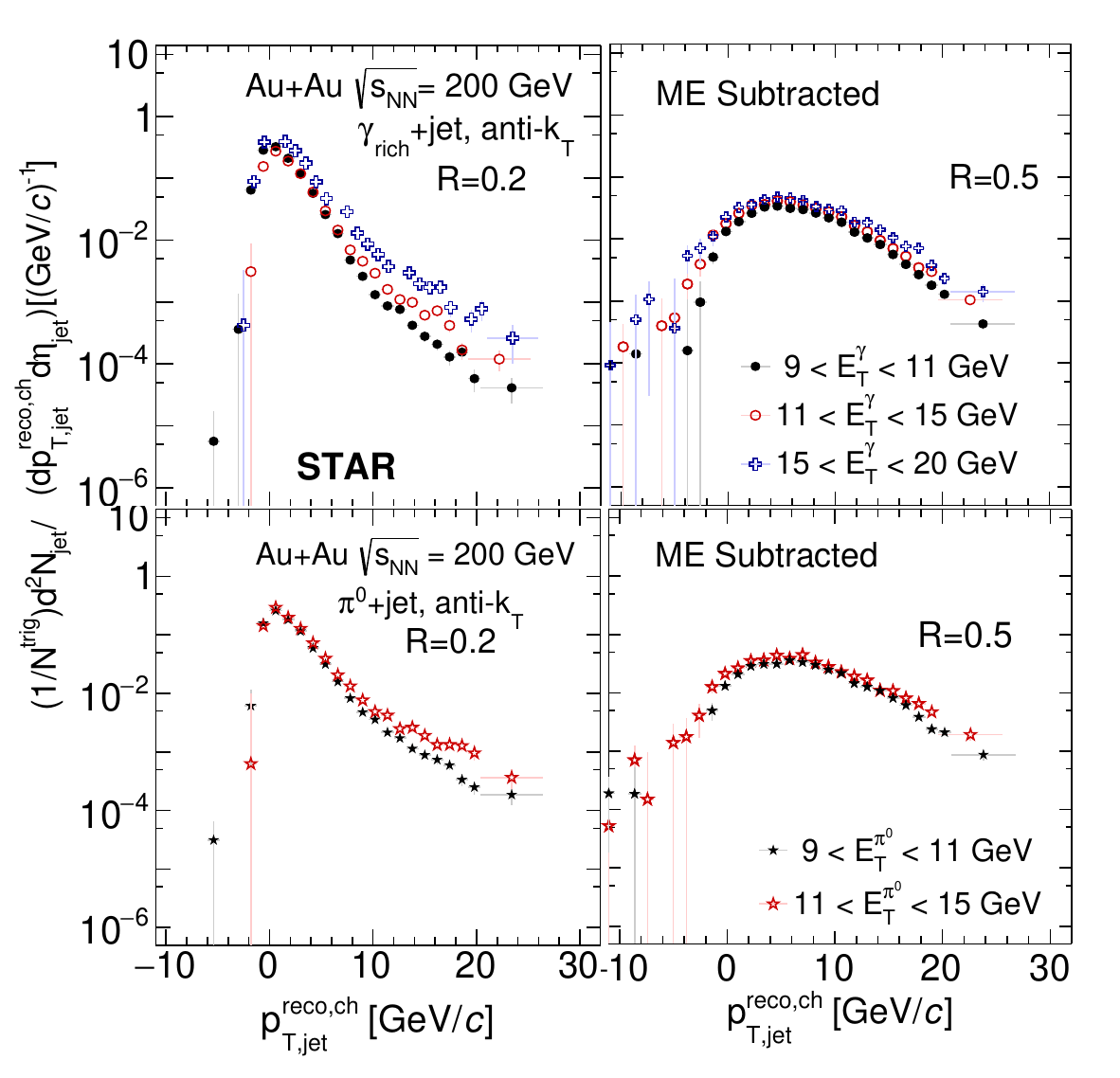}
\caption{(Color online) Recoil-jet distributions for central \AuAu\ collisions after subtraction of normalized ME distributions. Data are from Fig.~\ref{Fig:SEMEGjet} and ~\ref{Fig:SEMEPijet}. Upper: \gammarich\ triggers; lower: \pizero\ triggers. Left: $\rr=0.2$; right: $\rr=0.5$. Negative values after subtraction not shown, due to logarithmic vertical axis.}
\label{Fig:MESubRawSpec}
\end{figure}
%---

Figure~\ref{Fig:MESubRawSpec} shows the result of subtracting the normalized ME distributions from the SE distributions in Figs.~\ref{Fig:SEMEGjet} and ~\ref{Fig:SEMEPijet}. These are the distributions after correction for uncorrelated background yield, which are used in further analysis. Because the vertical axes have logarithmic scale, bins with negative values from the subtraction cannot be shown; however, these bins all have values consistent with statistical fluctuations. 

Since the subtraction at low and negative values of \pTreco\ corresponds to taking a small difference of two large numbers, large oscillations can arise if the two distributions are not aligned precisely. However, all difference distributions in Fig.~\ref{Fig:MESubRawSpec} are seen to be well-behaved at low \pTreco, without large oscillations or other non-monotonic features, providing independent validation of the $\rho$ alignment discussed in Sect.~\ref{sect:rhoalignment}. 

\section{Extraction of {\gammadir}+jet distributions}
\label{sect:GammajetConversion}

The \gammarich-trigger population comprises direct photons, decay photons from asymmetric \pizero\ and $\eta$ decays, and fragmentation photons. Discrimination of direct photons is based on the ansatz that direct photons have no near-side correlated hadron yield. This ansatz is applied to correct {\gammarich}+jet to \gammadir+jet distributions, using the \pizero+jet yield to measure the recoil jet distribution from background triggers~\cite{STAR:2016jdz} .

The \gammadir-triggered semi-inclusive recoil-jet yield, \Dgammadir(\pT), is defined as

\begin{equation}
\Dgammadir(\pTjet ) = \frac
{\Dgammarich(\pTjet) - \Rpurity\cdot\Dpizero(\pTjet)} 
{1 - \Rpurity}.
\label{Eq:ConvDirEq}
\end{equation}
\noindent
Here \Dgammadir(\pTjet), \Dgammarich(\pTjet), and \Dpizero(\pTjet) are the trigger-normalized recoil-jet yields for \gammadir, \gammarich, and \pizero\ triggers, respectively. The factor (1-\Rpurity) is the purity of \gammadir\ in the \gammarich\ sample.  The purity of the \pizero-trigger sample is estimated from simulation to be greater than $\approx 95$\%  for all \ETtrig\  bins, satisfying the high-purity condition necessary for this correction procedure. Systematic studies show negligible dependence of the fully--corrected recoil-jet spectrum on the value of \pizero\ purity between 95 and 100\%.

The calculation of \Dgammadir(\pTjet) can be applied to either the raw spectra before unfolding or the corrected (unfolded) recoil-jet spectra. For an analysis with limited statistics, subtraction before unfolding is preferable because unfolding may be unstable. For \AuAu\ collisions, the default method is to subtract the unfolded \pizero-triggered spectrum from the unfolded {\gammarich}+jet spectrum for all \ETtrig\ bins except \ETtrig=15-20~GeV. For \ETtrig=15-20~GeV, the \pizero-triggered recoil-jet spectrum is not statistically precise enough for unfolding to converge reliably, so the subtraction is carried out before unfolding. For \pp\ collisions, the default method for all \ETtrig\ bins is to subtract before unfolding. Where possible, the alternate method was checked for consistency.

The purity of \gammadir\ triggers in the \gammarich\ population is determined by applying the ansatz that \gammadir\ triggers do not have a correlated near-side yield of charged hadrons within relative azimuthal angle $\dphi<1.4$.  The fraction of background triggers is then determined statistically for each \ETtrig\ bin using the near-side correlated charged-hadron yields~\cite{STAR:2016jdz}.  

The ratio of correlated yields is defined as

\begin{equation}
\Rpurity = \frac{Y^{\rm near}_{\gammarich+h}}{Y^{\rm near}_{\pizero+h}},
\label{Eq:Rpurity}
\end{equation}

\noindent
where $Y^{\rm near}_{\gammarich+h}$ is the near-side yield ($\dphi \leq 1.4$ rad) of charged hadrons per \gammarich\ trigger and $Y^{\rm near}_{\pizero+h}$ is the near-side yield per \pizero\ trigger, after subtracting the uncorrelated background of charged hadrons. The value of \Rpurity\ is determined for different ranges of $\zT=\pTtrack/\ETtrig$, for $\zT>0.1$ and $\pTtrack>1.2$ \gev. The uncorrelated background is determined from the yield measured outside of the near--side peak region, whose size is varied to assess the systematic uncertainty of uncorrelated background subtraction. At low \zT, where the uncorrelated relative yield is high, this is the dominant uncertainty, while at high \zT\ the statistical error dominates. For each \ETtrig\ selection, the value of \Rpurity\ is determined as the average \Rpurity\ of the different \zT\ ranges. The uncertainty in \Rpurity\ includes the uncertainty in the uncorrelated background subtraction. 

All correlated charged tracks for \gammarich\ triggers are attributed to background in the trigger population (\pizero, single photons from \pizero\ and $\eta$ decays, and fragmentation photons). The subtraction of the recoil-jet yield due to background triggers (Eq.~\ref{Eq:ConvDirEq}) is carried out assuming that the correlated yield and the recoil-jet distribution associated with the background triggers is the same as that measured for \pizero\ triggers.  

We also assume that the recoil-jet distribution for high-\pT\ $\eta$-meson triggers is the same as that for \pizero-meson triggers, based on both simulation and $\eta$ measurements at RHIC~\cite{PHENIX:2006ujp,PHENIX:2013yhu}. This assumption applies to photons from \pizero\ and $\eta$ decays because the trigger requirement biases towards asymmetric decay, such that the decay photon carries most of the meson parent energy. This was verified in simulation by comparing the mean \ET\ of the meson parent, for which the decay photon falls within a given \ETtrig\ selection, to the mean \ET\ of symmetrically decaying \pizero\ which pass the TSP cut and fall into the same \ETtrig\ selection. For $9<\ETtrig<11$ and $11<\ETtrig<15$~GeV, the difference is approximately 5\%, while for $15<\ETtrig<20$~GeV the difference is around 3\%. However, the associated-hadron yield for fragmentation photons is not well known.  Therefore, the fragmentation photons are only subtracted to the extent that their near-side correlated hadron yields are similar to those of \pizero.  The magnitude of suppression of the fragmentation photon contribution due to this procedure can only be assessed by theoretical calculations. The uncertainty in the fragmentation photon contribution is thus model dependent, and determination of its value is beyond the scope of this paper.

For central \AuAu\ collisions, the purity of \gammadir\ in the \gammarich\ population varies between $67\pm3$\% and $84\pm4$\%, from lowest to highest \ETtrig. For \pp\ collisions, the purity of \gammadir\ varies between $43\pm5$\% and $53\pm7$\%, from lowest to highest \ETtrig.
The larger value of \gammadir\ purity in the central \AuAu\ data compared to  \pp\ data arises from the suppression of \pizero\ yield due to jet quenching in \AuAu\ collisions~\cite{PHENIX:2012jbv}. 
\section{{\ETtrig} resolution}
\label{sect:TrigRes}

The \gammadir\ and \pizero\ trigger particles are measured using the BEMC and BSMD detectors (Sect.~\ref{Sect:TrigMeasurement}). The measurement of \ETtrig\ is affected by the intrinsic energy resolution of the BEMC and by energy leakage to neighboring towers that is not accounted for by the clustering algorithm.  To quantify these effects, the Trigger Energy Scale (TES) and Trigger Energy Resolution (TER) are determined using the STAR GEANT simulation.

In this study, single \pizero\ and $\gamma$ particles are generated with a uniform \ET\ distribution in the range $6<\ET<30$~GeV, and with uniform spatial distribution on a regular grid in ($\eta,\phi$), with spacing of 0.6 radians (12 towers) and 0.3 radians (6 towers) for \pizero\ and $\gamma$ respectively. This approach enables multiple particles to be simulated per event without their signals overlapping in the BEMC. The larger spacing for \pizero\ prevents the overlap of decay photons from other \pizero. The generated events are then passed through the STAR GEANT3 simulation, with the detector configuration corresponding to the 2009 \pp\ run. 

The simulated BEMC showers are clustered using the same algorithm as is used for data, applying the same TSP cuts. Clusters are matched to the simulated particles in ($\eta,\phi$) phase space, based on the projected position of the particle on the face of the BEMC. 

We denote the value of \ET\ of a $\gamma$ or \pizero\ particle at the generated level as \eTgenParGa\ and \eTgenParPi; at the reconstructed level as \eTgenDetGa\ and \eTgenDetPi; and at the matched level as \eTgenMatGa\ and \eTgenMatPi. Note that the matched-level \ET\ is the generated-level \ET\ for a particle reconstructed within our selection ranges. 

%-------------------
% calculated TES/R
\begin{table}[htbp!]
  \centering
    \caption{TES and TER of triggers in the bins of \eTgenDet, from fitting the \qT\ distributions (Eq.~\ref{eq:qT}).  The uncertainty shown is statistical only.  The relative systematic uncertainty is approximately 3\% for TES and 5\% for TER. The average \ET\ for each selection is calculated using the smearing weights, assuming the trigger spectrum shape given by PYTHIA-6 STAR tune.}
  \begin{tabular}{l | c c c | c c c}
    \hline
    \multirow{3}{*}{\eTgenDet} & \multicolumn{3}{c}{$\mathbf{\pi^{0}}$} & \multicolumn{3}{c}{$\mathbf{\gamma}$} \\ 
    & TES & TER & $\langle{E_{\text{T}}}\rangle$ & TES & TER & $\langle{E_{\text{T}}}\rangle$\\ $\left[\mathrm{GeV}\right]$ & (\%) & (\%) & $\left[\mathrm{GeV}\right]$ & (\%) & (\%) & $\left[\mathrm{GeV}\right]$ \\
    \hline
    9 - 11 & $92.4\pm0.2$ & $9.1\pm0.1$ & 10.5 & $97.97\pm0.05$ & $8.12\pm0.03$ & 10.2\\
    11 - 15 & $94.4\pm 0.2$ & $8.4\pm 0.1$ & 11.8 & $97.77\pm 0.03$ & $7.83\pm 0.02$ & 12.7 \\
    15 - 20 &  &  & & $97.74\pm 0.03$ & $7.56\pm 0.02$ & 16.8\\
    \hline
  \end{tabular}	
  \label{table:TESTER}
\end{table}
%-------------------

The values of the TES and TER are determined from the ratio of the reconstructed and generated \ET\ for matched clusters, 

\begin{equation}
%\qT = E_{\rm T}^{\rm det-clust}/E_{\rm T}^{\rm part-match}. 
\qT = \frac{\eTgenDetGa}{\eTgenMatGa},
%\qT = \frac{\eTgenDet}{\eTgenMat};
\label{eq:qT}
\end{equation}

\noindent
similarly for \pizero. The \qT\ distribution is calculated by weighting the uniformly-generated particles by a physical \ET\ distribution. The resulting \qT\ distribution is fit with a Gaussian function, the mean of which is the TES and the RMS of which is the TER. 

Table~\ref{table:TESTER} gives the values of TES and TER for \pizero\ and $\gamma$ in the \ETtrig\ bins of this analysis. The TES values for \pizero\ are smaller than those for $\gamma$, due to larger probability of energy leakage into neighboring towers that is not recovered by the clustering algorithm. Table~\ref{table:TESTER} also gives values of $\langle{E_{\text{T}}}\rangle$ for different trigger selections, calculated using the smearing weights, assuming the trigger spectrum shape for \eTgenPar\ is that generated by PYTHIA-6 STAR tune. For \pizero\ triggers with $9<\ETtrig<11$~GeV, the mean \eTgenMat\ is higher than the mean \ET\ of the physical spectrum (9.8~GeV) in the interval $9<\ET<11$~GeV. However, for \pizero\ with $11<\ETtrig<15$~GeV, the mean \eTgenMat\ is lower than the mean \ET\ of the physical spectrum (12.2~GeV) due to the TSP selection bias toward wider \pizero\ showers. The TSP selection biases the \pizero\ population to lower \ETtrig, which has larger opening angle on average.

The trigger resolution arises from the intrinsic responses of the BEMC and BSMD, which at high \ET\ experience negligible influence from the overall event environment. The values of TES and TER should therefore be similar for \pp\ and central \AuAu\ collisions. This was studied by comparing \qT\ distributions for \pizero\ and $\gamma$ found in PYTHIA-6 STAR tune di-jet events, embedded in the 2009 \pp\ and the 2014 \AuAu\ data and analysed using the algorithm described in Sect.~\ref{Sect:TrigMeasurement}. For $\gamma$ triggers the event background does not significantly shift or broaden the \qT\ distribution. For \pizero\ triggers, there is a small (1--3\%) broadening of the resolution in the \AuAu\ events compared to \pp\ events, but no significant shift in the scale.

No correction for TES and TER is applied in the data analysis. For precise comparison of theoretical calculations to the measurements, the calculated distributions should therefore be smeared to account for the TES and TER effects. Weight factors are provided here for that purpose, to be applied bin-wise in \ETtrig\ to a theoretically calculated recoil-jet distribution. In order to be able to apply the weighting, the theory calculation should record recoil jets for triggers with \ET\ between 6 and 30 GeV, storing the recoil-jet spectrum in 1 GeV increments of \ETtrig. The weights $W$ are defined as the relative probability of a photon with calculated value \eTgenPar\ to contribute to a bin in measured \eTgenDet,

\begin{align}
\Wargs &= P(\eTgenDet | \eTgenPar) \nonumber \\
&= P( \eTgenPar | \eTgenDet ) 
\frac{P(\eTgenDet)}{P(\eTgenPar)},
\label{eq:Weights}
\end{align}

\noindent
where the last expression uses Bayes's Theorem.

%-------------------
% TES/R Matrices
\begin{figure}[htb!]
  \centering
  \includegraphics[width=0.99\textwidth]{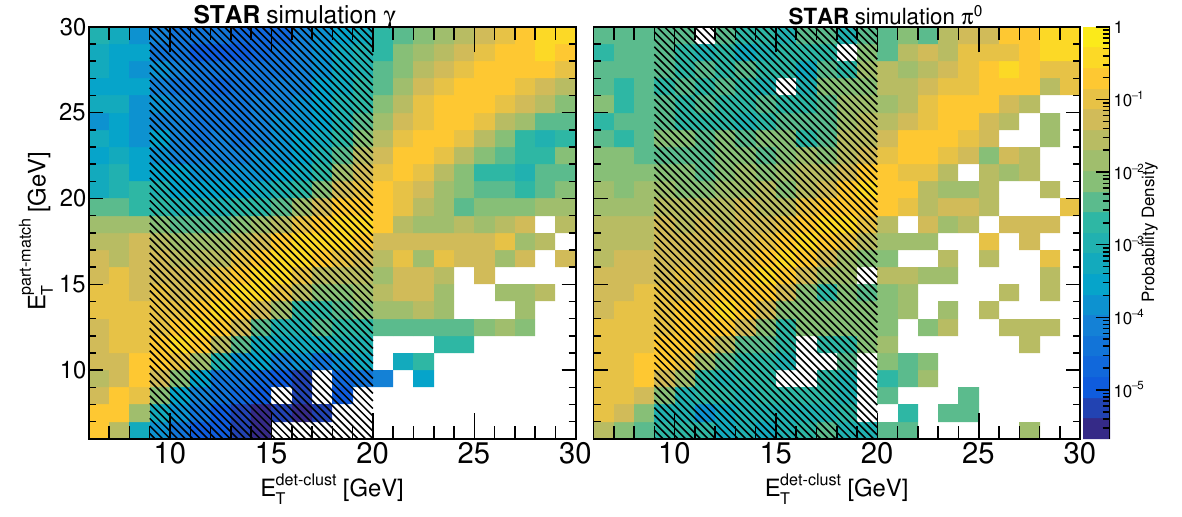}
  \caption{Distribution of \ETtrig\ at the matched (vertical) vs. reconstructed (horizontal) levels for \gammadir\ (left) and \pizero\ triggers (right).  The hashed boxes indicate the \ETtrig\ selection used in this analysis.} %\comment{PMJ}{Larger fonts for axis labels; make uniform with other Figs.}}
  %\comment{Comment to GPC}{TBD graphics will be improved: merge into 2-panel figure, adjust labels, and fonts.} \comment{DMA Feb 21}{Done}} 
  %\comment{PMJ Nov 11}{Figure needs work. Make notation consistent with text, improve info in legend (e.g. ``STAR simmulation'' not meaningful), merge into one 2-panel figure. Make hash pattern a bit heavier (but not too much heavier) - hard to see.} \reply{DMA Nov 21}{This will be done in parallel with GPC.}
  \label{Fig:EtSmearingMatrices}
\end{figure}

% old figure [Derek, 02.21.2023]
%\begin{figure}[htb!]
%  \centering
%  \includegraphics[width=0.49\textwidth]{Longpaper_PRC/Figs_LongPaper%/etMatrix_noShapeWeights.et650gam.d21m11y2022}
%  \includegraphics[width=0.49\textwidth]{etMatrix_noShapeWeights.et650pi0.d21m11y2022}
%  \caption{Distribution of \ETtrig\ at the matched (vertical) vs. reconstructed (horizontal) levels for \gammadir\ (left) and \pizero\ triggers (right).  The hashed boxes indicate the \ETtrig\ selection used in this analysis.  \comment{Comment to GPC}{TBD graphics will be improved: merge into 2-panel figure, adjust labels, and fonts.}}
%\comment{PMJ Nov 11}{Figure needs work. Make notation consistent with text, improve info in legend (e.g. ``STAR simmulation'' not meaningful), merge into one 2-panel figure. Make hash pattern a bit heavier (but not too much heavier) - hard to see.} \reply{DMA Nov 21}{This will be done in parallel with GPC.}
%  \label{Fig:EtSmearingMatrices}
%\end{figure}
%-------------------

Figure~\ref{Fig:EtSmearingMatrices} shows the 2-D correlation of matched (generated) \ET\ vs. reconstructed (detector-level) \ET, with a uniform (flat) distribution in \eTgenPar. Vertical slices in this distribution represent the probability distribution for a bin of generated-level \ET\ (\eTgenMat) to contribute to the bin of detector-level \ET (\eTgenDet). The weight factors \Wargs\ are calculated from the 2-D distribution in  using the last expression in Eq.~\ref{eq:Weights}, by projecting out the \eTgenMat\ distributions for a given selection of \eTgenDet\ (e.g. 11--15 GeV). This selection interval in \eTgenDet\ corresponds to the factor $P(\eTgenDet)$ in Eq.~\ref{eq:Weights}.

\begin{figure}[htb!]
  \centering
  \includegraphics[width=0.99\textwidth]{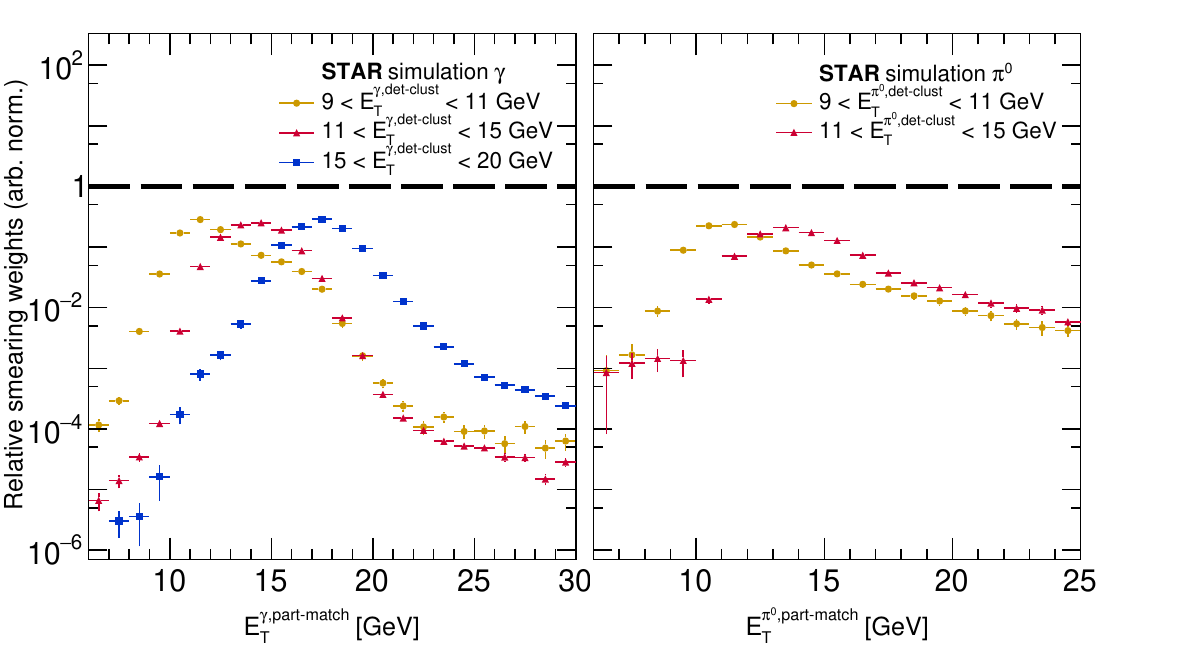}
  \caption{\ETtrig\ smearing weights for \gammadir\ (left panel) and \pizero\ triggers (right panel).}
  %\comment{PMJ}{Larger fonts; make uniform with other Figs. Vert axis label change to ``(arb. norm.)'' for space reasons.}}
  \label{Fig:SmearWeights}
\end{figure}

Since a uniform \eTgenPar\ distribution is used to generate Fig.~\ref{Fig:EtSmearingMatrices}, the factor $P(\eTgenPar)$ is simply a scaling factor that has no effect on the relative weighting.
The integral of \Wargs\ for each selection in \eTgenDet\ is therefore arbitrarily normalized to unit integral. This distribution is the relative weight factor to be applied as a function of \eTgenPar\ for a given selection in \eTgenDet. The distributions for both trigger types and all \ETtrig\ ranges are shown in Fig.~\ref{Fig:SmearWeights}.

%-------------------
 % smearing Weights

% old figure [Derek, 02.23.2023]
%\begin{figure}[htb!]
%  \centering
%  \includegraphics[width=0.49\textwidth]{smearingWeightsFromProjections.forPaper_rightLabel.et650gam.d30m11y2022.png}
%  \includegraphics[width=0.49\textwidth]{smearingWeightsFromProjections.forPaper_rightLabel_noEt1520.et650gam.d30m11y2022.png}
%  \caption{\ETtrig smearing weights for \gammadir\ (left panel) and \pizero\ triggers (right panel).\comment{Comment to GPC}{TBD graphics will be improved: merge into 2-panel figure, adjust labels and fonts, etc.}}
%  \label{Fig:SmearWeights}
%\end{figure}
%-------------------

%Then the weighted average is normalized to have the same integral as that of \eTgenPar\ over the same range as \eTgenDet.  Finally, the weights are the ratio of the normalized weighted average over \eTgenPar.

\begin{figure}[htbp!]
  \centering
  \includegraphics[width=0.99\textwidth]{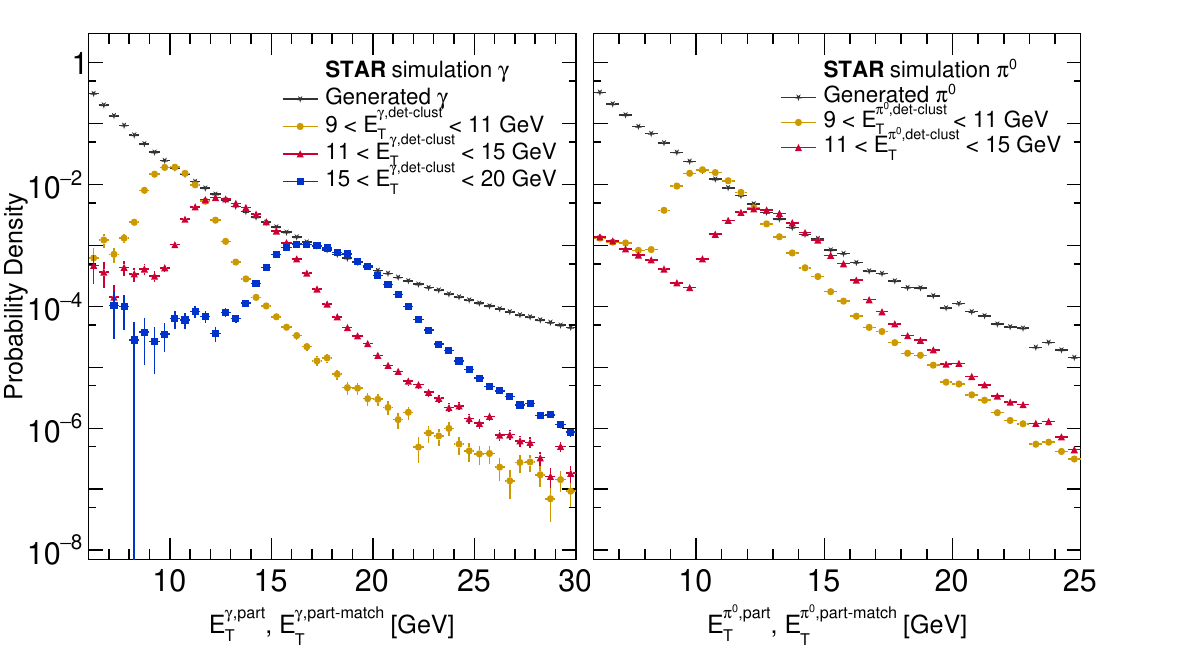}
  \caption{Generated vs. smeared \ETtrig\ distributions for \gammadir\ (left panel) and \pizero\ triggers (right panel).}
  %\comment{Jana}{Fig. 11 vs Fig. 12 use the same style when referring to left and right (left panel) vs Left: … and actually check across all figures that a uniform style is used} \comment{PMJ}{Seems to be a typo in this caption but I don't know what is missing - Derek/Saskia please check. Needs STAR label.}
  %\comment{Comment to GPC}{TBD graphics will be improved: merge into 2-panel figure, adjust labels, and fonts.} \comment{DMA Feb 24}{Done}} %\comment{SM}{These really need to be the shape-weighted distributions!}\comment{PMJ Nov 11}{Figure needs work. Make notation consistent with text, improve legend (e.g. ``STAR simulation'' not meaningful). Need larger fonts and axis labels. Why is vert axis "arbitrary units"? Can't this be expressed in absolute terms of cross section or per inelastic pp collision? Merge into one 2-panel figure. } \reply{NRS Nov13}{Done, except merging.}
  \label{Fig:TERPi0Gamma}
\end{figure}

As an example, Fig.~\ref{Fig:TERPi0Gamma} shows the effect of scaling
a physical trigger-\eTgenPar\ spectrum generated by PYTHIA by \Wargs\ from Fig.~\ref{Fig:SmearWeights}. The generated trigger spectra for photons (left) and \pizero\ (right) are shown as black stars. 
%The closed, colored markers show the reconstructed distributions \eTgenDet, normalized to have the same integral as the \eTgenPar\ over the reconstructed \ET\ selection range.  
The other markers show the resulting distributions of \eTgenMat\ for the different bins of \ETtrig, with the weight factors from Fig.~\ref{Fig:SmearWeights} applied to the \eTgenPar. These distributions have been normalized to conserve the number of triggers. For example, for the detector-level selection $11<\eTgenDet<15$~GeV, the integral over $6<\eTgenMat<30$~GeV is equal to the integral over $11<\eTgenPar<15$~GeV.

%-shape 1-D distributions of \ET\ at the generated, reconstructed, and matched levels for $\gamma$ and \pizero, in the \ETtrig\ bins of the analysis. The shape of the reconstructed \ETtrig\ distributions of the \pizero\ are steeper than the generated distributions due to the decreasing efficiency of the TSP cut with increasing \pizero\ \ET. 

%-------------------
% particle gun eTtrg distributions

\begin{figure}[htbp!]
  \centering
  \includegraphics[width=0.5\textwidth]{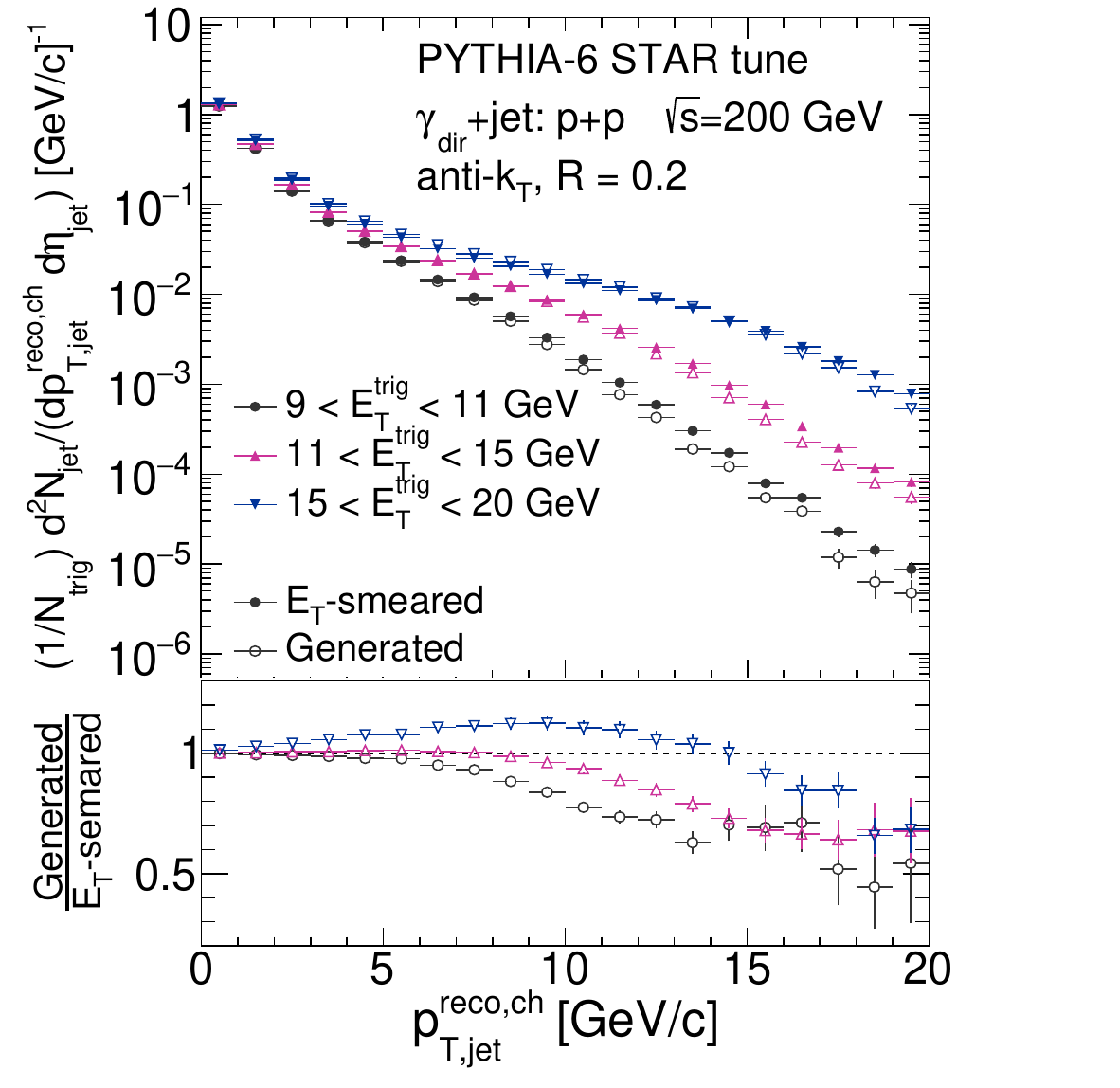}
  \includegraphics[width=0.49\textwidth]{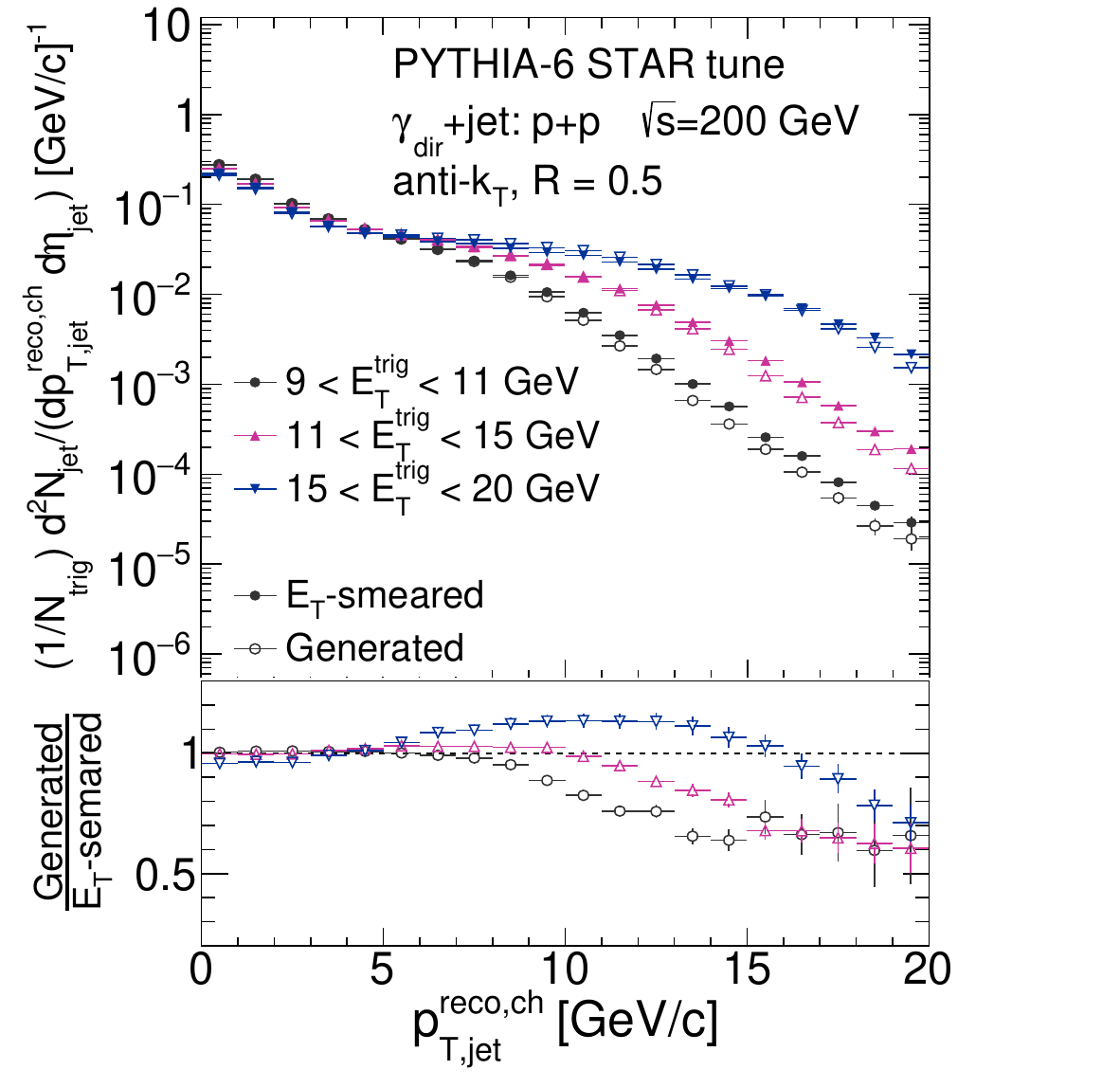}
  \caption{Effect of TES/TER on \gammadir-triggered recoil jets in \pp\ collisions at $\sqrts=200$ GeV with $\rr=0.2$ (left) and 0.5 (right), simulated using PYTHIA-6 STAR tune. Open markers denote the jet spectra without \ET\ weights, and filled markers denote the jet spectra after \ET\ reweighting.}
  \label{Fig:GjetEtSmrEff}
\end{figure}
%--------------------------------

%--------------------------------
% TES/R effect on pi0+jet
\begin{figure}[htb!]
  \centering
  \includegraphics[width=0.49\textwidth]{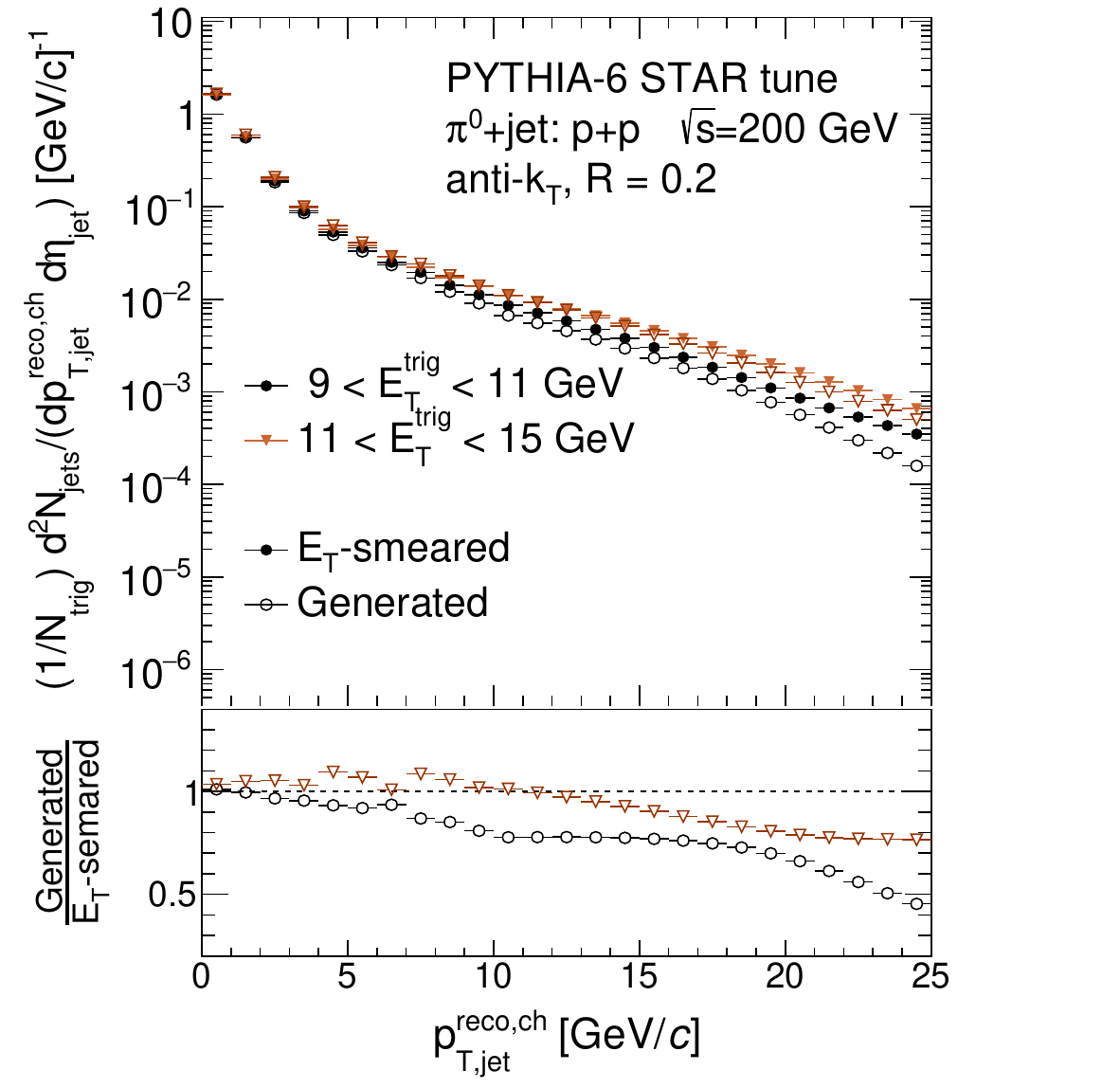}
  \includegraphics[width=0.5\textwidth]{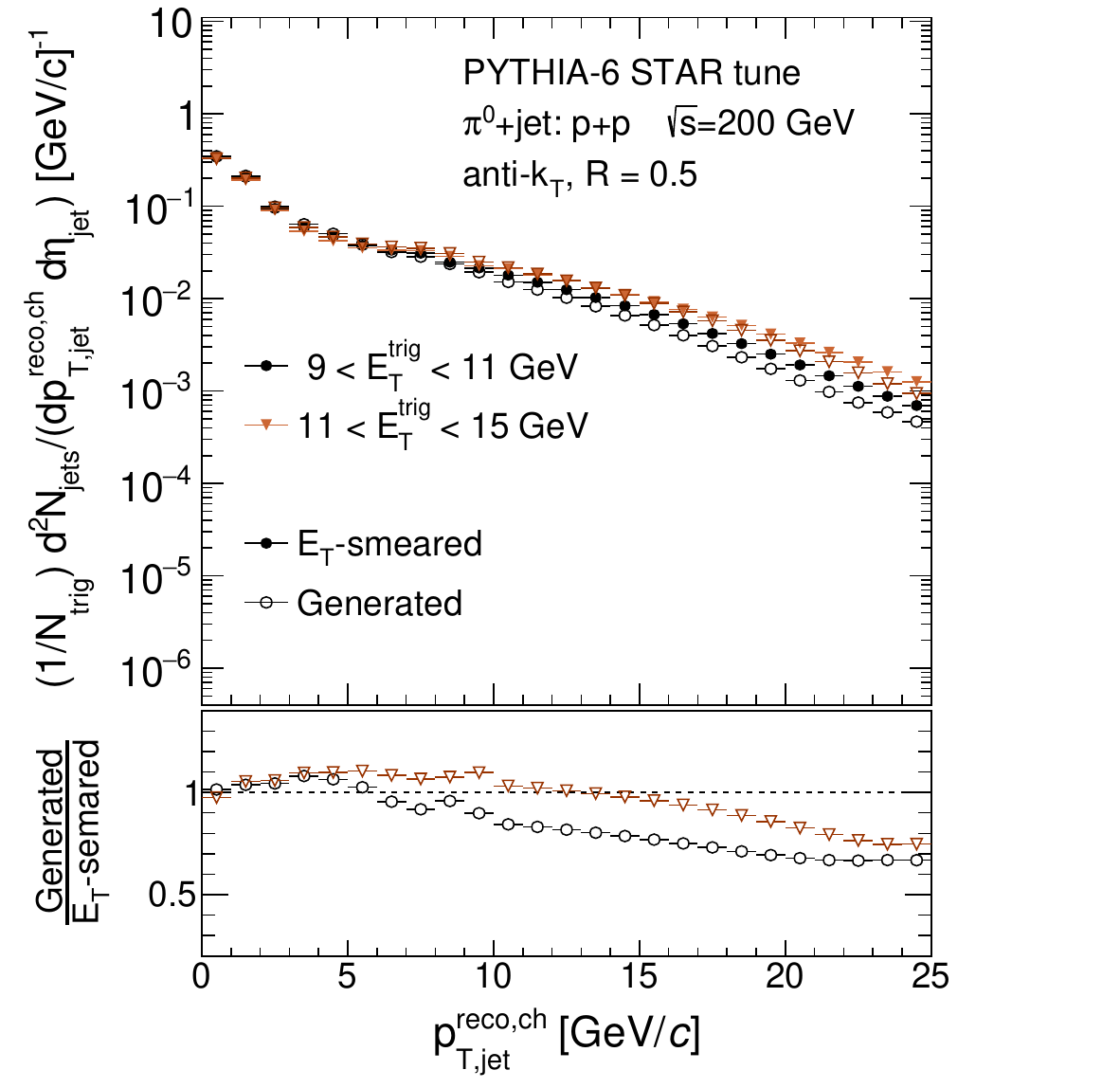}
  \caption{Same as Fig.~\ref{Fig:GjetEtSmrEff}, for \pizero\ triggers.}
  \label{Fig:Pi0jetEtSmrEff}
\end{figure}
%--------------------------------

To illustrate the effect of the TES/TER on the recoil-jet spectra, the weighting factors in Fig.~\ref{Fig:SmearWeights} are applied to \ETtrig-dependent jet spectra from PYTHIA-6 STAR tune.  Figures~\ref{Fig:GjetEtSmrEff} and \ref{Fig:Pi0jetEtSmrEff} show PYTHIA-6 STAR tune recoil-jet spectra before (open markers) and after the \ET\ weighting (filled markers).  For the open markers, \gammadir\ and \pizero\ triggers were selected in the \ETtrig\ ranges of this analysis. For the filled markers, \gammadir\ and \pizero\ were accepted for $6<\ETtrig<30$ GeV and then re-weighted for each \ETtrig\ selection, as described above. Note that simply restricting the \ETtrig\ range in order to match the sampled average \ET\ 
($\langle{E_{\rm{T}}}\rangle$ in Tab.~\ref{table:TESTER}) is not equivalent.

%--------------------------------------------------------------------------
\section{Corrections}
\label{sect:Corrections}

The distribution of trigger-correlated jet yield is obtained by subtracting the normalized ME distribution from the SE distribution, as discussed in Sect.~\ref{sect:RawDistr}. This correlated-yield distribution must then be corrected by unfolding the \pTjet-smearing due to instrumental effects in both \pp\ and central \AuAu\ collisions, and to background fluctuations in central \AuAu\ collisions. 

The unfolding procedure utilizes a response matrix, the map between particle-level and detector-level \pTjetch, which is generated from detailed simulations of the detector-level response (Sect.~\ref{sect:UnfoldingForm}). Determination of the response matrix requires a matching procedure between jets at the particle and detector levels, which is optimized separately for \pp\ and \AuAu\ collisions to obtain efficient matching while minimizing noise contributions. 

Not all particle-level jets have a corresponding detector-level partner in such a procedure. To account for this inefficiency, the response matrix is normalized to unity for each bin in \pTjetpart, integrated over \pTreco, and the unfolded distributions are then corrected for the \pTjetch-dependent efficiency that accounts for the fraction of particle-level jets with no matched partner at the detector level.

%========================
\subsection{Instrumental effects}
\label{sect:InstrumentEffects}

%------------------------------------
\subsubsection{\pp\ collisions}
\label{sect:ppInstrumentEffects}

Instrumental effects for \pp\ collisions at $\sqrts=200$ GeV are determined using simulated di-jet events generated by PYTHIA 6.42, Perugia 0 tune. These events are passed through the STAR GEANT3 simulation to produce detector-level events, which are embedded into real zero-bias \pp\ collision events at $\sqrts=200$ GeV from the 2009 RHIC run to emulate pile-up effects. 

The STAR Collaboration has measured the longitudinal and transverse distribution of charged hadrons within fully-reconstructed jets in \pp\ collisions at $\sqrts=200$ GeV~\cite{STAR:2022hqg} and 510 GeV~\cite{STAR:2019yqm}. Detector-level PYTHIA calculations agree well with these measurements for fully-reconstructed jets down to $\pTjet=8$ \gev, which corresponds to value of \pTjetch\ of about 5 \gev. The PYTHIA model therefore reproduces well the fragmentation of charged-particle jets over the full kinematic range of the \pp\ measurements in this analysis.

Jet reconstruction is carried out on both the particle-level and detector-level events. At detector level, track acceptance corresponds to that of primary particles (Sect.~\ref{Sec:DetectorDataset}), while at particle level the track acceptance is $|\eta|<1.0$, with no \pT\ cut. Recoil jets at the particle and detector levels are matched by requiring that the distance between jet axes $\delta\rr=\sqrt{\delta\varphi^{2}+\delta\eta^{2}}<\rr$, and that $0.5<\qTjet<1.3$, where $\qTjet=\pTjetdet/\pTjetpart$. The jet-matching efficiency is the fraction of particle-level jets which are matched to detector-level jets.  
%---
\begin{figure}[htbp]
\centering
\includegraphics[width=0.75\textwidth]{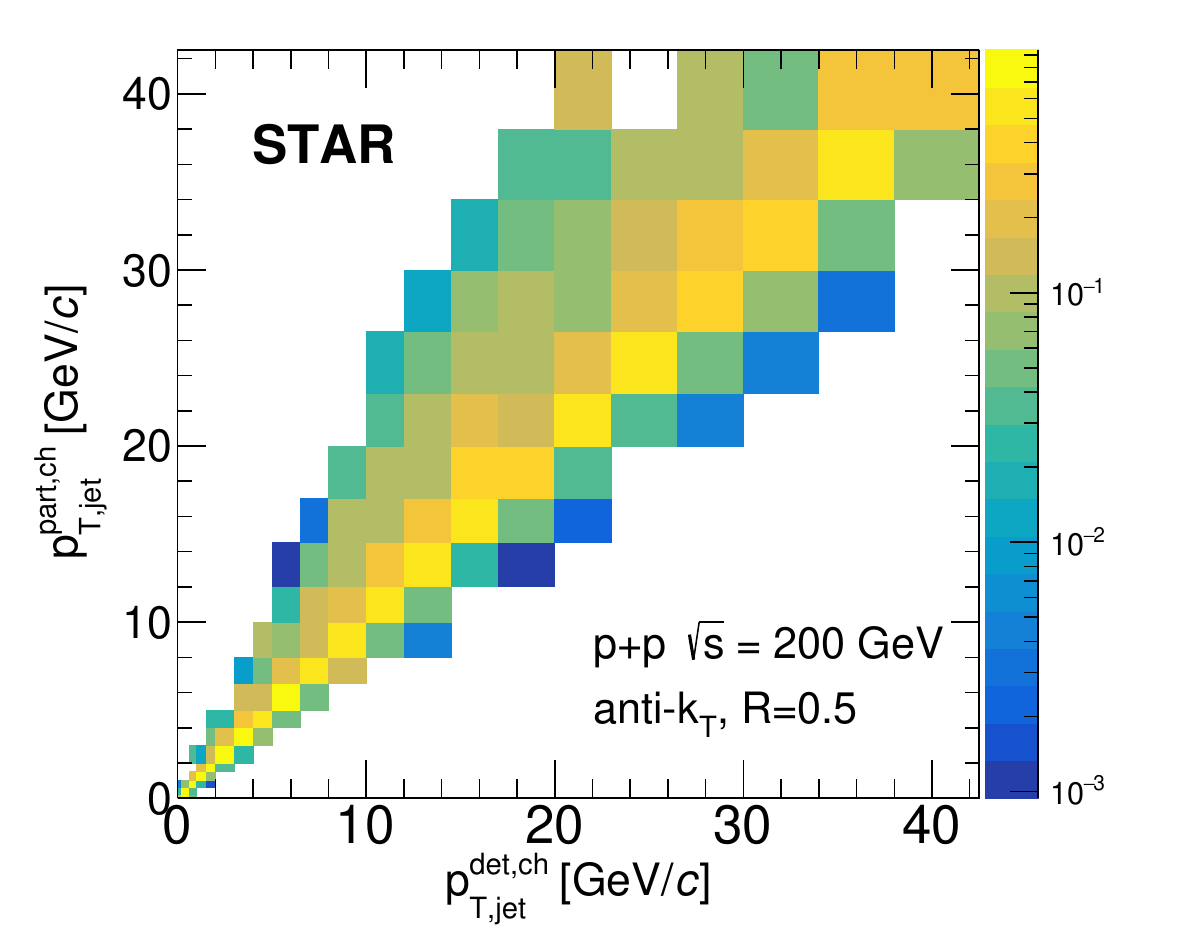}
\caption{Response matrix, \Rinstr, for charged-particle jets with $\rr=0.5$ in \pp\ collisions at $\sqrts=200$ GeV.}
\label{Fig:ppResponse}
\end{figure}
%---
%---
\begin{figure}[htbp]
\centering
\includegraphics[width=0.8\textwidth]{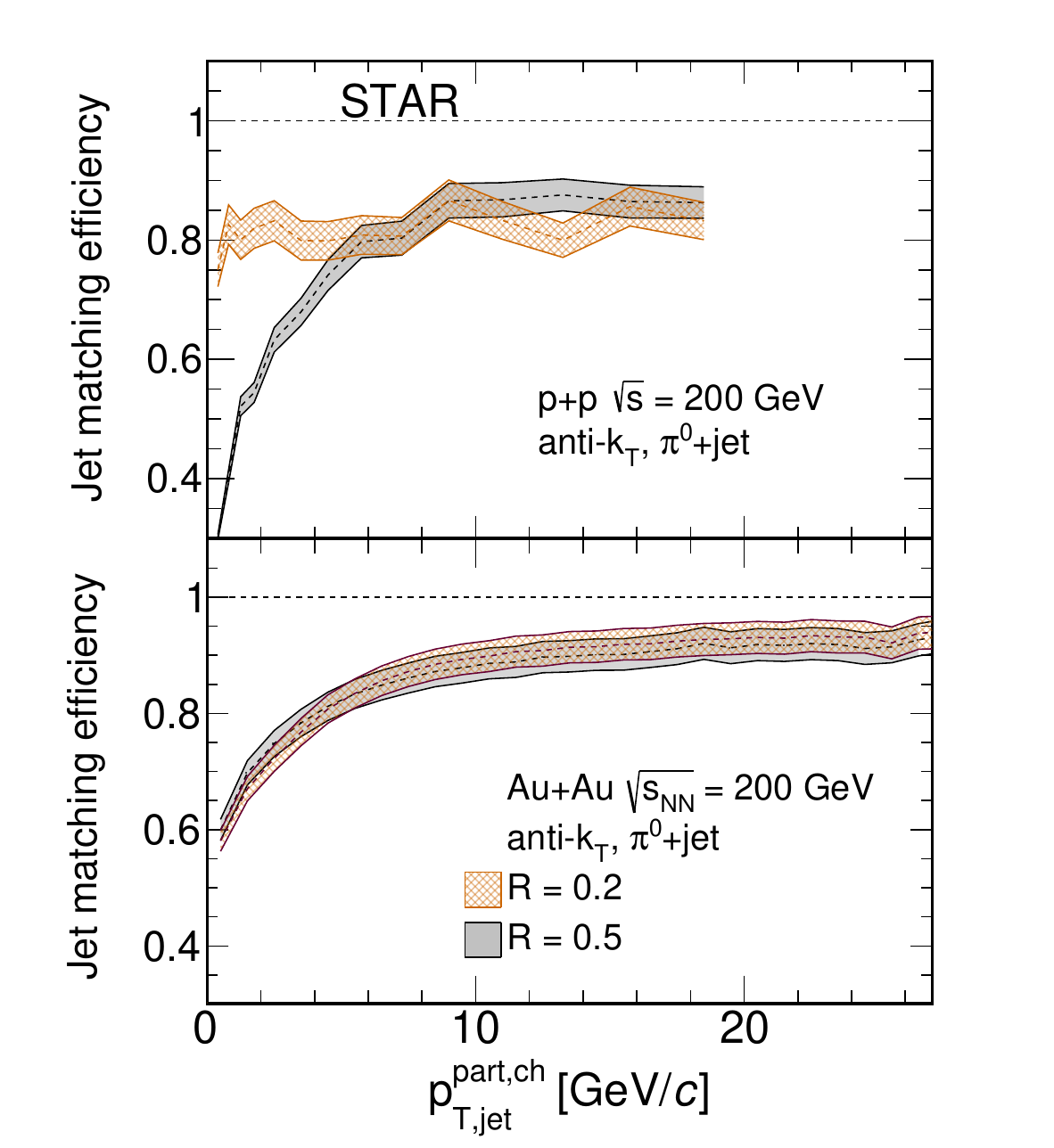}
\caption{Jet matching efficiency as a function of \pTjetpart, in \pp\ (upper) and central \AuAu\ (lower) collisions for recoil jets with $\rr=0.2$ and 0.5. The bands show the systematic uncertainties due to variation in the tracking efficiency. See text for discussion.}
\label{Fig:JetMatchEff}
\end{figure}
%---

Figure~\ref{Fig:ppResponse} shows \Rinstr, the response matrix for instrumental effects. The matrix is normalized such that the integral over \pTjetdet\ is unity for each bin in \pTjetpart. Figure~\ref{Fig:JetMatchEff}, upper panel, shows the jet-matching efficiency in \pp\ collisions for $\rr=0.2$ and 0.5.

The response matrix for \pp\ collisions without the \qTjet\ cut has contributions on both axes (low \pTjetdet or low \pTjetpart) that arise from jet splitting at either the particle or detector level, or both. The case of two close particle-level jets with only one nearby detector-level jet is accounted for by the jet reconstruction efficiency. For the case of one particle--level jet with two nearby detector--level jets, no correction is applied. For such occurrences in the actual data analysis, these detector--level split jets are corrected using the diagonal response in Fig.~\ref{Fig:ppResponse}. The closure test (Sec.~\ref{sect:Closure}), which includes these matching algorithm effects, validates the unfolding procedure, demonstrating that such splitting effects are not significant. 

%------------------------------------
\subsubsection{\AuAu\ collisions}
\label{sect:aaInstrumentEffects}

%The largest instrumental effect in this analysis is the tracking efficiency, which depends on the TPC occupancy and varies significantly between \pp\ collisions and central \AuAu\ collisions. 

The charged-particle tracking efficiency for central \AuAu\ collisions is determined by embedding detector-level simulated tracks for pions, kaons, and protons into real \AuAu\ events. The charged-particle tracking efficiency corresponds to the weighted average of efficiencies for individual species, with the weight based on measurements of the relative inclusive particle yield for each species~\cite{Abelev:2008ab}.

The instrumental response matrix \Rinstr\ is calculated using PYTHIA 8.185 to generate \pizero-triggered and \gammadir-triggered events for \pp\ collisions at $\sqrts=200$ GeV. These particle-level events are passed through a fast simulator incorporating a parameterization of the charged-particle tracking response for \AuAu\ collisions, with track acceptance $|\eta|<1.0$ and no cut on \pT. Jet reconstruction is run on the resulting detector-level particles. Matching between particle-level and detector-level recoil jets is carried out as described above for \pp\ collisions, except that \pTjetdet\ is only required to be at least 15\% of \pTjetpart, consistent with~\cite{Adamczyk:2017yhe}. Figure~\ref{Fig:JetMatchEff}, lower panel, shows the jet-matching efficiency for central \AuAu\ collisions. 

The jet-matching efficiencies for \pp\ and \AuAu\ collisions in Fig.~\ref{Fig:JetMatchEff} differ significantly. These differences arise from the different criteria for matching particle- and detector-level jets, with more stringent matching requirements for \pp\ collisions. In addition, the jet-matching efficiency for $\rr=0.2$ in \pp\ collisions approaches the single charged-particle tracking efficiency and therefore has a different shape than for $\rr=0.5$. We note that the different matching criteria also generate differences in \Rinstr, with opposite effect. However, these are intermediate steps in the analysis; it is only the combination of unfolding and the efficiency correction that has absolute meaning, and ideally should not be dependent upon the specific choice of analysis parameters. The residual dependence on the specific matching criteria is accounted for in the systematic uncertainty.

%----
\begin{figure}[htbp]
\centering
\includegraphics[width=0.65\textwidth]{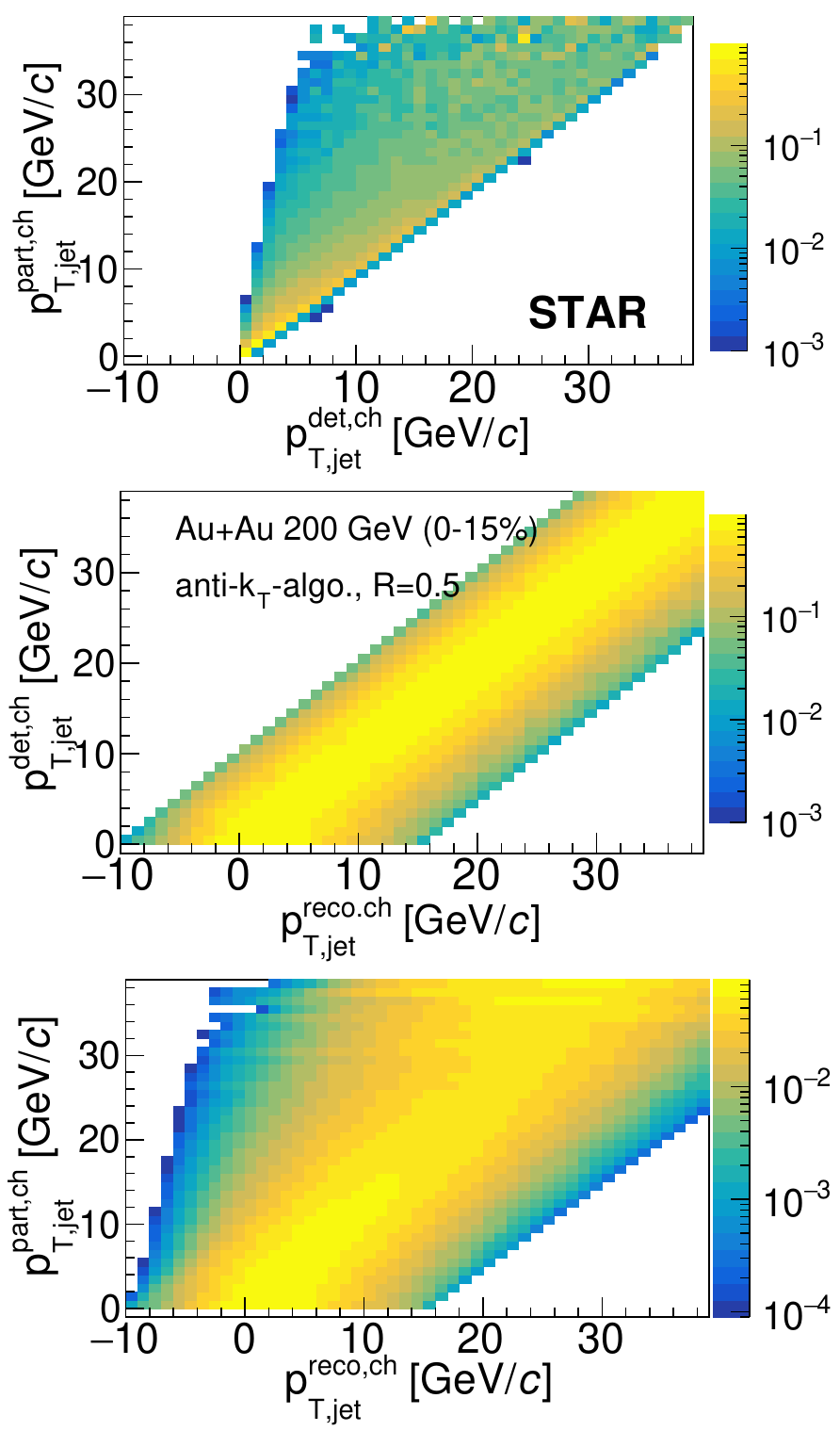}
\caption{Response matrices for recoil jets with $\rr=0.5$ in central \AuAu\ collisions with \pizero\ trigger. Top: instrumental response matrix \Rinstr; 
middle: background fluctuation response matrix (SP embedding) \Rbkgd; 
bottom: total response matrix \Rtot. 
}
\label{Fig:AuAuRespMatrix}
\end{figure}
%---
 
Figure~\ref{Fig:AuAuRespMatrix}, upper panel, shows \Rinstr\ for recoil jets with $\rr=0.5$ in central \AuAu\ collisions. The other panels are discussed below. 

%---------------------------------
\subsubsection{Jet energy scale and resolution}
\label{sect:JESJER}

While the full instrumental response matrix, \Rinstr, is used to correct the measured recoil-jet spectrum via unfolding, it is nevertheless valuable to characterize the instrumental response qualitatively. The instrumental response is similar in this analysis to that reported in Ref.~\cite{Adamczyk:2017yhe}, for both \pp\ and \AuAu\ collisions.

For charged-particle jets in the range $5<\pTreco<30$ \gev, the distribution characterizing jet energy resolution (JER) has RMS $\approx25$\%, comprising a peak with width 5-10\%, with an asymmetric tail to low jet energy, with negligible \rr-dependence. The jet energy scale (JES)
uncertainty is 5\%, likewise with negligible \rr-dependence.

%========================
\subsection{Uncorrelated background: \AuAu\ collisions}
\label{Sect:UncorrBkgd}

The effect of uncorrelated background fluctuations on \pTjetdet\ is determined using the embedding procedure described in Ref.~\cite{Adamczyk:2017yhe}. Three different classes of jet model are used for systematic variation: PYTHIA-generated jets at the detector level; the same but with azimuthal correlation between trigger \pizero\ and the EP due to non-zero \vtwo\ of the trigger hadron; and ``Single Particle'' (SP) jets in which one particle carries \pTreco\ and is collinear with the jet centroid. 

As shown in Refs.~\cite{Adam:2015doa,Adamczyk:2017yhe,STAR:2020xiv}, the background response matrices for inclusive and semi--inclusive jet distributions calculated by embedding PYTHIA-generated jets and SP jets are very similar, demonstrating that jet reconstruction in the heavy-ion environment is largely independent of the specific fragmentation distribution of the jet population. This in turn indicates that corrections based on this procedure are valid for inclusive and semi-inclusive distributions of quenched jets, whose fragmentation distribution is {\it a priori} unknown. We utilize SP embedding for the primary analysis, because it provides the clearest  criteria for matching particle and detector-level jets.

The uncorrelated background distribution is per definition the same for \gammadir-triggered and \pizero-triggered datasets, except for a possible EP bias in the \pizero\ case. However, the effect of EP bias in central \AuAu\ collisions has been shown to be negligible for charged--hadron triggers~\cite{Adamczyk:2017yhe}; the same was found in this analysis. No significant difference is observed in the response matrices for \gammadir-triggered and \pizero-triggered events. We therefore focus on the response matrix for \pizero-triggered events, which is used to unfold both the \gammadir-triggered and \pizero-triggered distributions. Figure~\ref{Fig:AuAuRespMatrix}, middle panel, shows \Rbkgd\ for \pizero-triggered central \AuAu\ collisions using SP embedding, for $\rr=0.5$.

%========================
\subsection{Unfolding}
\label{sect:Unfolding}

%----------------------------
\subsubsection{Formalism}
\label{sect:UnfoldingForm}

The \pTreco\ distribution is first corrected for uncorrelated yield by subtracting the normalized ME distribution (Sect.~\ref{sect:RawDistr}) giving the vector $M$, the measured distribution of correlated yield. $M$ is related to the Truth distribution $T$ according to

\begin{equation}
M(\pTreco) = \Rtotarg \times T(\pTjetpart),
\label{Eq:Smearing}
\end{equation}

\noindent
where \Rtot\ represents the cumulative response matrix from both the instrumental and heavy-ion background fluctuations. We assume that \Rtot\ factorizes\cite{Adamczyk:2017yhe}, such that

\begin{equation}
\Rtotarg =  \Rbkgdarg \times \Rinstrarg.
\label{Eq:Rtot}
\end{equation}

\noindent
Equation~\ref{Eq:Smearing} is then approximately inverted using regularized unfolding methods~\cite{Hocker:1995kb,DAgostini:1994fjx}. We utilize Singular Value Decomposition (SVD)~\cite{Hocker:1995kb} and Iterative Bayesian unfolding~\cite{DAgostini:1994fjx} as implemented in the RooUnfold package~\cite{Adye:2011gm}. %{roounfold}. 

%--------------------------------------------------------

\subsubsection{\pp\ collisions}
\label{sect:ppUnfolding}

For \pp\ collisions, uncorrelated background  is negligible in the \pTjet-range considered in the analysis. Only corrections for the instrumental response therefore need to be unfolded; i.e. $\Rtot\equiv\Rinstr$. Determination of \Rinstr\ is described in Sect.~\ref{sect:ppInstrumentEffects}. 

Iterative Bayesian unfolding is used for \pp\ collisions. The unfolding is regularized by restricting the number of iterations to $n_{\rm iter} = 4$ for $\rr=0.2$ and $n_{\rm iter} = 3$ for $\rr=0.5$. These values were determined by the inflection point in the distribution of $\chi^{2}/$NDF from comparison of the backfolded distribution and the raw data, and adding one iteration.  The prior distributions used in the unfolding correspond to the \pizero-triggered recoil-jet distributions from the simulation described in Sect.~\ref{sect:ppInstrumentEffects}.
% The systematic uncertainty on the dependence of the results on this choice of $n_{\rm iter}$ is determined by varying the regularization $n_{\rm iter} \pm 1$.

%----------------------------
\subsubsection{{\AuAu} collisions}
\label{sect:AuAuUnfolding}

Figure~\ref{Fig:AuAuRespMatrix}, bottom panel, shows the combined response matrix for \pizero-triggered central \AuAu\ collisions. Prior distributions correspond to \pizero-triggered and \gammadir-triggered recoil-jet distributions generated by PYTHIA, modified by suppression of the recoil-jet yield to account for jet quenching~\cite{Adamczyk:2017yhe}. 
 
 For Iterative Bayesian unfolding, regularization corresponds to truncation of the number of iterations at values between 2 and 5. SVD unfolding is regularized by truncating the expansion at a value of the stabilization term ($k$). This value is varied in the range 2-4, and is optimized using the value of $\chi^{2}/$NDF from comparing the unfolded distribution multiplied by \Rtot\ (``backfolding'', analogous to Eq.~\ref{Eq:Smearing}) with the measured distribution $M$.

The \pTreco\ distributions after ME subtraction (Fig.~\ref{Fig:MESubRawSpec}) constitute the input distributions for unfolding. The distributions for $\rr=0.2$ drop rapidly in the region $\pTreco<0$, and the full distribution is used. However, for $\rr=0.5$ the subtracted distributions extend to $\pTreco<-10$ \gev, with data points that oscillate around zero with large variance (note that negative values are not shown in Fig.~\ref{Fig:MESubRawSpec}, due to the logarithmic vertical axis). 

The region $\pTreco<0$ of the $\rr=0.5$ distributions therefore has limited information content, and inclusion of that range gives unstable unfolding  results. Unfolding for $\rr=0.5$ therefore only utilizes data from the distributions in Fig.~\ref{Fig:MESubRawSpec} for $\pTreco>0$.
\section{Systematic uncertainties}
\label{sect:SysUncert}

Systematic uncertainties are determined following the procedures described in Ref.~\cite{STAR:2016jdz,Adamczyk:2017yhe}. Contributions to the systematic uncertainty are as follows:

\begin{itemize}
    \item instrumental effects: uncertainty in tracking efficiency and resolution;
    \item unfolding: uncertainty due to choice of regularization, prior, and algorithm;
    \item uncorrelated background yield, \AuAu\ collisions: uncertainty in boundary of normalization region for measuring \fME;
    \item \gammadir\ purity: uncertainty in measured value of \Rpurity\ (Eq.~\ref{Eq:Rpurity}); 
    \item fragmentation model: variation in choice of Monte-Carlo model used to calculate the response matrix and jet-matching efficiency.
\end{itemize}

\noindent 

For each source of uncertainty, alternative unfolding solutions are generated by varying the corresponding parameter or algorithm. The variation with the largest relative difference to the baseline solution is assigned as the systematic uncertainty for that source. Variants are excluded if unfolding does not converge or if there is anomalously large $\chi^{2}/\text{NDF}$ between the back-folded solution and the raw data. 

%----------------------------------
\subsection{\pp\ collisions}
\label{sect:ppSysUncert}
 
The systematic variations applied for \pp\ collisions are as follows:

\begin{itemize}

\item For the unfolding uncertainty, $n_{\text{iter}}$ is varied $\pm 1$ relative to its baseline value. Several L\'evy functions are used as alternative priors~\cite{Adamczyk:2017yhe}. 

\item For \gammadir-triggered distributions, \Rpurity\ is varied within its measured uncertainties (see Sect.~\ref{sect:GammajetConversion}).  Near the limit of the kinematic range for jets recoiling from \gammadir\ triggers, the uncertainty on the recoil-jet yields can be very large, as the yields can be consistent with zero.

\item For the instrumental uncertainty, an alternative unfolding procedure is carried out in which \Rinstr\ is constructed from a sample of \gammadir- and \pizero-triggered events generated by PYTHIA 8.185, with a parameterized instrumental response applied as described below. The resulting \Rinstr\ matrix is used to unfold the corresponding recoil-jet distributions. The parameterized instrumental response is then varied: the tracking efficiency is varied by $\pm 4\%$ (absolute), and an alternative fit to the one described above is used for the tracking resolution.

\item The fragmentation model for the calculation of \Rinstr\ is varied by using \gammadir-triggered events generated by HERWIG-7 for \pp\ collisions at $\sqrts=200$ GeV with default tune~\cite{HERWIG:2008}. The difference between PYTHIA and HERWIG-derived corrections is found to be independent of \ETtrig. The jet--finding efficiency (Fig.~\ref{Fig:JetMatchEff}) may also depend upon the fragmentation model, which was likewise explored by comparing PYTHIA and HERWIG-derived corrections. The total uncertainty from the fragmentation--model dependence of both \Rinstr\ and jet--finding efficiency is negligible for $\rr=0.2$ over the full measured range of \pTjet. For $\rr=0.5$, this uncertainty is 15\% at $\pTjet=3$~\gev, 6\% at $\pTjet=5$~\gev, and negligible for $\pTjet>6$~\gev.

\end{itemize}

Due to the limited statistical precision of the embedding dataset described in Sect. \ref{sect:ppInstrumentEffects}, 
%and the inability to freely adjust the instrumental response of this simulation, 
the instrumental systematic uncertainty for \pp\ collisions is evaluated using a parameterization of the embedding dataset instrumental response, which can be varied to represent systematic variations. The effects of these variations are explored using \pizero-triggered recoil-jet spectra generated by PYTHIA 8.185.

The parametrization has two components: tracking resolution $\sigma_{\text{trk}}$, and tracking efficiency $\epsilon_{\text{trk}}$. The tracking resolution is quantified by a polynomial fit to the ratio of the difference between reconstructed and matched particle-level \pT\ to the matched particle-level \pT, with functional form

\begin{equation}
    \sigma_{\pT} = \sigma_{0} + \sigma_{1} \pT + \sigma_{2} \pT^{2},
\end{equation}

\noindent where $\sigma_{i}$ are constants.
%with values $\{\sigma_{i}\} = \{0.0045, 0.0070, 0.0013\}$.
The tracking efficiency is quantified by a fit to the ratio of the number of reconstructed tracks over the number of generated particles, with functional form

\begin{equation}
    \epsilon_{\text{trk}} \left( \pTtrack \right) = \epsilon_{0} + \epsilon_{1} e^{-k_{1} \pTtrack} + \epsilon_{2} e^{-k_{2} \pTtrack^{2}}.
\end{equation}

\noindent 
%where $\epsilon_{i}$ and $k_{i}$ are constants with values $\{\epsilon_{i}\} = \{0.82, -2.18, 0.18\}$ and $\{k_{i}\} = \{13, 0.67\}$. 
The second-order term in $\epsilon_{\text{trk}}$, and the dependence of $\epsilon_{\text{trk}}$ on reconstructed \pT\ rather than particle-level \pT, account for differences in the treatment of pileup and secondary decays in the PYTHIA 8.185 and embedding datasets. In the former, no pileup was simulated and all short-lived particles were allowed to decay, while the opposite holds for the latter. Note that the response matrix and jet-matching efficiency of the PYTHIA 8.185 dataset were not used to correct the measured data, but only to assess the instrumental systematic uncertainty.

% pp systematic uncertainties (pi0)
\begin{table}
  \centering
\caption{Significant systematic uncertainties for \pizero\ triggers in \pp\ collisions, in representative \pTjetch\ intervals. The cumulative uncertainty is the sum in quadrature of the individual contributions.}
  \begin{tabular}{| c | c | c || c | c || c |}
    \hline 
    \multicolumn{6}{|c|}{\pp; Trigger: \pizero} \\
    \hline
    \ETtrig\ & \rr\ & \pTjet & \multicolumn{3}{|c|}{Systematic uncertainty (\%)} \\
    \cline{4-6}
    [GeV] & & [\gev] & instr & unfold & cumulative \\
    \hline
    \multirow{6}{*}{[9,11]} & \multirow{3}{*}{0.2} & [5,10] & 7 & 2 & 7 \\
    & & [10,15] & 9 & 2 & 9 \\
    & & [15,20] & 12 & 4 & 12 \\
    \cline{2-6}
    & \multirow{3}{*}{0.5} & [5,10] & 6 & 4 & 8 \\
    & & [10,15] & 10 & 4 & 11 \\
    & & [15,20] & 14 & 7 & 16 \\
    \hline
    \multirow{6}{*}{[11,15]} & \multirow{3}{*}{0.2} & [5,10] & 7 & 1 & 7 \\
    & & [10,15] & 8 & 2 & 8 \\
    & & [15,20] & 10 & 3 & 11 \\
    \cline{2-6}
    & \multirow{3}{*}{0.5} & [5,10] & 5 & 5 & 8 \\
    & & [10,15] & 9 & 5 & 10 \\
    & & [15,20] & 11 & 9 & 14 \\
    \hline
  \end{tabular}
  \label{Tab:PpSysUncertaintyPi0}  
\end{table}

% pp systematic uncertainties (gamma)
\begin{table}
  \centering
    \caption{Significant systematic uncertainties for \gammadir\ triggers in \pp\ collisions, in representative \pTjetch\ intervals. The cumulative uncertainty is the sum in quadrature of the individual contributions.}
  \begin{tabular}{| c | c | c || c | c | c || c |}
    \hline 
    \multicolumn{7}{|c|}{\pp; Trigger: \gammadir} \\
    \hline
    \ETtrig\ & \rr\ & \pTjet & \multicolumn{4}{|c|}{Systematic uncertainty (\%)} \\
    \cline{4-7}
    [GeV] & & [\gev] & instr & unfold & purity & cumulative \\
    \hline
    \multirow{2}{*}{[9,11]} & 0.2 & [5,10] & 11 & 2 & 80 & 81 \\
    \cline{2-7}
    & 0.5 & [5,10] & 14 & 15 & 81 & 83 \\
    \hline
    \multirow{4}{*}{[11,15]} & \multirow{2}{*}{0.2} & [5,10] & 9 & 2 & 21 & 23 \\
    & & [10,15] & 17 & 101 & 100 & 143 \\
    \cline{2-7}
    & \multirow{2}{*}{0.5} & [5,10] & 7 & 8 & 2 & 10 \\
    & & [10,15] & 12 & 12 & 16 & 23 \\
    \hline
    \multirow{4}{*}{[15,20]} & \multirow{2}{*}{0.2} & [5,10] & 3 & 2 & 13 & 14 \\
    & & [10,15] & 2 & 16 & 2 & 17 \\
    & & [15,20] & 8 & 13 & 15 & 21 \\
    \cline{2-7}
    & \multirow{2}{*}{0.5} & [5,10] & 4 & 2 & 12 & 13 \\
    & & [10,15] & 3 & 2 & 12 & 16 \\
    & & [15,20] & 2 & 6 & 15 & 16 \\
    \hline
  \end{tabular}
  \label{Tab:PpSysUncertaintyGam}  
\end{table}

Tables~\ref{Tab:PpSysUncertaintyPi0} and~\ref{Tab:PpSysUncertaintyGam} show the largest systematic uncertainties for \pp\ collisions. The cumulative systematic uncertainty of the recoil-jet yield is the quadrature sum of uncertainties from each component. The large systematic uncertainties in some bins in Tab.~\ref{Tab:PpSysUncertaintyGam} are due to the rapid change in yield in these bins, which are at the limit of statistical significance in this analysis. We nevertheless specify an uncertainty for such bins, since a factor $\approx2$ uncertainty still provides a meaningful yield limit.

%------------------------------------------------
\subsection{\AuAu\ collisions}
\label{sect:aaSysUncert}

The uncertainty associated with the tracking efficiency is assessed by varying the absolute efficiency by $\pm4$\%. Corrections for the track \pT-resolution and weak decays are found to be significantly smaller than those associated with the tracking efficiency, consistent with Ref.~\cite{Adamczyk:2017yhe}. No systematic uncertainty due to the \pT-resolution or weak decays is assigned.

The uncertainty due to the choice of unfolding algorithm was assessed by comparing the results of the SVD algorithm to the alternative Iterative Bayesian algorithm. Uncertainty due to the choice of prior was determined by using a modified PYTHIA-generated prior (Sect.~\ref{sect:AuAuUnfolding}) and a Levy 
function~\cite{Adamczyk:2017yhe}). The uncertainty associated with the choice of regularization for SVD unfolding was assessed by varying the value of $k_{\text{reg}}$ $\pm1$ relative to the default value. Varying the choice of regularization for Iterative Bayesian unfolding generates negligible change. 

Table~\ref{Tab:SEMENorm} and Figs.~\ref{Fig:SEMEGjet} and \ref{Fig:SEMEPijet} show the nominal \fME\ normalization regions. To assess the uncertainty associated with this choice, the upper limit of this region was varied by 1~\gev.

The value of \Rpurity\ is determined from the near-side correlation yields of \pizero\ vs. \gammarich\ triggers (Sect.~\ref{sect:GammajetConversion}). The systematic uncertainty in \Rpurity\ is estimated by varying the range of $\zT = \pTtrack / \ETtrig$ of the tracks counted in the near-side correlation measurement~\cite{STAR:2016jdz}.  By varying \Rpurity\ within its uncertainty, the corresponding uncertainty on the recoil-jet yields is determined.  
%Note that this uncertainty depends on \ETtrig. 
 
The effects of varying the fragmentation model used to determine \Rtot\ for central \AuAu\ collisions were studied in Ref.~\cite{Adamczyk:2017yhe}, and were found to be small relative to other systematic effects. In the measurement of h+jet correlations in \PbPb\ collisions at \sqrtsNN=2.76 TeV~\cite{Adam:2015doa}, variations in jet fragmentation model, including a simulation of the effects of jet quenching, likewise resulted in sub-leading systematic uncertainties for recoil jets in the kinematic range of this analysis. We therefore do not consider this contribution further here for central \AuAu\ measurements.

Correction for jet-finding efficiency (Fig.~\ref{Fig:JetMatchEff}) is applied to the unfolded distribution. For \pp\ collisions, the fragmentation-model dependence of the efficiency is explored by comparing HERWIG and PYTHIA calculations (Sect.~\ref{sect:ppSysUncert}) and found to be negligible for $\pTjetch>6$ \gev. Due to the multi-hadron nature of jets, similar response is expected in central \AuAu\ collisions, where quenching effects will, if anything, increase the jet constituent multiplicity, thereby increasing the jet--finding efficiency relative to the single-track efficiency. We therefore assign the same systematic uncertainty for the fragmentation-pattern dependence of jet--finding efficiency for central \AuAu\ collisions as that found for \pp\ collisions, which provides a conservative estimate.

% AuAu systematic uncertainties (pi0)
\begin{table}
  \centering
    \caption{Significant systematic uncertainties for \pizero\ triggers in central \AuAu\ collisions, in representative \pTjetch\ intervals. Cumulative uncertainty is the quadrature sum of the individual contributions.}
  \begin{tabular}{| c | c | c || c | c | c | c || c |}
    \hline 
    \multicolumn{8}{|c|}{Central \AuAu; Trigger: \pizero} \\ \hline
    \ETtrig\ & \rr\ & \pTjet & \multicolumn{5}{|c|}{Systematic uncertainty (\%)} \\
    \cline{4-8}
    [GeV] &  & [\gev] & instr & unfold  & ME norm & \dpT & cumulative \\	\hline
    \multirow{6}{*}{[9,11]} & \multirow{3}{*}{0.2} & [5,10] &  4  &  11 & 17& 18 & 27 \\
    & & [10,15] &  5  &  20 & 2 & 9 & 22\\
    & & [15,20] &  5  &  22 & 4 & 13  & 26 \\ \cline{2-8}
    & \multirow{3}{*}{0.5} & [5,10] &  3  &  34 & 7 & 5 & 35 \\
    & & [10,15] &  7  &  36 & 2 & 4 & 36 \\
    & & [15,20] &  8  &  24 & 2 & 9 & 26 \\ \hline
    \multirow{6}{*}{[11,15]} & \multirow{3}{*}{0.2} & [5,10] &  4  &  36 & 14 & 10 & 40 \\
    & & [10,15] &  5  &  29 & 8 & 11 & 32 \\
    & & [15,20] &  5  &  38 & 7 & 7 & 39 \\ \cline{2-8}
    & \multirow{3}{*}{0.5} & [5,10] &  6  &  38 & 4 & 6 & 39 \\
    & & [10,15] &  8  &  29 & 3 & 3 & 30 \\
    & & [15,20] &  8  &  22 & 2 & 7 & 24 \\ \hline
  \end{tabular}
  \label{Tab:SysUncertPizero}  
\end{table}

% AuAu systematic uncertainties (gamma)
\begin{table}
  \centering
    \caption{Significant systematic uncertainties for \gammadir\ triggers in central \AuAu\ collisions, in representative \pTjetch\ intervals. Cumulative uncertainty is the quadrature sum the individual contributions.}
  \begin{tabular}{| c | c | c || c | c | c | c  |c  || c |}
    \hline 
    \multicolumn{9}{|c|}{Central \AuAu; Trigger: \gammadir} \\ \hline
    \ETtrig\ & \rr\ & \pTjet & \multicolumn{6}{|c|}{Systematic uncertainty (\%)} \\
    \cline{4-9}
    [GeV] &  & [\gev] & instr & unfold  & ME norm & \dpT & purity & cumulative \\	\hline
    \multirow{2}{*}{[9,11]} & 0.2 & [5,10] &  4  &  30 & 18 & 14 & 12 & 40 \\ \cline{2-9}
    & 0.5 & [5,10] &  5  &  30 & 16 & 3 & 5 & 34 \\ \hline
    \multirow{4}{*}{[11,15]} & \multirow{2}{*}{0.2} & [5,10] &  4  &  36 & 17 & 10 & 3 & 41 \\
    & & [10,15] &  5  &  20 & 10 & 15 & 16 & 31 \\ \cline{2-9}
    & \multirow{2}{*}{0.5} & [5,10] &  2  &  36 & 10 & 8 & 2& 38 \\
    & & [10,15] &  5  &  20 & 11 & 3 & 11 & 26 \\ \hline
    \multirow{4}{*}{[15,20]} & \multirow{2}{*}{0.2} & [5,10] &  4  &  26 & 12 & 11 & 10 & 32 \\
    & & [10,15] &  5  &  32 & 11 & 10 & 23 & 42 \\ \cline{2-9}
    & \multirow{2}{*}{0.5} & [5,10] &  3  &  44 & 12 & 9& 5 & 46 \\
    & & [10,15] &  4  &  42 & 10 & 6 & 9 & 44\\ \hline
  \end{tabular}
  \label{Tab:SysUncertGammadir}  
\end{table}

Tables~\ref{Tab:SysUncertPizero} and \ref{Tab:SysUncertGammadir} show the systematic uncertainties for central \AuAu\ collisions. The cumulative systematic uncertainty in the recoil-jet yield is the quadrature sum of the uncertainties from each individual source. 
\section{Closure test}
\label{sect:Closure}

A closure test is used to validate the analysis algorithm. The test is based on detector-level events which are embedded into real data to model the effects of uncorrelated background and then subject to the full analysis chain that is used for the real data analysis, including unfolding and estimation of systematic uncertainties. Corrected distributions from these fully analysed events are then compared to those from the initial particle-level events (``Truth''). Successful closure corresponds to the agreement of these two distributions within uncertainties.

%------------------
\subsection{\pp\ collisions}
\label{subsect:ClosurePP}

The \pp\ closure test was carried out using events generated  by PYTHIA 6.426 Perugia 0 tune for \pp\ collisions at $\sqrts=200$ GeV, selected by a di-jet trigger condition. Detector-level events from this population are embedded into STAR 2009 zero-bias \pp\ data (hybrid events). Jets are reconstructed in the same manner as data using both the particle-level and detector-level hybrid events. The detector-level recoil-jet distributions are modified to correspond to the trigger statistics of the measured data. 

% [Text as of version 6] The \pp\ closure test was carried out using events generated  by PYTHIA 6.426 Perugia 0 tune for \pp\ collisions at $\sqrts=200$ GeV, selected by a di-jet trigger condition. Detector-level events from this population are embedded into STAR 2009 zero-bias \pp\ data (hybrid events). Jets are reconstructed in the same manner as data using both the particle-level and detector-level hybrid events. The detector-level recoil jet distributions are modified to correspond to the trigger statistics of the measured data. 

The detector-level event population is divided into two sub-samples of roughly equal size. These sub-samples correspond to the two orientations of the STAR  magnetic field in which data were taken: Full-Field (FF) and Reverse Full-Field (RFF). The closure test utilizes the RFF sub-sample for calculating \Rtot\ and the FF sub-sample serving as pseudo-data for validation. Jet reconstruction, unfolding, and determination of systematic uncertainties, are carried on the RFF sub-sample in the same way as is done for real data.

% [Text as of version 6] The detector-level event population is divided randomly into two equal sub-samples, corresponding to the two orientations of the STAR  magnetic field: Full-Field (FF) and Reverse Full-Field (RFF). The closure test utilizes the RFF sub-sample for calculating \Rtot\ and the FF sub-sample serving as pseudo-data for validation. Jet reconstruction, unfolding, and determination of systematic uncertainties, are carried on the RFF sub-sample in the same way as is done for real data.

%Jet reconstruction is carried out on the FF detector-level events. The raw jet distribution is then unfolded using the response matrix and jet-matching efficiency trained on the RFF sub-sample. The instrumental response is applied, as it is known by GEANT simulation, and then corrected, as quantified by the same simulation. Thus, the tracking efficiency and momentum resolution are not varied, as they are not tested by the closure test. Unfolding is carried out five times, each iteration varying the unfolding parameters as was done to assess the \pp\ unfolding systematic uncertainty. Finally, the five unfolding solutions are averaged together and are compared against the true distribution (i.e. the modified FF particle-level recoil jet spectrum).

% pp closure test
\begin{figure}[htbp]
\centering
\includegraphics[width=0.49\textwidth]{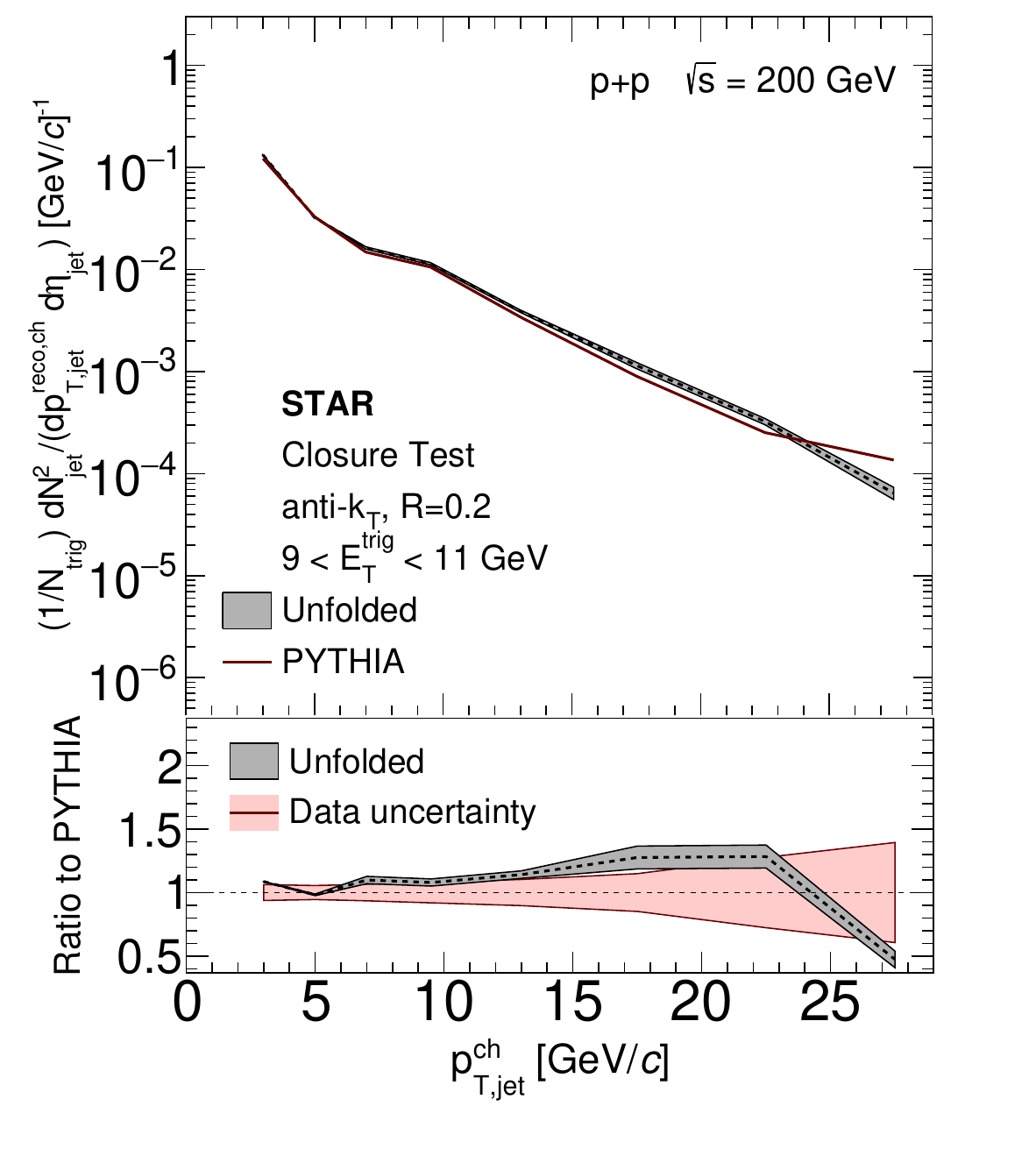}
\includegraphics[width=0.49\textwidth]{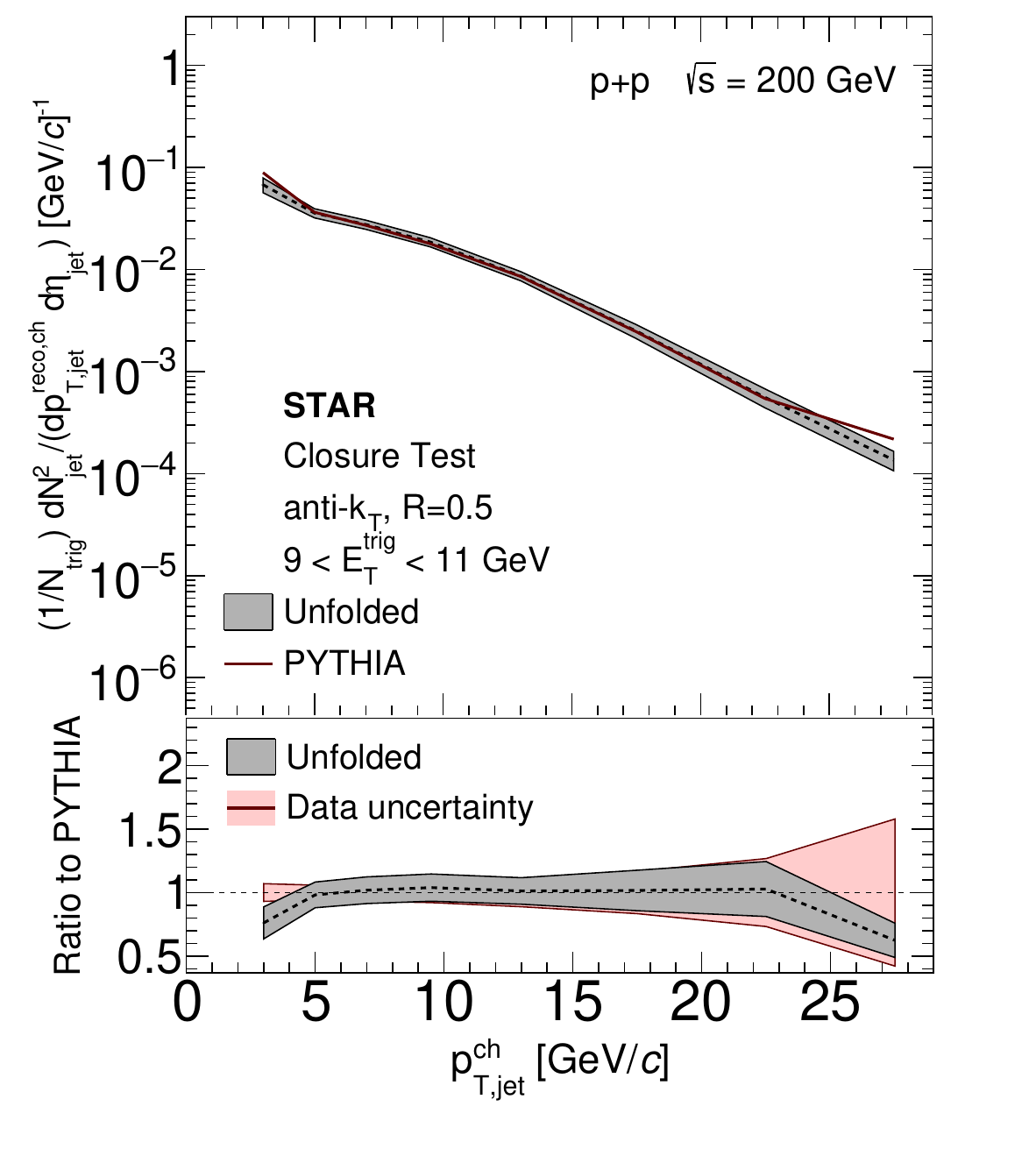}
\caption{Closure test for \pp\ collisions: \pizero-triggered events with $9<\ETtrig<11$ GeV; recoil jets with $\rr=0.2$ (left) and 0.5 (right). Upper panel: semi-inclusive recoil-jet distributions. Dotted line shows unfolded detector-level distribution. Band on unfolded detector-level represents systematic uncertainty. Red line shows particle-level jet distribution. Band on particle-level represents systematic precision of the data. Lower panel: ratio of unfolded detector- and particle-level distributions.}
\label{Fig:PpClosureTest}
\end{figure}

Figure~\ref{Fig:PpClosureTest} shows corrected distributions for \pizero-triggered events with $9<\ETtrig<11$ GeV and recoil jets with $\rr=0.2$ and 0.5, compared to the corresponding Truth (particle-level) distributions whose statistical errors have been modified to match the systematic precision of the data. The unfolded and Truth distributions agree within the systematic precision of the measurement, thereby validating the \pp\ analysis chain for this choice of kinematics. Similarly good closure is observed for triggers with $11<\ETtrig<15$ and $15<\ETtrig<20$ GeV, and with the roles of the FF and RFF data sub-samples reversed. This closure study validates the \pp\ analysis chain over the full reported kinematic range.

%-----------------------------------
\subsection{\AuAu\ collisions}
\label{subsect:ClosureAuAu}

The closure test for central \AuAu\ collisions utilizes PYTHIA-generated events for \pp\ collisions at $\sqrts=200$ GeV which contain a \pizero\ trigger in the range $9<\ETtrig<11$ GeV. The trigger statistics are the same as those of the real data.  Detector-level events are generated with the same fast simulation approach used in Sect.~\ref{sect:aaInstrumentEffects} and are embedded in central \AuAu\ events recorded with a MB trigger. The full analysis chain is then applied to these hybrid events, including jet reconstruction, subtraction of ME, and corrections. Systematic uncertainties are determined by the same procedures as used for real data.

%------
\begin{figure}[htbp]
\centering
\includegraphics[width=0.49\textwidth]{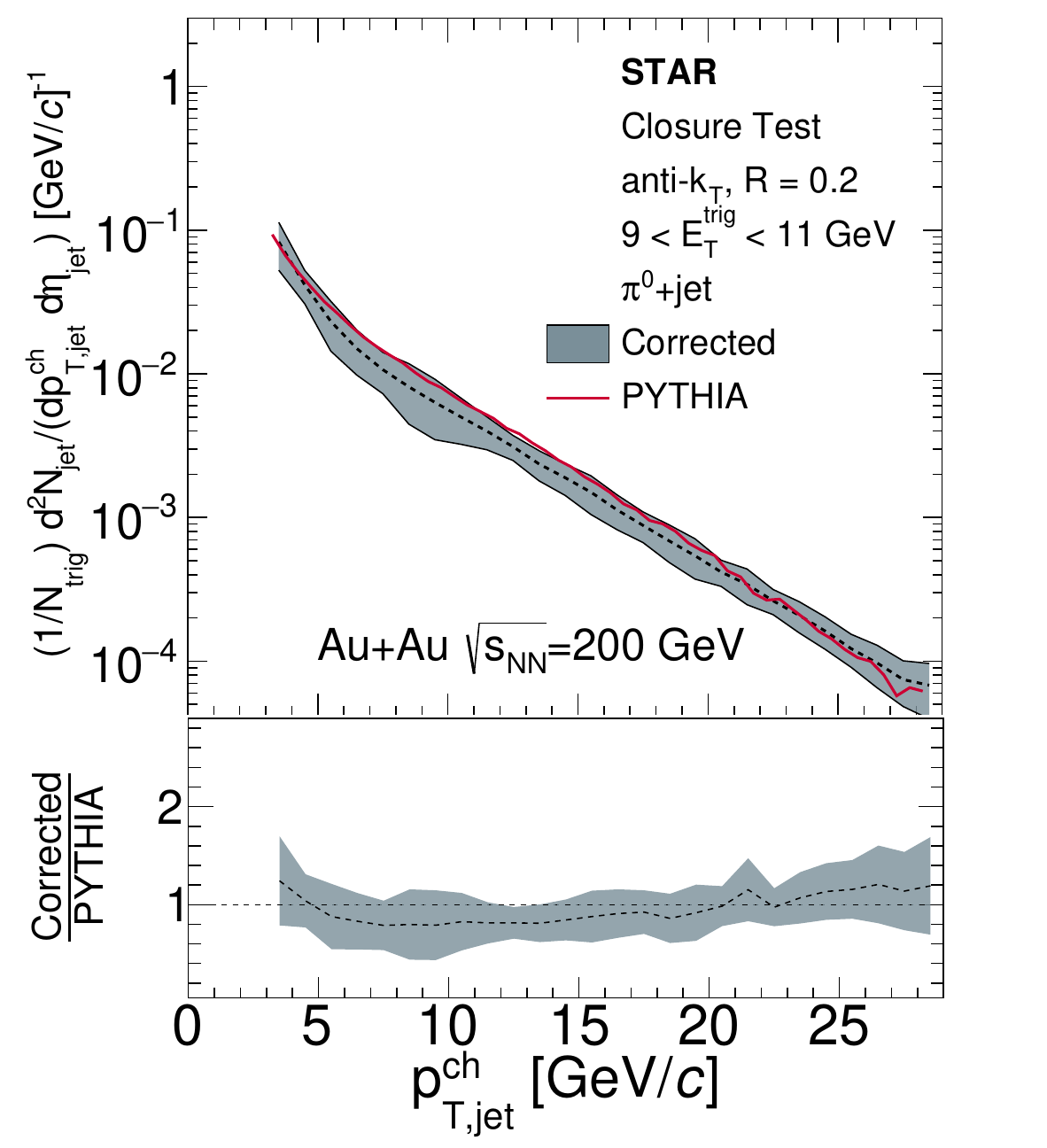}
\includegraphics[width=0.49\textwidth]{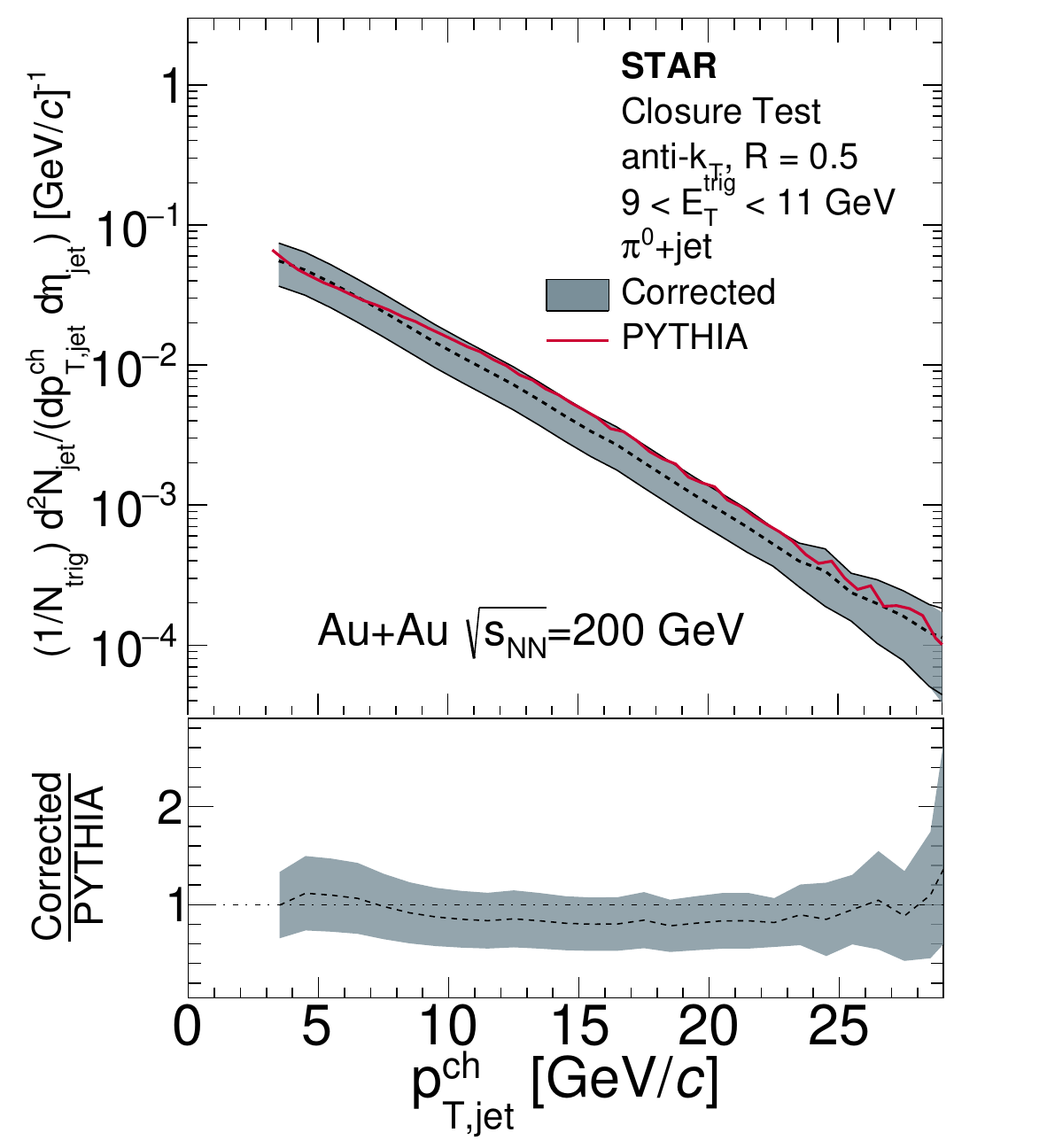}
\caption{Closure test for central \AuAu\ collisions: \pizero-triggered events with $9<\ETtrig<11$ GeV; recoil jets with $\rr=0.2$ (left) and 0.5 (right). Upper panel: semi-inclusive recoil-jet distributions. Red line shows PYTHIA truth distribution (particle-level). Lower panel: ratio of data and PYTHIA truth distributions.}
\label{Fig:AuAuClosureTest}
\end{figure}
%------

Figure~\ref{Fig:AuAuClosureTest} shows corrected semi-inclusive recoil-jet distributions for such hybrid events, for $\rr=0.2$ and 0.5. The ratio of corrected data to the PYTHIA truth distributions in the lower panel is consistent with unity within the systematic uncertainty, over the full reported range of \pTjetch. This agreement, corresponding to closure,  validates the analysis chain for central \AuAu\ collisions.
\section{Results}
\label{sect:Results}
 
This section reports the fully-corrected semi-inclusive recoil jet distributions measured in \pp\ and central \AuAu\ collisions, and compares them using the observables defined in Sect.~\ref{Sect:Observables}. Comparison to theoretical calculations are also presented. See the companion Letter~\cite{STAR:2023pal} for additional discussion.

%----------------------------------------------
\subsection{Corrected recoil \pTjetch\ distributions}
\label{sect:CorrDist}

%-----
\begin{figure}[htbp]
\centering
\includegraphics[width=0.75\textwidth]{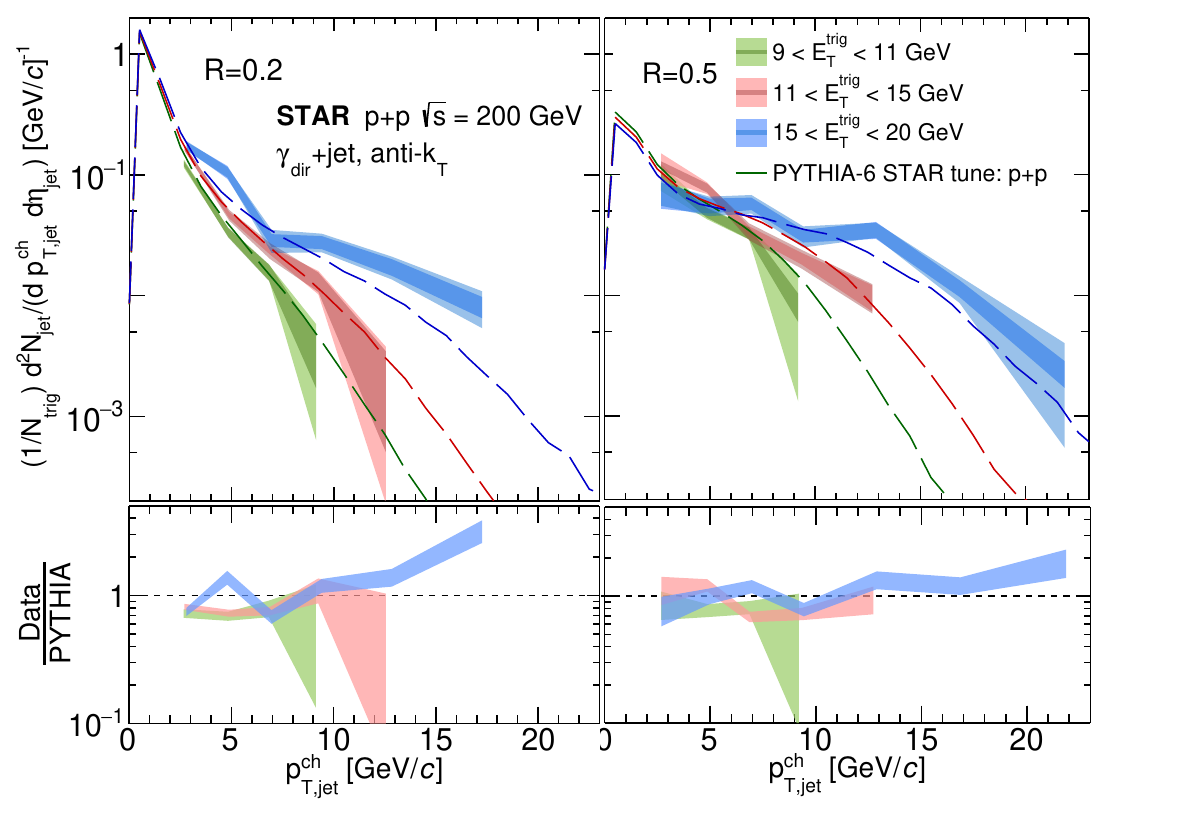}
\includegraphics[width=0.75\textwidth]{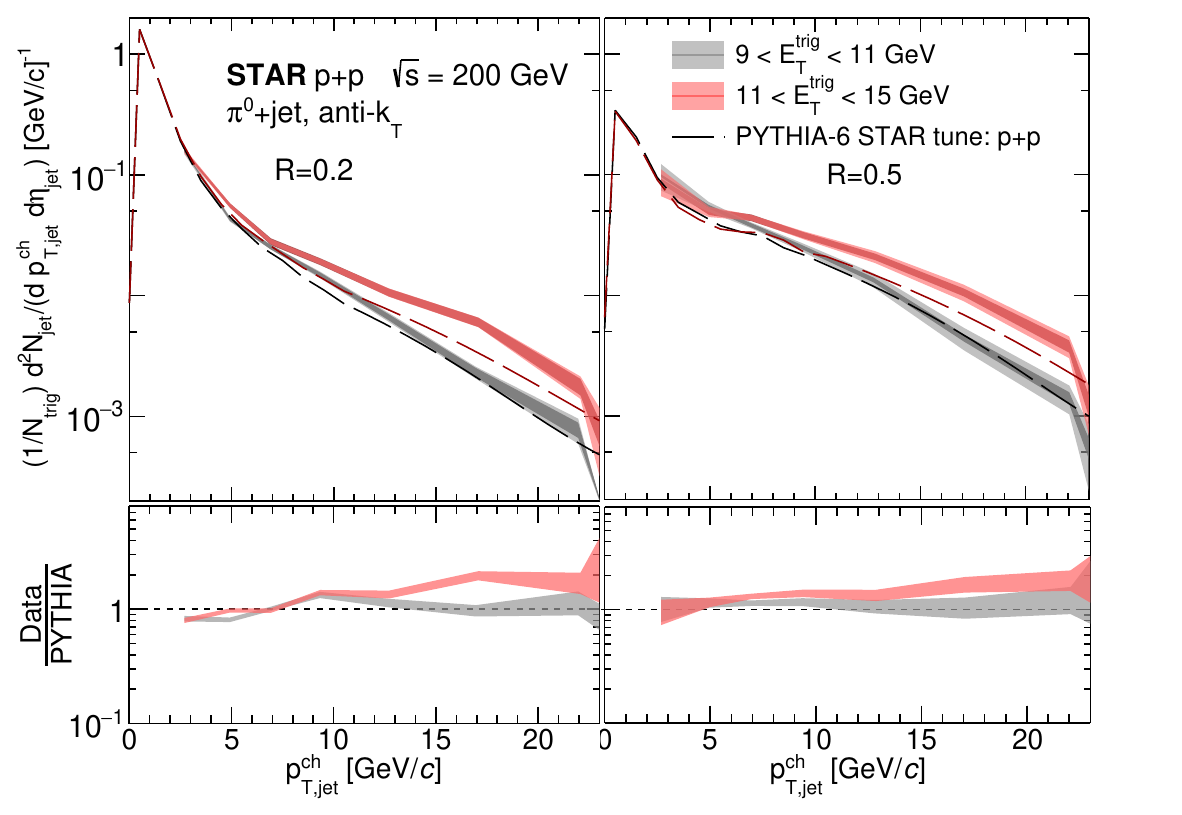}
\caption{Corrected semi-inclusive charged-particle recoil-jet distributions for \gammadir-triggered (upper) and \pizero-triggered (lower) \pp\ collisions at $\sqrts=200$ GeV, in \ETtrig\ bins. Left: $\rr=0.2$; right: $\rr=0.5$. Dark bands are statistical error; light bands are systematic uncertainty. The ratios in the lower sub-panels show the systematic uncertainties only. Dashed lines show the same distributions calculated using  PYTHIA 6, STAR tune~\cite{STAR:2019yqm}.}
\label{fig:CorrDistpp}
\end{figure}
%-----

Figure~\ref{fig:CorrDistpp} shows fully-corrected \pTjetch\ distributions for recoil jets with $\rr=0.2$ and 0.5, in \gammadir-triggered and \pizero-triggered \pp\ collisions at $\sqrts=200$ GeV. The distributions are shown as continuous bands rather than distinct points, to indicate the strong off-diagonal covariance of the systematic uncertainty~\cite{Adamczyk:2017yhe}. The underlying distributions are nevertheless binned, with horizontal coordinate in each bin determined by its weighted centroid~\cite{Lafferty:1994cj}.

Higher \ETtrig\ correlates with a harder recoil-jet spectrum, as expected if higher \ETtrig\ corresponds to larger momentum transfer in the partonic interaction. PYTHIA-6 STAR tune calculations are also shown~\cite{STAR:2019yqm}. For $\rr=0.2$, PYTHIA largely reproduces the measured distributions within $\approx30\%$, except for the \gammadir-triggered data for $15<\ETtrig<20$ GeV. For $\rr=0.5$, PYTHIA reproduces the data except for \pizero\ triggers in $11<\ETtrig<15$ GeV.

%---------------
\begin{figure}[htbp]
\centering
\includegraphics[width=0.75\textwidth]{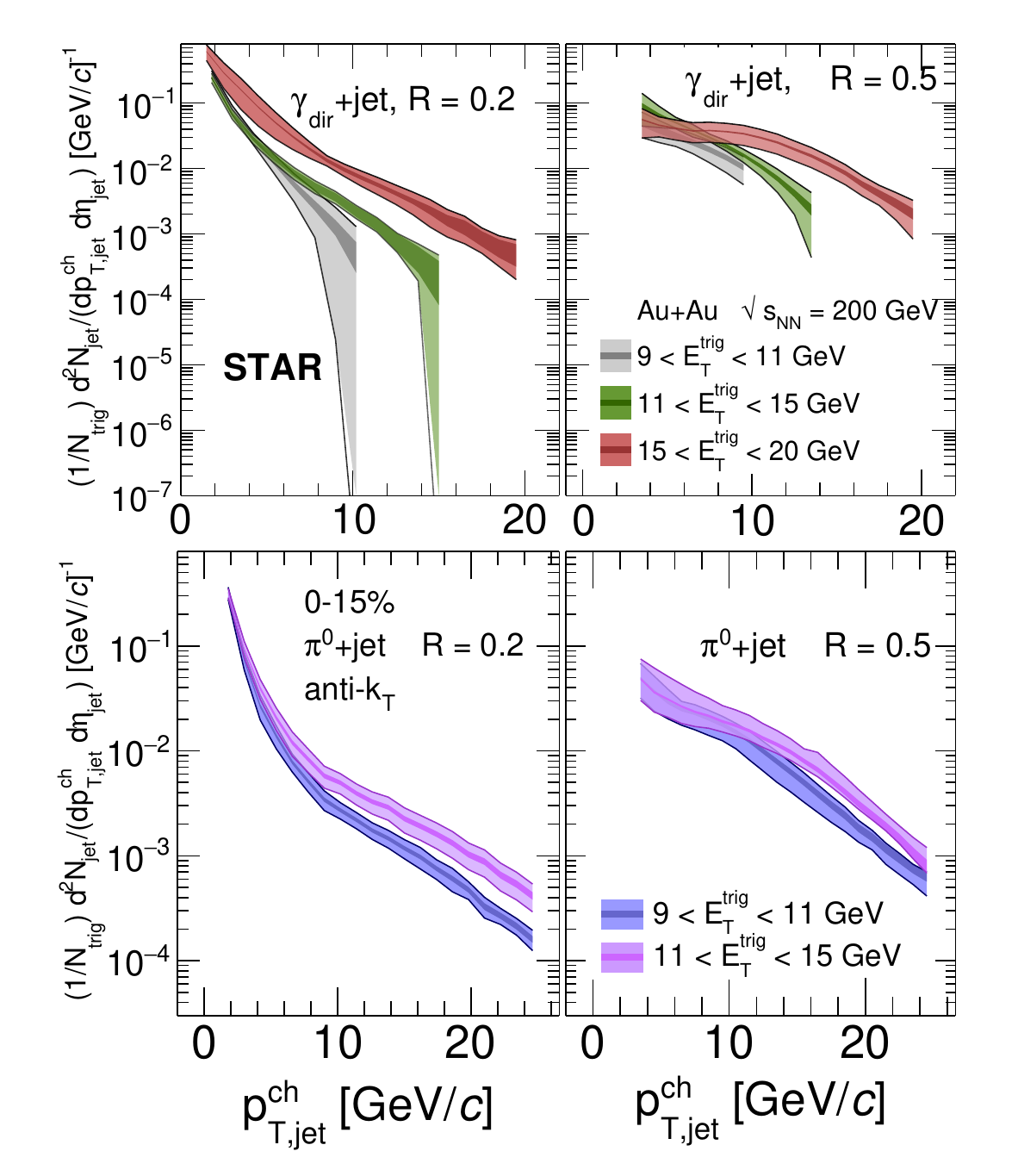}
\caption{Corrected semi-inclusive charged-particle recoil-jet distributions for \gammadir-triggered (upper) and \pizero-triggered (lower) \AuAu\ collisions at $\sqrtsNN=200$ GeV, in \ETtrig\ bins. Left: $\rr=0.2$; right: $\rr=0.5$. Dark bands are statistical error; light bands are systematic uncertainty. }
\label{fig:CorrDistAuAu}
\end{figure}
%---------------

Figure~\ref{fig:CorrDistAuAu} shows the same distributions for \AuAu\ collisions at $\sqrtsNN=200$ GeV. A hardening of the recoil spectra with larger \ETtrig\ is likewise observed, qualitatively similar to that for \pp\ collisions (Fig.~\ref{fig:CorrDistpp}). The recoil-jet spectra extend to larger \pTjetch\ for \pizero\ than for \gammadir\ triggers, as expected, due to the stronger constraint on kinematic balance for \gammadir\ triggers. 

The \pizero-triggered spectra for $15<\ETtrig<20$ GeV are not shown, due to their limited statistical precision arising from the stringent TSP cut required for high \pizero\ purity (Fig.~\ref{Fig:TSP}). While the \pizero-triggered distributions in this \ETtrig\ selection are sufficient for the background subtraction needed to obtain the \gammadir-triggered distributions, they are not precise enough for good convergence of the unfolding to obtain corrected \pizero-triggered spectra.

%---------------------------------------------------
\subsection{Recoil-jet yield modification: \IAA}
\label{subsect:IAA}
  
%----
\begin{figure}[htbp]
\centering
\includegraphics[width=0.99\textwidth]{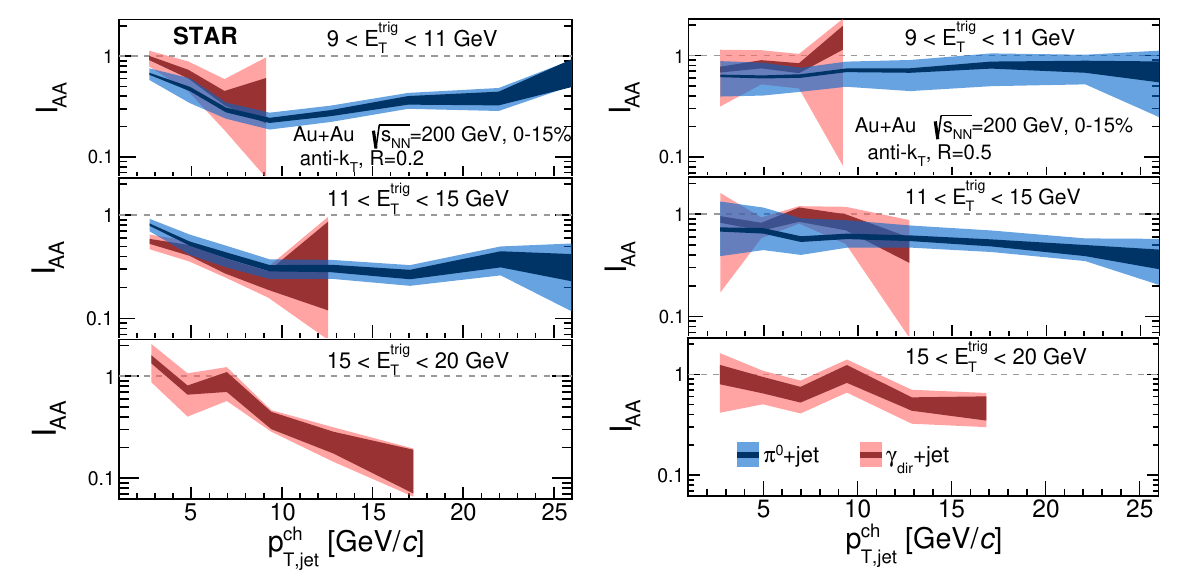}
\caption{\IAA\ for \gammadir- and \pizero-triggered recoil jet distributions measured in  \pp\ (Fig.~\ref{fig:CorrDistpp}) and central \AuAu\ collisions (Fig.~\ref{fig:CorrDistAuAu}) at $\sqrtsNN=200$ GeV. Left: $\rr=0.2$; right: $\rr=0.5$. Top: $9<\ETtrig<11$ GeV; middle: $11<\ETtrig<15$ GeV;  bottom: $15<\ETtrig<20$ GeV. }
\label{Fig:IAAdata}
\end{figure}  
%---- 
  
Figure~\ref{Fig:IAAdata} shows \IAA\ (Eq.~\ref{Eq:IAA}) as a function of \pTjetch\ for \gammadir- and \pizero-triggered recoil jet distributions. For $\rr=0.2$, \IAA\ in all cases decreases with increasing \pTjetch\ for $\pTjetch<10$ \gev\ and is significantly below unity for $\pTjetch>10$ \gev, indicating medium-induced jet energy loss. The value of \IAA\ is larger for $\rr=0.5$ than for $\rr=0.2$ for $\pTjetch>10$ \gev. These features suggest that energy of small-\rr\ jets at high \pTjetch\ is recovered at low \pTjetch\ and at large angles to the jet axis. 

In the upper and middle panels of Fig.~\ref{Fig:IAAdata}, the \IAA\ distributions for \gammadir\ and \pizero\ triggers are consistent within uncertainties. This is notable in light of the difference expected in the recoil-jet populations in terms of quark/gluon fraction and path-length distribution (Sect.~\ref{Sect:SemiInclDistr}).

The ATLAS collaboration has measured the medium-induced yield suppression in central \PbPb\ collisions at $\sqrtsNN=5.02$ TeV for an inclusive jet population and for a population of jets recoiling from a direct photon with $\ET>50$ GeV~\cite{ATLAS:2023iad}. Larger yield suppression is observed in the region $\pTjet>100$ \gev\ for the inclusive population, which has a larger fraction of gluon-initiated jets relative to photon-recoil jets. This observation is in contrast to the measurements shown in Fig.~\ref{Fig:IAAdata}, where similar yield suppression is found for two jet populations that are expected to comprise significantly different fractions of quark and gluon-initiated jets (Sect.~\ref{Sect:SemiInclDistr}).

Yield suppression corresponding to $\IAA<1$ arises from the combined effect of medium-induced jet energy loss and the jet spectrum shape.  Phenemenological analysis of the ATLAS measurements~\cite{ATLAS:2023iad} suggests that the mean energy loss is larger for the inclusive jet population, which has a larger fraction of gluon jets. For the STAR measurement presented here, Figs.~\ref{fig:CorrDistpp} and \ref{fig:CorrDistAuAu} show that the recoil-jet spectrum shape is steeper for \gammadir\ than for \pizero\ triggers in the same \ETtrig\ bin, so that a similar value of \IAA\ for the two triggers corresponds to larger mean energy loss for the \pizero-triggered recoil-jet population, which is likewise expected to have a larger fraction of gluon jets (Fig.~\ref{Fig:qgFrac}). Model calculations of these ATLAS and STAR measurements within a single theoretical framework, incorporating jet quenching and a realistic model of the QGP, would provide new insight into the role played by color charge in the physical processes underlying jet quenching.

%---------------------
\subsection{Jet shape modification: {\Rbroadening}}
\label{subsect:Jetshape}

%-----
\begin{figure}[htbp]
\centering
\includegraphics[width=0.8\textwidth]{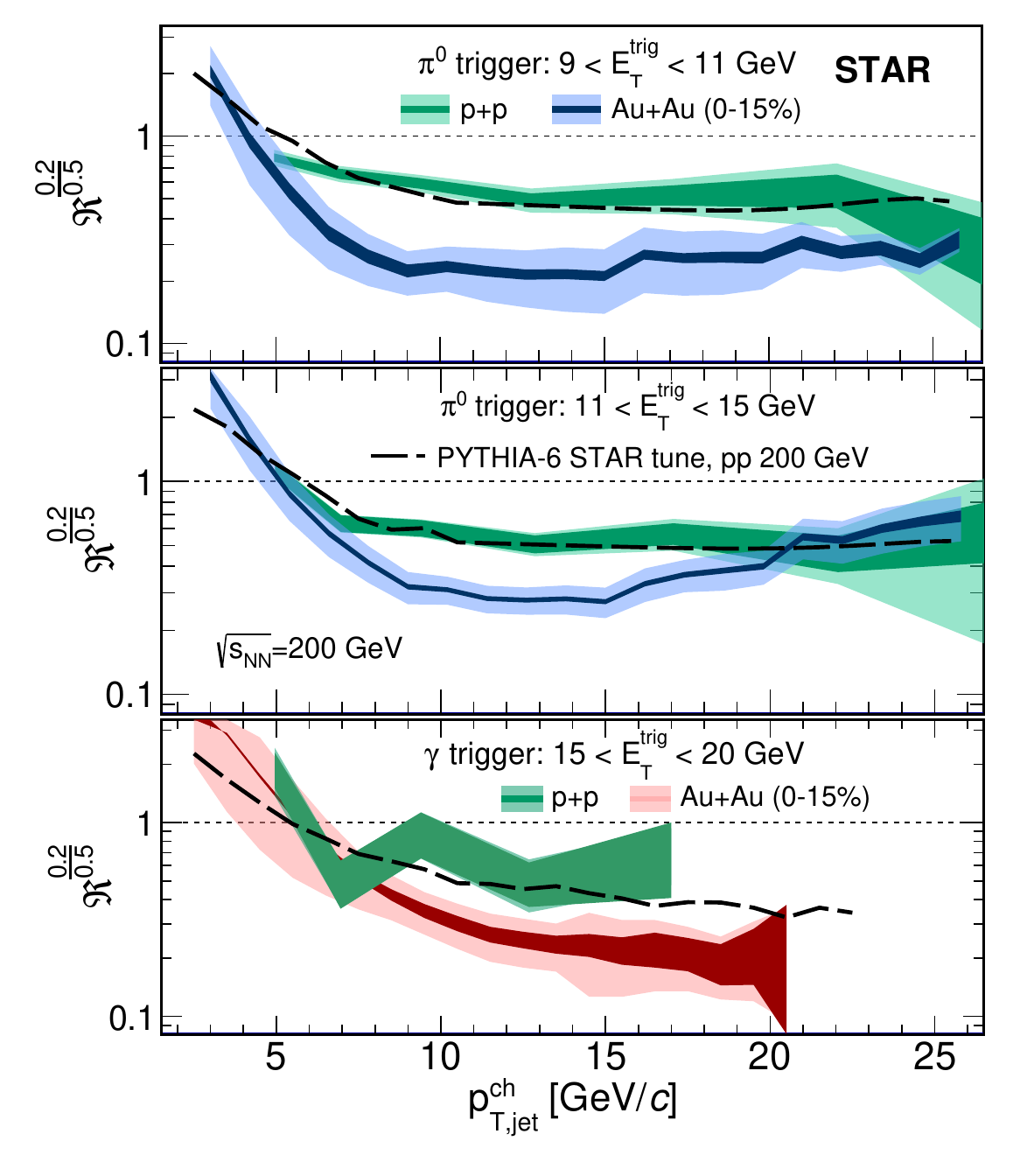}
\caption{Recoil-jet yield ratio \Rbroadening\ for $\rr=0.2$ and 0.5 measured in \pp\ and \AuAu\ collisions. Top panel: \pizero\ trigger, $9<\ETtrig<11$ GeV; middle panel: \pizero\ trigger, $11<\ETtrig<15$ GeV; bottom panel: \gammadir\ trigger, $15<\ETtrig<20$ GeV. Dark bands are statistical error; light bands are systematic uncertainty. Also shown are the same distributions calculated using PYTHIA-6 STAR tune for \pp\ collisions. }
\label{Fig:YieldRatioPi0Gamma}
\end{figure}
%-----

The observable \Rbroadening\ (Eq.~\ref{Eq:Intrajet}) probes the jet transverse profile. Figure~\ref{Fig:YieldRatioPi0Gamma} shows distributions of \Rbrtwofive\ measured with \pizero\ and \gammadir\ triggers in \pp\ and central \AuAu\ collisions. The systematic uncertainty of the ratio takes into account the correlated uncertainties in numerator and denominator. The systematic uncertainty due to instrumental effects largely cancels in the ratio. 

%\subsubsection{{\pp} collisions}

For \pp\ collisions, \Rbrtwofive\ is less than unity for $\pTjetch\gtrsim7$ \gev, reflecting the transverse energy profile of jets in vacuum. Figure~\ref{Fig:YieldRatioPi0Gamma} also shows distributions of \Rbrtwofive\ for \pp\ collisions calculated by PYTHIA-6 STAR tune~\cite{STAR:2019yqm}, which are consistent with the data within uncertainties. A similar level of agreement of PYTHIA calculations with data has been found for QCD and model calculations of \Rbrtwofive\ in \pp\ collisions for inclusive jet~\cite{ALICE:2013yva,ALICE:2019qyj,ALICE:2019wqv,ALICE:2023ama,CMS:2014nvq,CMS:2020caw,Dasgupta:2016bnd} and semi-inclusive recoil-jet production~\cite{Adam:2015doa,ALICE:2023jye}. 

The shape of the \Rbrtwofive\ distribution  in \pp\ collisions has been observed to differ markedly at low \pTjetch\ for inclusive and semi-inclusive jet production~\cite{ALICE:2023jye}. For inclusive jet production, \Rbroadening\  decreases as \pTjet\ decreases~\cite{ALICE:2019qyj,ALICE:2019wqv,ALICE:2023ama,Dasgupta:2016bnd}, reflecting the well-established broadening of the transverse jet profile for the inclusive jet population at low \pTjet. In contrast, Fig.~\ref{Fig:YieldRatioPi0Gamma} shows that, for these semi-inclusive distributions \Rbrtwofive\ increases as \pTjetch\ is reduced, with the increase well-described by the PYTHIA-6 STAR tune calculation, consistent with observations in Ref.~\cite{ALICE:2023jye}. QCD calculations and PYTHIA-6 likewise reproduce well the opposite trend for inclusive jet production~\cite{ALICE:2019qyj,ALICE:2019wqv,ALICE:2023ama}. 

We explore here whether these features can be accounted for by the consideration of elementary QCD processes, focusing on \Rbrtwofive\ for \pizero-triggered data which has larger dynamic range than the \gammadir-triggered data. The \pizero\ trigger is typically the leading fragment of the ``trigger jet,'' which is not measured in the analysis. However, \pTjet\ of the trigger jet is necessarily greater than \pTtrig\ of the trigger \pizero. In an LO picture, in which jet production is a $2\rightarrow2$ process, the recoil jet must therefore also have a value of \pTjet\ that is larger than \pTtrig. 

We therefore consider the observable introduced in Ref.~\cite{ALICE:2023jye},

\begin{equation}
\ztilde=\frac{\pTtrig}{\pTjetch}, 
\label{eq:ztilde}
\end{equation}

\noindent
which is the ratio of \pT\ of the \pizero\ trigger to \pTjetch\ of the recoil jet. Taking into accounting the fact that the recoil jet is measured with charged particles only, LO production corresponds predominantly to $\ztilde<1.5$. Initial-\kT\ effects will provide additional smearing of \ztilde, which we disregard here.

Figure~\ref{Fig:YieldRatioPi0Gamma} shows that the growth in \Rbrtwofive\ with decreasing \pTjetch\ occurs predominantly in the region $\ztilde>1.5$ (i.e. low \pTjetch), where LO production is suppressed. Higher-order production corresponds to additional gluon radiation, which has an angular distribution relative to the initiating-parton direction. Such additional radiation may cause jet reconstruction with $\rr=0.2$ to find two separate jets, while jet reconstruction with $\rr=0.5$ finds only one. 

This semi-inclusive measurement simply counts recoil jets in an event (Eq.~\ref{eq:hJetDefinition}), and the case of multiple correlated jets per trigger is therefore accounted for. This jet-splitting effect may account for the increase in \Rbrtwofive\ towards low \pTjetch\ in \pp\ collisions. The effect occurs only in the semi-inclusive measurement, where an external \pT\ scale is imposed by the value of \pTtrig, in contrast to inclusive jet production that has no such external scale.

Validation of this picture can be carried out using QCD calculations which exhibit the same features, and by additional jet substructure measurements. If validated, this mechanism may be used to generate a population of initially ``wide jets'' in heavy-ion collisions, in order to explore the interplay between jet substructure and jet quenching~\cite{Casalderrey-Solana:2019ubu}.

Figure~\ref{Fig:YieldRatioPi0Gamma} shows that for central \AuAu\ collisions, \Rbrtwofive\ is additionally suppressed relative to its value for \pp\ collisions for \pizero\ triggers in the range $7<\pTjetch<20$ \gev, and for \gammadir\ triggers in the range $\pTjetch>8$ \gev. The additional suppression corresponds to medium-induced jet broadening. This observation of intra-jet broadening, combined with the measurement of \IAA\ values close to unity for $\rr=0.5$ (Fig.~\ref{Fig:IAAdata}), suggests that the typical angular scale of medium-induced energy momentum transfer due to jet quenching is less than 0.5 radians. See the companion Letter~\cite{STAR:2023pal} for additional discussion.

%----------------------------------------------
\subsection{Comparison with theoretical calculations}

%---
\begin{figure*}[htb!]
\centering
\includegraphics[width=0.95\textwidth]{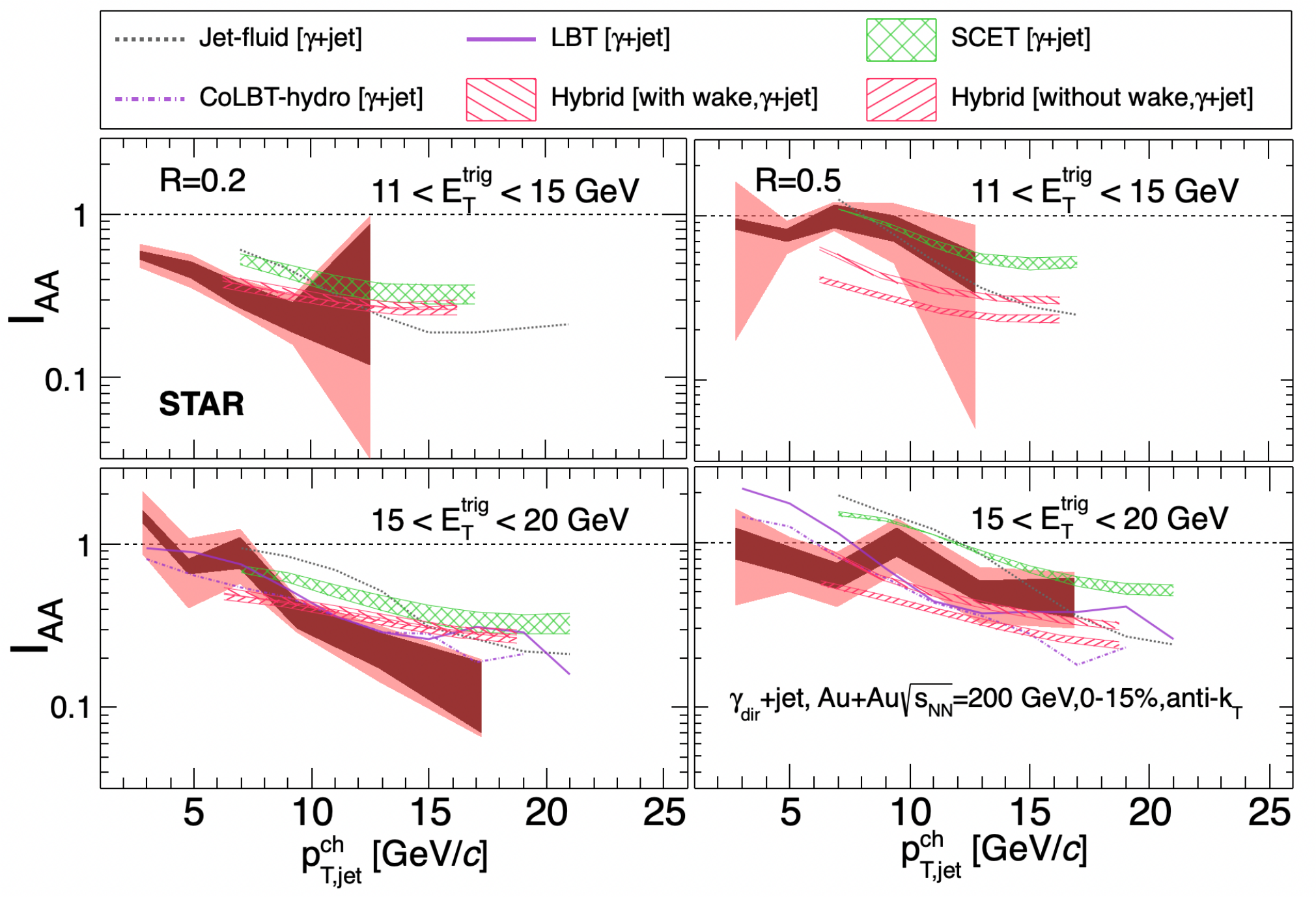}
\vspace{-2mm}
\caption{(Color online) Comparison of \gammadir-triggered \IAA\ distributions from Fig.~\ref{Fig:IAAdata} and the model calculations described in Sect.~\ref{Sec:Theory}.}  
\label{Fig:IAAgammadir}
\end{figure*}
\noindent
%----

Figure~\ref{Fig:IAAgammadir} shows \gammadir-triggered \IAA\ distributions from Fig.~\ref{Fig:IAAdata}, together with the model calculations described in Sect.~\ref{Sec:Theory}. The calculations are largely consistent with each other within a factor $\approx2$ for $\rr=0.2$, but exhibit significantly broader spread amongst themselves for $\rr=0.5$. This may arise from different implementations of medium response in the various models, which could preferentially affect larger \rr. 

For $\rr=0.2$, calculations agree well with data for $11<\ETtrig<15$ GeV but predict less yield suppression than observed for $15<\ETtrig<20$ GeV and \pTjetch>10 \gev. For $\rr=0.5$, all models except SCET predict greater yield suppression than observed for $11<\ETtrig<15$ GeV, while for $15<\ETtrig<20$ GeV the calculations are widely scattered and bracket the data. 

To summarize the level of agreement of the calculations with the \gammadir-triggered \IAA\ distributions, agreement is found in limited regions of phase space, but none of the calculations reproduces well the full set of measured distributions. However, these comparisons cannot identify directly the most significant modification needed for each model to be consistent with both these data and other related data, which requires broader community discussion.

%---
\begin{figure*}[htb!]
\centering
\includegraphics[width=1.0\textwidth]{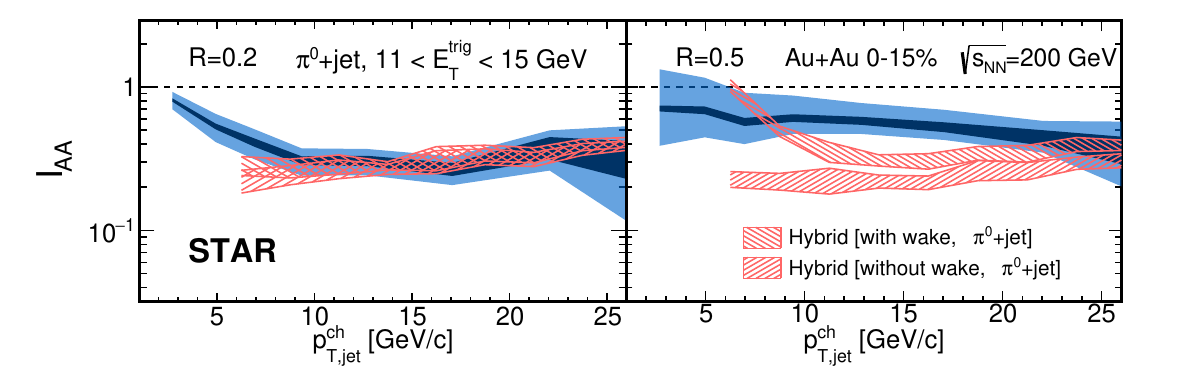}
\vspace{-2mm}
\caption{(Color online) Comparison of \pizero-triggered \IAA\ distributions (Fig.~\ref{Fig:IAAdata}) and theoretical calculations (Sect.~\ref{Sec:Theory}).}  
\label{Fig:IAApizero}
\end{figure*}
\noindent
%----

Figure~\ref{Fig:IAApizero} shows \pizero-triggered \IAA\ distributions from Fig.~\ref{Fig:IAAdata}, together with Hybrid model calculations described in Sect.~\ref{Sec:Theory}. Both model variants (wake and no wake) agree well with the data for $\rr=0.2$ but predict significantly larger medium-induced yield suppression for $\rr=0.5$. We return to this point below.

%---
\begin{figure}[htb!]
\centering 	 	
\includegraphics[width=0.95\textwidth]{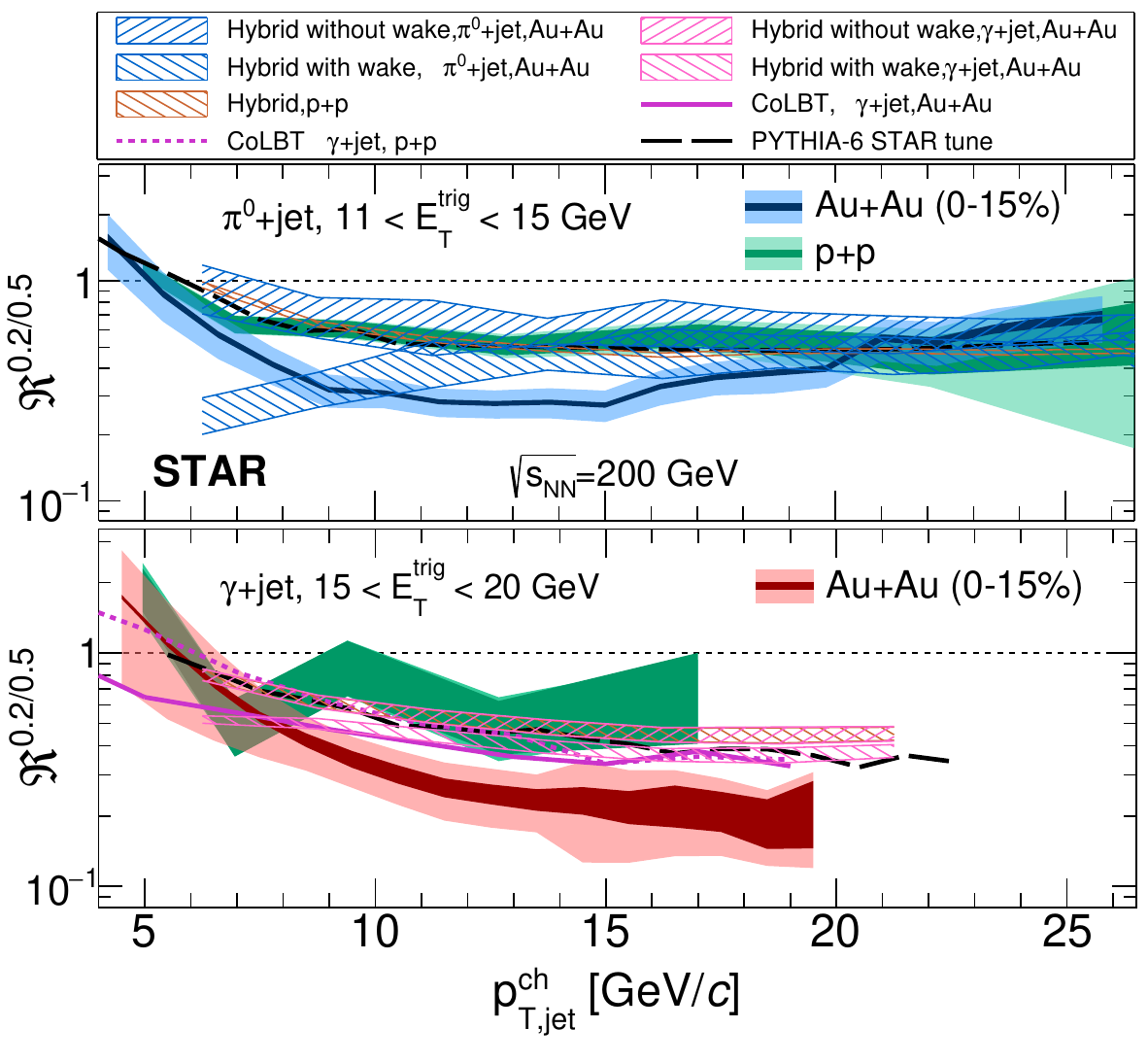}
\caption{(Color online) Selected \Rbrtwofive\ distributions from Fig.~\ref{Fig:YieldRatioPi0Gamma}, compared to theory calculations described in Sect.~\ref{Sec:Theory}.} 
\label{Fig:Rratio} 
\end{figure} 
%---

Figure~\ref{Fig:Rratio} shows \Rbrtwofive\ distributions for \gammadir\ and \pizero\ triggers in both \pp\ and central \AuAu\ collisions from Fig.~\ref{Fig:YieldRatioPi0Gamma}. Also shown are theory calculations incorporating jet quenching for central \AuAu\ collisions, described in Sect.~\ref{Sec:Theory}, and a PYTHIA calculation for \pp\ collisions. For $\pTjetch>10$ \gev, the jet quenching calculations are largely consistent with each other, and with the PYTHIA calculation for \pp\ collisions, indicating little to no in-medium modification of the \Rbrtwofive\ distribution predicted by the jet quenching models. 

However, the data are seen to exhibit significantly smaller values of \Rbrtwofive\ in central \AuAu\ compared to \pp\ collisions over much of the measured \pTjetch\ range, indicating significant in-medium broadening of jet transverse structure which is not captured by current jet quenching models. Similar intra-jet broadening due to quenching, which is not captured by the model calculations, is observed for \IAA\ with \pizero\ triggered in Fig.~\ref{Fig:IAApizero}.

%_______________________________________________________
\subsection{Comparison with previous results}
\label{sect: hJetcompare}

This section compares \pizero-triggered measurements from this analysis with previously published measurements of $h$+jet correlations in \AuAu\ collisions at $\sqrtsNN=200$ GeV~\cite{Adamczyk:2017yhe}. These analyses differ in their choice of centrality, trigger, and reference spectrum for measuring yield suppression. Nevertheless, they address similar physics questions, and such differences may have only secondary effect on the physics results. This comparison provides a cross-check of the two analyses.

%--------
\begin{figure}[htbp]
\centering
\includegraphics[width=0.70\textwidth]{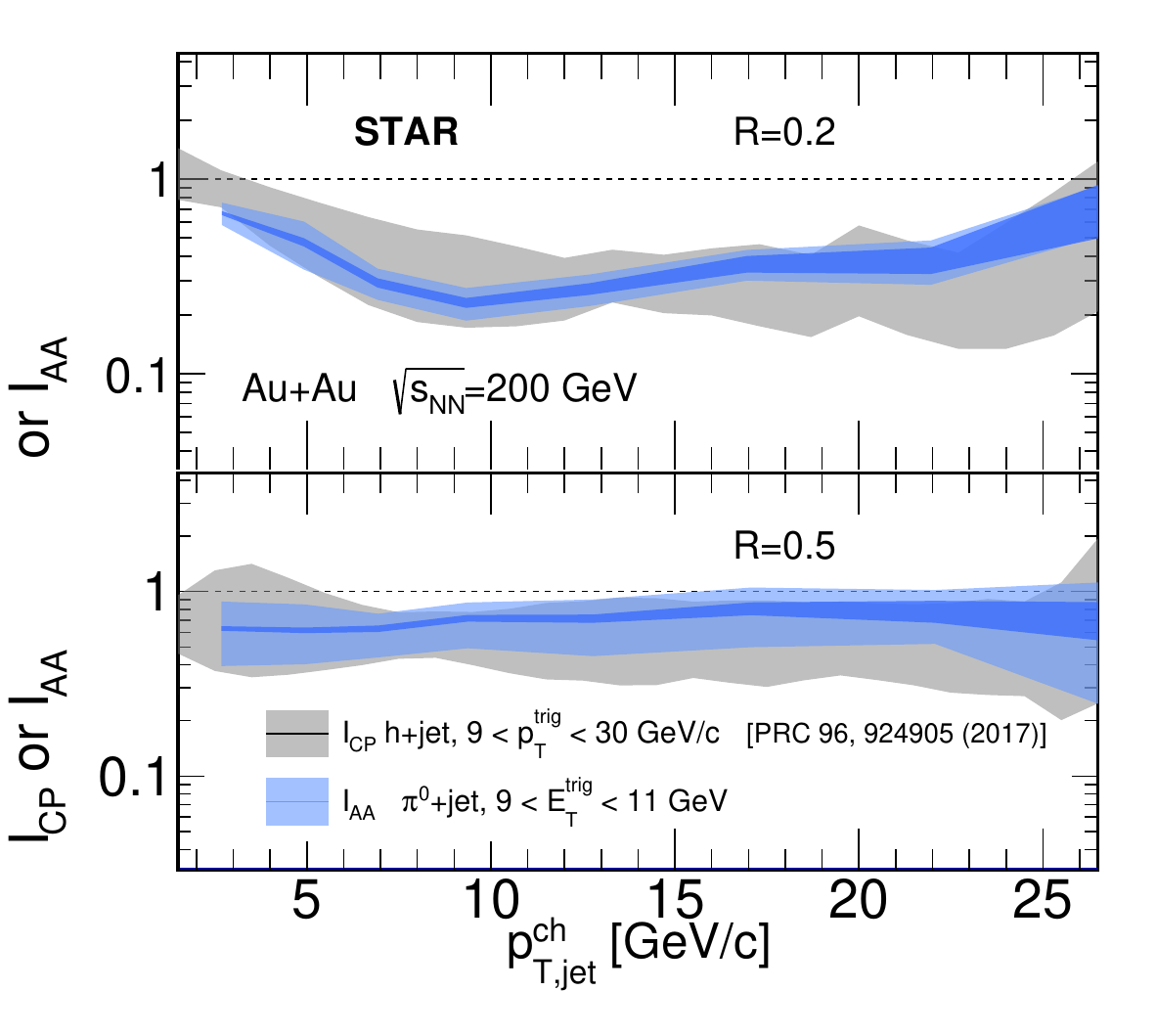}
\caption{Comparison of \IAA\ from \pizero-triggered distributions in this analysis with \ICP\ from hadron-triggered distributions~\cite{Adamczyk:2017yhe}, for \rr=0.2 (upper) and 0.5 (lower).}
\label{Fig:pi0hjetcompIAA}
\end{figure}
%--------

%--------
\begin{figure}[htbp]
\centering
\includegraphics[width=0.70\textwidth]{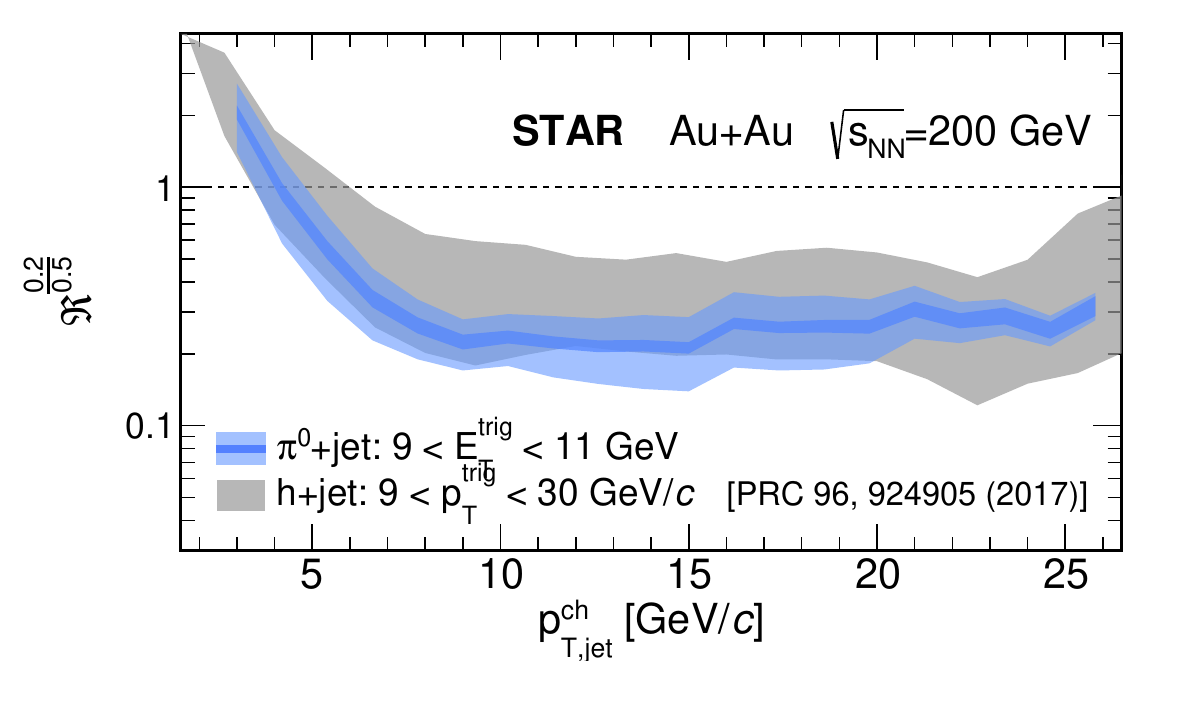}
\caption{Same as Fig.~\ref{Fig:pi0hjetcompIAA} for \Rbrtwofive.
}
\label{Fig:pi0hjetcompRbroad}
\end{figure}
%--------

Figure~\ref{Fig:pi0hjetcompIAA}  compares the \pizero-triggered distribution of \IAA\   with  \ICP\ (ratio of recoil-jet distributions in central and peripheral \AuAu\ collisions) from Ref.~\cite{Adamczyk:2017yhe}, for $\rr=0.2$ and 0.5, while Fig.~\ref{Fig:pi0hjetcompRbroad} compares \Rbrtwofive\ in the two analyses. The systematic uncertainty and statistical error are significantly smaller in this analysis, for both observables. The two analyses are consistent within uncertainties.
\section{Summary}
\label{sect:Summary}

We have reported the first measurement of semi-inclusive distributions of charged-particle jets recoiling from \gammadir\ and \pizero\ triggers in \pp\ and central \AuAu\ collisions at $\sqrtsNN=200$ GeV. Fully-corrected recoil-jet distributions are measured for $\rr=0.2$ and 0.5, with $9<\ETtrig<20$ GeV. The \pizero-triggered measurements are consistent with results in a previous publication that utilized charged-hadron triggers in central and peripheral \AuAu\ collisions, with significantly improved uncertainties.

Recoil-jet yields in \pp\ collisions are broadly described by QCD-based calculations using the standard PYTHIA event generator, with deviations of up to 30\% found in limited regions of phase--space. Marked medium-induced recoil-jet yield suppression is observed for $\rr=0.2$ for both triggers, with significantly less suppression observed for $\rr=0.5$. Measurements of the transverse structure of jets using this \rr-dependence provide new insight into the angular redistribution of jet energy due to quenching processes.

The semi-inclusive recoil yield in \pp\ collisions is observed to have significantly different \pTjetch-dependence than that of the inclusive jet population at low \pTjetch, similar to a recent measurement at the LHC. This phenomenon provides a new, QCD-based tool to engineer initial jet shapes for the further exploration of jet quenching.

Comparison to theoretical calculations of jet quenching reveals that current models do not correctly describe the broadening of intra-jet structure due to quenching. The reported data provide significant new constraints on the theoretical description of in-medium jet modification.
\section{Acknowledgments}
\label{sect:Acknowledge}

{\it Acknowledgments- } We thank
Shanshan Cao,
Tan Luo, Guang-You Qin,
Abhijit Majumder,
Daniel Pablos,
Krishna Rajagopal,
Chathuranga Sirimanna,
Xin-Nian Wang,
and Ivan Vitev
for providing theoretical calculations.
%STAR acknowledgement
We thank the RHIC Operations Group and RCF at BNL, the NERSC Center at LBNL, and the Open Science Grid consortium for providing resources and support.  This work was supported in part by the Office of Nuclear Physics within the U.S. DOE Office of Science, the U.S. National Science Foundation, National Natural Science Foundation of China, Chinese Academy of Science, the Ministry of Science and Technology of China and the Chinese Ministry of Education, the Higher Education Sprout Project by Ministry of Education at NCKU, the National Research Foundation of Korea, Czech Science Foundation and Ministry of Education, Youth and Sports of the Czech Republic, Hungarian National Research, Development and Innovation Office, New National Excellency Programme of the Hungarian Ministry of Human Capacities, Department of Atomic Energy and Department of Science and Technology of the Government of India, the National Science Centre and WUT ID-UB of Poland, the Ministry of Science, Education and Sports of the Republic of Croatia, German Bundesministerium f\"ur Bildung, Wissenschaft, Forschung and Technologie (BMBF), Helmholtz Association, Ministry of Education, Culture, Sports, Science, and Technology (MEXT) and Japan Society for the Promotion of Science (JSPS).

\newpage

\bibliographystyle{apsrev4-2}

% pmj Jan 8: use same bib file as long paper
%\bibliography{GammaJet_references}

\bibliography{references}

%=========================================================
 
\end{document}

%% file: authorlist_short.tex
%_____________________________Author list and affilitions 

\affiliation{Academia Sinica}
\affiliation{Abilene Christian University, Abilene, Texas   79699}
\affiliation{AGH University of Krakow, FPACS, Cracow 30-059, Poland}
\affiliation{Argonne National Laboratory, Argonne, Illinois 60439}
\affiliation{American University in Cairo, New Cairo 11835, Egypt}
\affiliation{Ball State University, Muncie, Indiana, 47306}
\affiliation{Brookhaven National Laboratory, Upton, New York 11973}
\affiliation{University of Calabria \& INFN-Cosenza, Rende 87036, Italy}
\affiliation{University of California, Berkeley, California 94720}
\affiliation{University of California, Davis, California 95616}
\affiliation{University of California, Los Angeles, California 90095}
\affiliation{University of California, Riverside, California 92521}
\affiliation{Central China Normal University, Wuhan, Hubei 430079 }
\affiliation{University of Illinois at Chicago, Chicago, Illinois 60607}
\affiliation{Chongqing University, Chongqing, 401331}
\affiliation{Creighton University, Omaha, Nebraska 68178}
\affiliation{Czech Technical University in Prague, FNSPE, Prague 115 19, Czech Republic}
\affiliation{Technische Universit\"at Darmstadt, Darmstadt 64289, Germany}
\affiliation{National Institute of Technology Durgapur, Durgapur - 713209, India}
\affiliation{ELTE E\"otv\"os Lor\'and University, Budapest, Hungary H-1117}
\affiliation{Frankfurt Institute for Advanced Studies FIAS, Frankfurt 60438, Germany}
\affiliation{Fudan University, Shanghai, 200433 }
\affiliation{Guangxi Normal University, Guilin, 541004}
\affiliation{University of Heidelberg, Heidelberg 69120, Germany }
\affiliation{University of Houston, Houston, Texas 77204}
\affiliation{Huzhou University, Huzhou, Zhejiang  313000}
\affiliation{Indian Institute of Science Education and Research (IISER), Berhampur 760010 , India}
\affiliation{Indian Institute of Science Education and Research (IISER) Tirupati, Tirupati 517507, India}
\affiliation{Indian Institute Technology, Patna, Bihar 801106, India}
\affiliation{Indiana University, Bloomington, Indiana 47408}
\affiliation{Institute of Modern Physics, Chinese Academy of Sciences, Lanzhou, Gansu 730000 }
\affiliation{University of Jammu, Jammu 180001, India}
\affiliation{Kent State University, Kent, Ohio 44242}
\affiliation{University of Kentucky, Lexington, Kentucky 40506-0055}
\affiliation{Lanzhou University,Lanzhou, Gansu, 730000}
\affiliation{Lawrence Berkeley National Laboratory, Berkeley, California 94720}
\affiliation{Lehigh University, Bethlehem, Pennsylvania 18015}
\affiliation{Max-Planck-Institut f\"ur Physik, Munich 80805, Germany}
\affiliation{Michigan State University, East Lansing, Michigan 48824}
\affiliation{National Institute of Science Education and Research, HBNI, Jatni 752050, India}
\affiliation{National Cheng Kung University, Tainan 70101 }
\affiliation{Nuclear Physics Institute of the CAS, Rez 250 68, Czech Republic}
\affiliation{The Ohio State University, Columbus, Ohio 43210}
\affiliation{Panjab University, Chandigarh 160014, India}
\affiliation{Purdue University, West Lafayette, Indiana 47907}
\affiliation{Rice University, Houston, Texas 77251}
\affiliation{Rutgers University, Piscataway, New Jersey 08854}
\affiliation{University of Science and Technology of China, Hefei, Anhui 230026}
\affiliation{South China Normal University, Guangzhou, Guangdong 510631}
\affiliation{Sejong University, Seoul, 05006, South Korea}
\affiliation{Shandong University, Qingdao, Shandong 266237}
\affiliation{Shanghai Institute of Applied Physics, Chinese Academy of Sciences, Shanghai 201800}
\affiliation{Southern Connecticut State University, New Haven, Connecticut 06515}
\affiliation{State University of New York, Stony Brook, New York 11794}
\affiliation{Instituto de Alta Investigaci\'on, Universidad de Tarapac\'a, Arica 1000000, Chile}
\affiliation{Temple University, Philadelphia, Pennsylvania 19122}
\affiliation{Texas A\&M University, College Station, Texas 77843}
\affiliation{University of Texas, Austin, Texas 78712}
\affiliation{Tsinghua University, Beijing 100084}
\affiliation{University of Tsukuba, Tsukuba, Ibaraki 305-8571, Japan}
\affiliation{University of Chinese Academy of Sciences, Beijing, 101408}
\affiliation{United States Naval Academy, Annapolis, Maryland 21402}
\affiliation{Valparaiso University, Valparaiso, Indiana 46383}
\affiliation{Variable Energy Cyclotron Centre, Kolkata 700064, India}
\affiliation{Warsaw University of Technology, Warsaw 00-661, Poland}
\affiliation{Wayne State University, Detroit, Michigan 48201}
\affiliation{Wuhan University of Science and Technology, Wuhan, Hubei 430065}
\affiliation{Yale University, New Haven, Connecticut 06520}

\author{B.~E.~Aboona}\affiliation{Texas A\&M University, College Station, Texas 77843}
\author{J.~Adam}\affiliation{Czech Technical University in Prague, FNSPE, Prague 115 19, Czech Republic}
\author{L.~Adamczyk}\affiliation{AGH University of Krakow, FPACS, Cracow 30-059, Poland}
\author{I.~Aggarwal}\affiliation{Panjab University, Chandigarh 160014, India}
\author{M.~M.~Aggarwal}\affiliation{Panjab University, Chandigarh 160014, India}
\author{Z.~Ahammed}\affiliation{Variable Energy Cyclotron Centre, Kolkata 700064, India}
\author{D. M. Anderson}\affiliation{Texas A\&M University, College Station, Texas 77843}
\author{E.~C.~Aschenauer}\affiliation{Brookhaven National Laboratory, Upton, New York 11973}
\author{S.~Aslam}\affiliation{Indian Institute Technology, Patna, Bihar 801106, India}
\author{J.~Atchison}\affiliation{Abilene Christian University, Abilene, Texas   79699}
% ...
